\documentclass[useAMS,usenatbib]{mn2e}
\usepackage{epsfig}

\def\eSF{\mbox{$\epsilon_{\rm SF}$}}
\def\dML{\mbox{$\nabla_{\ell} \Upsilon$}}
\def\Re{\mbox{$R_{\rm eff}$}}
\def\sig{\mbox{$\sigma$}}
\def\sigc{\mbox{$\sigma_0$}}

\def\Ie{\mbox{$I_{\rm eff}$}}
\def\Msun{\mbox{$M_\odot$}}
\def\Lsun{\mbox{$L_\odot$}}
\def\Ysun{\mbox{$\Upsilon_\odot$}}
\def\ML{\mbox{$M/L$}}
\def\Yst{\mbox{$\Upsilon_{*}$}}
\def\YstB{\mbox{$\Upsilon_{*,B}$}}
\def\Ydyn{\mbox{$\Upsilon_{\rm dyn}$}}

\def\mst{\mbox{$M_{*}$}}
\def\age{\mbox{$t_{\rm gal}$}}
\def\Zsun{\mbox{$Z_{\odot}$}}
\def\LB{\mbox{$L_{B}$}}

\def\drot{\mbox{$\delta_{\rm rot}$}}

\def\MB{\mbox{$M_{B}$}}
\def\vmax{\mbox{$V_{\rm max}$}}
\def\lsim{\mathrel{\rlap{\lower3.5pt\hbox{\hskip0.5pt$\sim$}}
    \raise0.5pt\hbox{$<$}}}                
\def\gsim{~\rlap{$>$}{\lower 1.0ex\hbox{$\sim$}}}
\def\MLga{\mbox{$\ML\propto\LB^\gamma$}}
\def\fDMRe{\mbox{$f_{\rm DM}(\Re)$}}
\def\fDM{\mbox{$f_{\rm DM}$}}

\def\mtot{\mbox{$M_{\rm tot}$}}
\def\kms{\mbox{\,km~s$^{-1}$}}
\def\rhoDM{\mbox{$\langle \rho_{\rm DM} \rangle$}}
\def\Lcrit{\mbox{$\log\LB\sim 10.4 \, \Lsun$}}
\def\Mcrit{\mbox{$\log\mst\sim 11 \, \Msun$}}

\title[Central $M/L$ and DM fraction in early-type galaxies]{Central mass-to-light ratios and dark matter fractions in early-type galaxies}

\author[Tortora et al.]{\noindent
C.~Tortora$^{1,2}$\thanks{E-mail: ctortora@na.astro.it},
N.R.~Napolitano$^{2}$, A.J.~Romanowsky$^{3}$,
M.~Capaccioli$^{4,5}$, G.~Covone$^{4,6}$
\\~\\
$^1$ INAF -- Osservatorio Astrofisico di Catania, Via S. Sofia 78,
I-95123 - Catania, Italy \\
$^2$ INAF -- Osservatorio Astronomico di
Capodimonte, Salita Moiariello, 16, 80131 - Napoli, Italy\\
$^3$ UCO/Lick Observatory, University of California, Santa Cruz,
CA 95064, USA\\
$^4$ Dipartimento di Scienze Fisiche, Universit\`{a} di Napoli
Federico
II, Compl. Univ. Monte S. Angelo, 80126 - Napoli, Italy\\
$^5$ INAF -- VSTceN, Salita Moiariello 16, 80131 - Napoli, Italy\\
$^6$ INFN -- Sezione di Napoli\\}

\begin{document}
\date{Accepted  Received }
\pagerange{\pageref{firstpage}--\pageref{lastpage}} \pubyear{xxxx}
\maketitle

\label{firstpage}
\begin{abstract}
Dynamical studies of local elliptical galaxies and the Fundamental
Plane point to a strong dependence of the total mass-to-light
ratio on luminosity with a relation of the form $M/L \propto
L^{\gamma}$. The ``tilt'' $\gamma$ may be caused by various
factors, including stellar population properties (metallicity, age
and star formation history), IMF, rotational support, luminosity
profile non-homology and dark matter (DM) fraction. We evaluate
the impact of all these factors using a large uniform dataset of
local early-type galaxies from \citet{PS96}. We take particular
care in estimating the stellar masses, using a general star
formation history, and comparing different population synthesis
models. We find that the stellar \ML\ contributes little to the
tilt. We estimate the total \ML\ using simple Jeans dynamical
models, and find that adopting accurate luminosity profiles is
important but does not remove the need for an additional tilt
component, which we ascribe to DM. We survey trends of the DM
fraction within one effective radius, finding it to be roughly
constant for galaxies fainter than $M_B \sim -20.5$, and
increasing with luminosity for the brighter galaxies; we detect no
significant differences among S0s and fast- and slow-rotating
ellipticals. We construct simplified cosmological mass models and
find general consistency, where the DM transition point is caused
by a change in the relation between luminosity and effective
radius. A more refined model with varying galaxy star formation
efficiency suggests a transition from total mass profiles
(including DM) of faint galaxies distributed similarly to the
light, to near-isothermal profiles for the bright galaxies. These
conclusions are sensitive to various systematic uncertainties
which we investigate in detail, but are consistent with the
results of dynamics studies at larger radii.
\end{abstract}


\section{Introduction} \label{sec:intro}

Early-type galaxies (ETGs) are the most massive stellar systems in
the Universe, containing much of the cosmic budget of visible and
dark matter (DM). They include elliptical (E) and lenticular (S0)
galaxies and form a nearly uniform class of objects: usually red,
old and with only traces of cold gas and active star formation.
The striking regularities in their properties include strong
correlations between size (e.g. the effective radius, \Re) and the
surface brightness therein (\Ie; \citealt{1977ApJ...218..333K}),
and between kinematics (the central velocity dispersion
$\sigma_0$) and luminosity ($L$; \citealt[hereafter FJ]{FJ76}).

The two relations above merge into the so-called Fundamental Plane
(FP; \citealt{DD87, Dressler87}), i.e. a relation between the
(logarithm of) \sigc, \Re\ and \Ie\ of ETGs. The FP can be
interpreted in terms of the virial theorem of relaxed systems,
according to which $2T+U=0$ where $U$ is the potential energy and
$T$ the kinetic energy. This can be re-written in terms of
observed quantities as approximately $L \propto \sigma^{2} \Re$.
However, the FP is found observationally to be $L \propto
\sig^{\eta} \Re^{\alpha}$ with $\alpha \neq 1$ and $\eta \neq 2$,
i.e. a different orientation of the plane in the space of the
logarithmic quantities with respect to the virial prediction. This
\emph{tilt} of the FP provides insight for the formation and
structure of ETGs, and can be interpreted as a variation of the
total mass-to-light ratio ($M/L$) with $L$ (\citealt{Dressler87})
with the simplest parameterization as a power law, $M/L \propto
L^{\gamma}$. The slope, $\gamma$, of this relation could be driven
by one or more different factors: a variation in stellar $M/L$
(due to metallicity or age gradient or change in IMF), a variation
in the DM content, non-homology, rotational support, etc. (see
e.g. \citealt{1997A&A...320..415B,2006NewAR..50..447D,Graves09}).
It is of considerable importance to disentangle these factors
using high-quality data at low redshift, in order to use the FP as
a guide to galaxy evolution in different environments and cosmic
epochs (e.g.
\citealt{2000ApJ...543..131K,Bernardi2003,2005MNRAS.360..693R,2007ApJ...655...30V}).

Many studies over the years have attempted to decode the FP tilt
(e.g.
\citealt{RC93,1995ApJ...445...55H,1995ApJ...453L..17P,Pahre98,1998AJ....116.1606P};
\citealt[hereafter PS96]{PS96};
\citealt{1997MNRAS.287..221G,Graham98,1998MNRAS.301.1001S,1999MNRAS.304..225M,Bertin2002,2002MNRAS.332..901N,
2005A&A...443..133R,dSA06,2007ApJ...665L.105B,2008ApJ...684..248B,Gargiulo09}).
The emerging consensus is that stellar populations account for a
minor fraction of the tilt (e.g. \citealt[hereafter
T+04]{TBB2004}; \citealt[hereafter C+06]{Cappellari06};
\citealt{2008MNRAS.386.1781P,2008arXiv0807.3829L,Graves09}; see
however \citealt{2008ApJ...678L..97J}), with the major contributor
yet to be firmly identified--which would have ramifications for
galaxy formation models (e.g.
\citealt{1995ApJ...451..525C,1996MNRAS.280L..13L,1997MNRAS.284..327K,1998ApJ...496..713B,2000MNRAS.316..786F,2000ApJ...545..181M,2002MNRAS.335..335C,2003MNRAS.340..398D,borr+03,2003MNRAS.342L..36G,2003MNRAS.342..501N,2004MNRAS.349.1052E,2005MNRAS.360L..50A,2005ApJ...632L..57O,2006MNRAS.373..503O,2005MNRAS.362..184B,2006MNRAS.369.1081B,2006MNRAS.370.1445D,2006ApJ...641...21R,2006ApJ...643...14S,2007MNRAS.376.1711A,2008arXiv0806.3974H}).

An additional complication is that the tilted FP may not be flat,
with claims made for curvature
\citep{2006ApJ...638..725Z,2008ApJ...685..875D,HydeBern08}, and
projections of the FP showing a bend at a characteristic magnitude
of $M_B \sim -20.5$ (see Section~\ref{sec:sample}). This
transition concords with more general findings of a discontinuity
in ETG properties at a similar luminosity (e.g.
\citealt{2008IAUS..244..289N,Coccato09}).

Of particular interest is the FP contribution from the central DM
content, since connecting luminous galaxies to their DM haloes is
one of the key ingredients in modern recipes for galaxy assembly.
The DM component of the FP tilt is currently a bone of contention,
with findings alternatively of negligible impact
(\citealt[hereafter G+01]{Gerhard01}; T+04), and of primary
importance (\citealt{Padmanabhan04,2006ApJ...648..826L}; C+06).
More generally, several lines of evidence point to a small but
significant fraction of DM ($f_{\rm DM}$) inside 1~\Re{} (e.g.
G+01; \citealt{2007ApJ...667..176G,2008MNRAS.383.1343W}), but the
galaxies probed are typically very bright Es with $\sigma_0 >
200$~\kms. As data have become available on more ``ordinary''
ETGs, there are suggestions that their DM properties vary strongly
with luminosity, with perhaps even a dichotomy following the
classic division of faint, discy, fast-rotating galaxies and
bright, boxy, slow-rotators (\citealt{Cap+02,R+03};
\citealt[hereafter N+05]{Nap05};
\citealt{2005ApJ...623L...5F,2008MNRAS.383..857F}; C+06;
\citealt{vdeB07,D+07,Covone09,Nap08,Ruszkowski+09}).

The purpose of this paper is to reexamine all the plausible factors
that could contribute to the FP tilt.  We will not be directly
studying the FP, but rather will survey systematic trends in the
central properties of ETGs which factor into the apparent
$M/L$ variation, $\gamma$.
Our study springs from the classic data-set of central photometry and
kinematics of PS96, which is one of the largest such catalogues
of local galaxies covering a wide range of luminosities.
Our analysis consists of two parts, wherein we independently determine
the stellar component of the $M/L$ by stellar populations models,
and the total $M/L$ by dynamical models.
We take particular care in considering realistic star formation histories,
stellar populations modelling systematics, and dynamical contributions from DM.

We briefly describe the galaxy sample in Section~\ref{sec:sample}
and analyze the stellar populations in
Section~\ref{sec:MtoL_star}. We determine the dynamical masses in
Section~\ref{sec:dynM} and infer the DM fractions in
Section~\ref{sec:DM}. Section~\ref{sec:av_DM} considers some
implications for galaxy formation, and
Section~\ref{sec:conclusions} draws conclusions and considers
future prospects.
\begin{figure*}
\epsfig{file=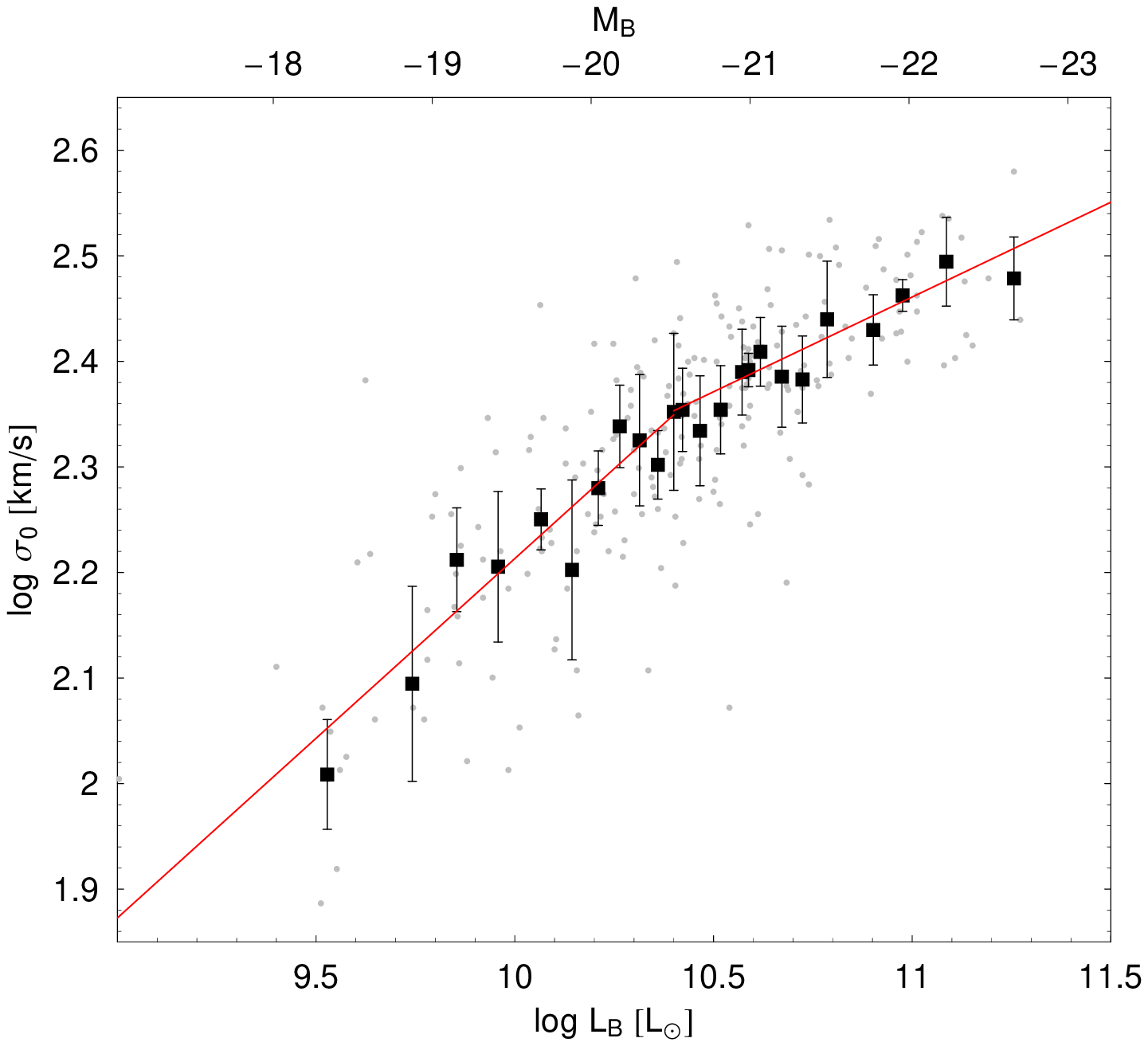, width=0.49\textwidth}\hspace{0.2cm}
\epsfig{file= 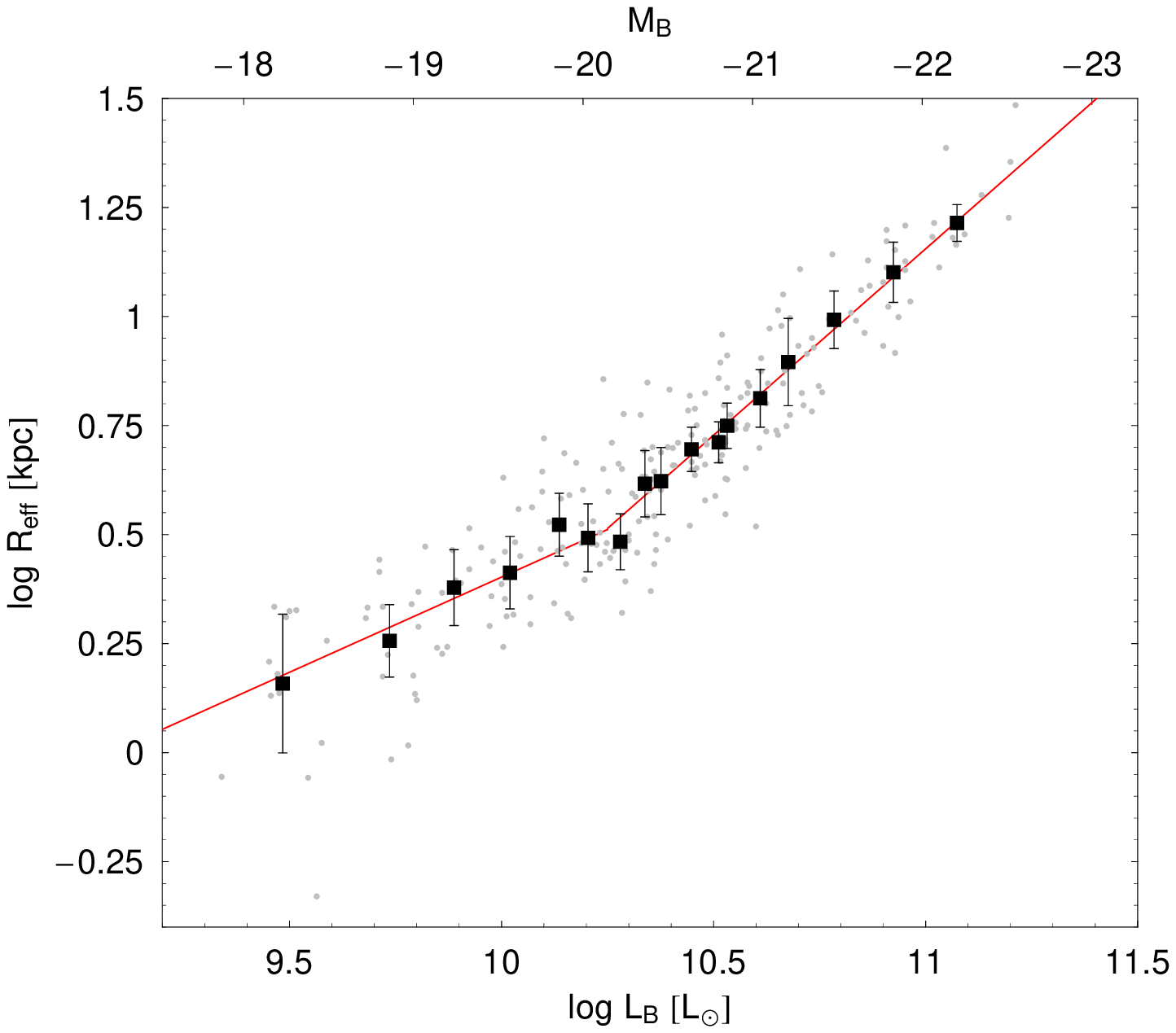, width=0.49\textwidth}
\caption{Correlations among photometric and kinematical properties
of Es in our galaxy sample.  The small points show individual
galaxies, while the large points with error bars show binned
averages.  The solid lines show linear fits, carried out
separately for the bright and faint subsamples (divided at $M_B
\sim -20.5$, i.e. $\log \LB\ \sim 10.4 \, \rm \Lsun$). {\it Left
Panel}. FJ relation. {\it Right Panel}. Size-luminosity relation.}
\label{fig:Sample}
\end{figure*}

\section{Sample}\label{sec:sample}
Our data-set of local ETGs is drawn from
PS96\footnote{Downloadable at {\tt
http://vizier.cfa.harvard.edu/viz
-bin/VizieR?-source=J/A+A/309/749}.}. This is currently one of the
largest homogeneous samples of local ETGs available in the
literature containing both photometry and kinematics of the galaxy
central regions, and the only one including information on the
peak rotation velocity (\vmax). As discussed in PS96, the colours
(extinction- and K-corrected) are measured within 1~\Re, the
central velocity dispersions \sigc\ are recovered from long slit
spectra\footnote{The authors do not report detailed information
about their measurement set-up but, as extensively adopted in
literature analyses of this dataset, we will interpret \sigc\ as
the luminosity-weighted velocity dispersion within a circular
aperture of radius \Re/8.} and \vmax\ is defined as the quadratic
sum of the maximum rotation on the major and minor axes. Since we
are interested in fitting spectral energy distributions (SEDs), we
select galaxies with at least two measured colours (most of the
selected galaxies have four colours)\footnote{Apparent total
magnitudes are on average slightly brighter than those in RC3
catalog (\citealt{RC3}) by $-0.05\pm 0.1 \, \rm mag$, while the
differences of $B-V$ and $U-B$ colours with those in RC3 are $0.00
\pm 0.03$ and $-0.02 \pm 0.03$.}. Selecting also for galaxies
brighter than $M_{B}=-16$, we recover $\approx 400$ galaxies among
which, following the PS96 classification, $\approx 55\%$ are bona
fide Es (their subsample 1), $\approx 30\%$ are type S0 and Sa
(their subsample 5), and the remaining $\approx 15\%$ are dusty
objects, interacting galaxies, dwarf spheroidals, compact, dwarf,
low-luminosity and peculiar ellipticals, etc. (subsamples 2, 3, 4
and 6). For the main purposes of this paper, we will use subsample
1 (hereafter ``Es'') and 5 (hereafter ``S0s''), thus incorporating
335 galaxies, or $\approx 85\%$ of the PS96 sample.

In all the following, we use a cosmological model with
$(\Omega_{m},\Omega_{\Lambda},h)=(0.3,0.7,0.7)$, where $h =
H_{0}/100 \, \textrm{km} \, \textrm{s}^{-1} \, \textrm{Mpc}^{-1}$
(\citealt{WMAP}), corresponding to a universe age of $t_{\rm
univ}=13.5 \, \rm Gyr$\footnote{The distance scale is critical
for normalizing the luminosities and $M/L$s.
The distance moduli ($m-M$) from PS96, rescaled to $h= 0.7$,
are on average lower those reported in the RC3 catalog
(\citealt{RC3}) by $-0.11^{+0.26}_{-0.20} \, \rm mag$
(uncertainties are $25^{th}$ and $75^{th}$ percentiles), while
agreeing closely with estimates from \cite{Tonry2001}
(shifted by $-0.06$ mag to correct to the Cepheid distance scale; \citealt{2003ApJ...583..712J})
which differs by $0.00^{+0.18}_{-0.16} \, \rm mag$.}.

E galaxies populate a restricted region in the colour-magnitude
diagram, the so-called ``red sequence'', with a colour range of
$B-V \sim 0.9-1$. S0s span a wider range of colours (i.e., $B-V
\sim 0.7-1$) and are fainter than Es on average. The two
subsamples follow similar FJ relations  $L \propto \sigc^{\eta}$,
including a characteristic magnitude ($M_{B} \sim -20.5$) where
the relation clearly changes its slope (see left panel of
Fig.~\ref{fig:Sample}). For the E sample, $\eta = 2.9 \pm 0.5$ and
$5.6 \pm 1.2$ for the faint and bright galaxies, respectively.

This FJ ``dichotomy'' has been reported elsewhere
\citep{MG05,Forbes08}, and seems related to systematic changes
seen in other optical properties, e.g. the Kormendy relation
\citep{1992MNRAS.259..323C,2008arXiv0805.0961N}, the
\cite{Sersic68} index (\citealt{Caon93}; \citealt[hereafter
PS97]{PS97}; \citealt{Graham98,GG03}) and the size-magnitude
relation
\citep{Shen2003,Bernardi+07,Lauer+07,2007MNRAS.377..402D,2008arXiv0810.4922H}.
The latter relation is illustrated for our sample Es in the right
panel of Fig.~\ref{fig:Sample}, where the faint and bright
galaxies have fitted slopes of $0.85 \pm 0.07$ and $0.43 \pm
0.11$, respectively.

We will see that a characteristic luminosity scale is also found to
characterize other correlations of ETG parameters.

\section{Stellar mass-to-light ratio}\label{sec:MtoL_star}

One of the key aspects of our analysis is determining each sample
galaxy's stellar $M/L$, \Yst, which we do by fitting model SEDs to
the observed galaxy colours. Although photometric modelling may
seem less powerful than detailed spectroscopic fits, most
spectroscopic samples are restricted to the very central regions
of galaxies and may be very biassed indicators of the stellar
populations on scales of $\sim \Re$.

In Section~\ref{sec:stellarmodels} we describe the modelling procedure and
present the recovered stellar populations properties for the sample galaxies.
We report the implications for the size-mass relation in Section~\ref{sec:size_relations},
and for trends in \Yst\ in Section~\ref{sec:stellarML}.

\subsection{Stellar populations modelling procedure}
\label{sec:stellarmodels}

We create a
set of synthetic stellar spectra using the prescription of \citet[hereafter BC03]{BC03},
which encompasses a wide range of initial
metallicities and ages. A \cite{Salpeter55} or \cite{Chabrier01,Chabrier02,Chabrier03}
initial mass function (IMF) is assumed, with initial masses
$m$ in the range $0.1-100$. The two IMFs do not influence the
colours, but basically affect the \Yst\ estimates,
which are $\approx 1.8$ times larger with a Salpeter IMF than with a Chabrier IMF.

To generate a more general and realistic star formation history
(SFH), we convolve the BC03 ``single burst'' models with an
exponentially-decaying star formation (SF) rate with time $t$,
$\propto e^{-t/\tau}$, where $\tau$ is a characteristic time
scale. The choice of BC03 is dictated mostly by its versatility
and ability to span the stellar parameter space (metallicities and
ages) but it is not the only prescription available on the market.
We test for the presence of any modelling systematics by mainly
checking two different popular prescriptions, \citet[hereafter
BdJ01]{BdeJ2001} and \citet[hereafter M05]{Maraston05} in
Appendix~\ref{sec:appA}.

\begin{figure}
\centering \epsfig{file=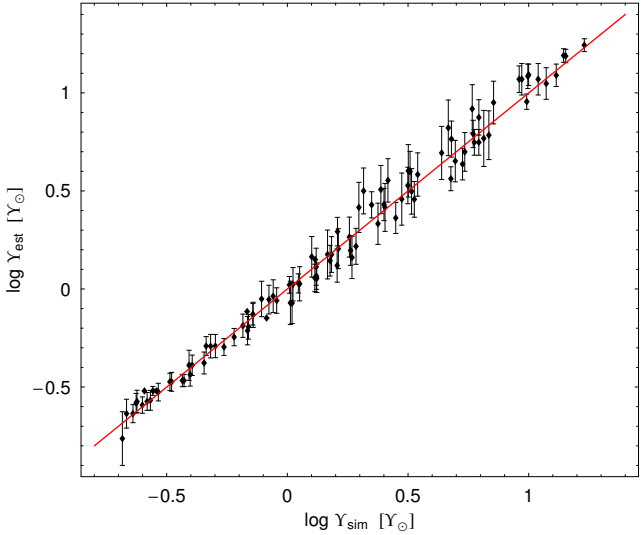,width=0.48\textwidth}
\caption{Monte Carlo simulation of the stellar populations fitting
procedure, where the estimated stellar \ML\ values $\Upsilon_{\rm
est}$ are plotted against the intrinsic model values
$\Upsilon_{\rm sim}$. The solid line is the one-to-one relation.
The input value of \Yst\ is reproduced with no systematic bias,
and the scatter of the recovered values is $\sim$~10\%.}
\label{fig:sim_ML}
\end{figure}

For each galaxy, we fit synthetic spectra to the observed colours
($U-B$, $B-V$, $V-R$ and $V-I$, after convolving the spectra with
the appropriate filter bandpass functions), allowing us to
estimate the age (\age), metallicity ($Z$), $\tau$, \YstB\, and
hence the stellar mass, \mst$=\YstB \times \LB$ (hereafter we will
always quote luminosity and \ML\ values in the $B$-band, even if
not specified). In detail, we build a set of synthetic colours
with $Z \in (0.008,~0.02,~0.05)$\footnote{Lower metallicities
would have produced \age\ larger than the age of the universe in
our assumed cosmology ($\sim 13.5$ Gyr).}, $\tau \in (0.1-5) \,
\rm Gyr$ and \age\ up to $t_{\rm univ}$\footnote{We allowed
$\sim10\%$ scatter around this value in order to account for some
intrinsic uncertainty in the age estimates, and thus some of the
estimated ages might be slightly larger than 13.5 Gyr.}. The
fitting procedure consists of generating 100 Monte Carlo
realizations of the observed galaxy colour sets assuming Gaussian
errors of 0.05 mag per colour, and minimizing a $\chi^2$~statistic
between the modelled and observed colours for each realization.
The overall best-fit model parameters and their uncertainties are
defined as the median and scatter of these 100 best
fits\footnote{Typical $1 \sigma$ uncertainties on the estimated
\Yst\ are $\sim 10-20\%$. In general in this paper, we will fit
{\it medians} rather than {\it means} in order to be more robust
to outliers.}.

Our synthetic modelling procedure is more general than the
extensively used ``simple stellar population'' (SSP) model where a
galaxy is approximated as experiencing a single burst of star
formation (i.e. $\tau=0$;
\citealt{2000AJ....120..165T,Maraston05}; see also
Appendix~\ref{sec:appA}). Instead we leave \age, $\tau$ and $Z$
all as free parameters in order to better represent the wide
variety of SFH expected both observationally and theoretically
(see e.g. \citealt{deLucia06,2007ApJ...660L..47N}). The allowed
ranges in the parameters will then be larger than in the more
simplified SSP case because of the well known degeneracies among
them (\citealt{Gavazzi02}; BC03).

To test the reliability of our modelling technique and the
intrinsic parameter scatter, and to check for the presence of
spuriously-generated correlations, we run a suite of Monte Carlo
simulations. We extract 100 simulated galaxy spectra from our BC03
SED libraries with random \age, $Z$ and $\tau$ (i.e. with no
correlation among these parameters), and apply our generalized fit
procedure---comparing the recovered parameters with the input
model values. We find that \Yst\ is recovered well, with a scatter
of $\sim 10\%$ (see Fig. \ref{fig:sim_ML}). Similar consistency is
found for \age, $Z$ and $\tau$, which have on average larger
scatter: $\sim 20\%$, $\sim 30\%$ and $\sim 30\%$ respectively. We
check for spurious correlations using a Spearman rank test
(\citealt{Nrecip}), finding that $\tau$ vs \age\ show no
correlation at the $95\%$ confidence level, but that \age\ and $Z$
are weakly correlated, which is a common effect in stellar
populations analyses, However, this \age{}-$Z$ degeneracy does not
affect the \Yst\ inference, which is our primary concern.

\begin{figure*}
\psfig{file= 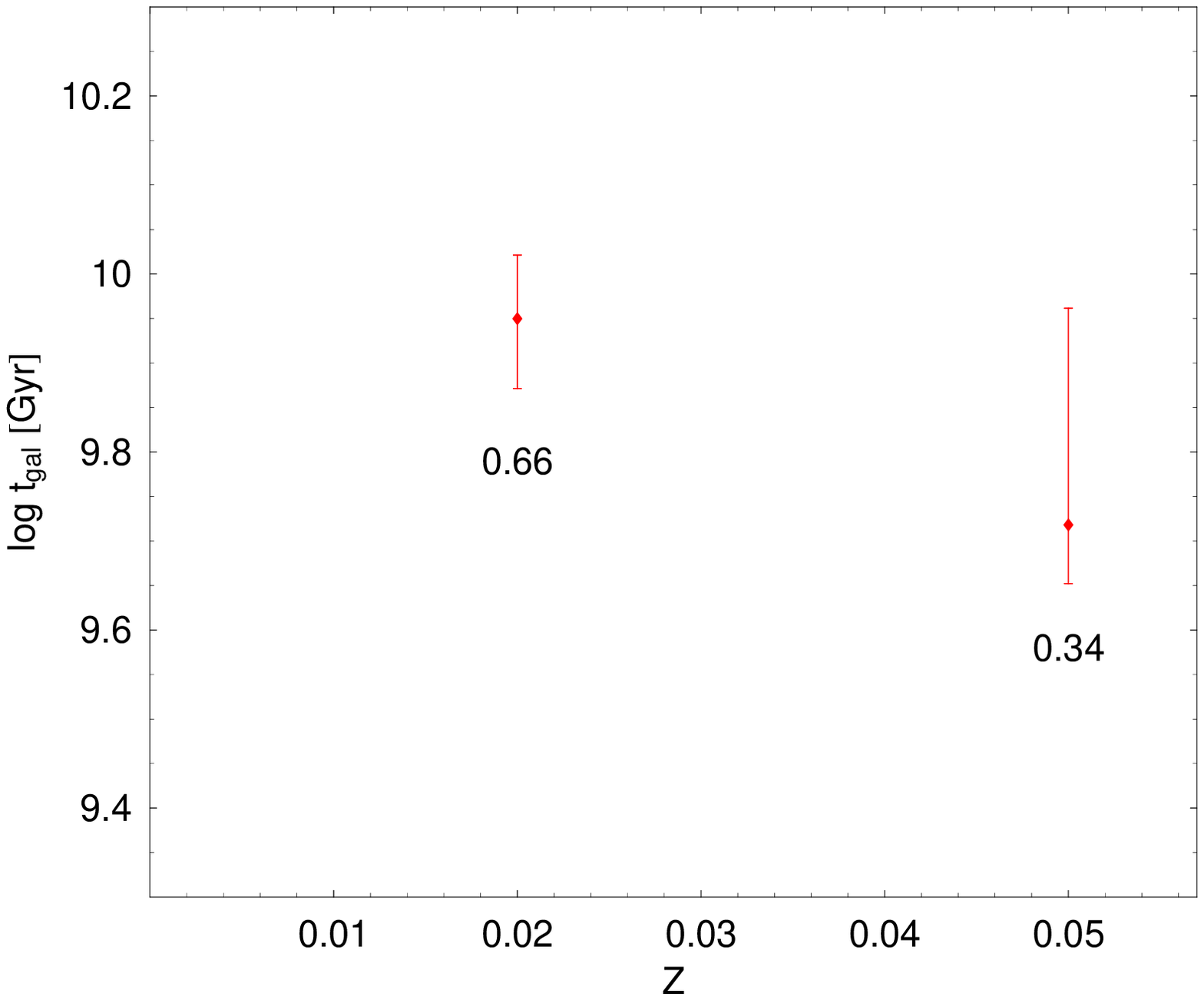, width=0.3\textwidth}
\hspace{1.5cm}\psfig{file=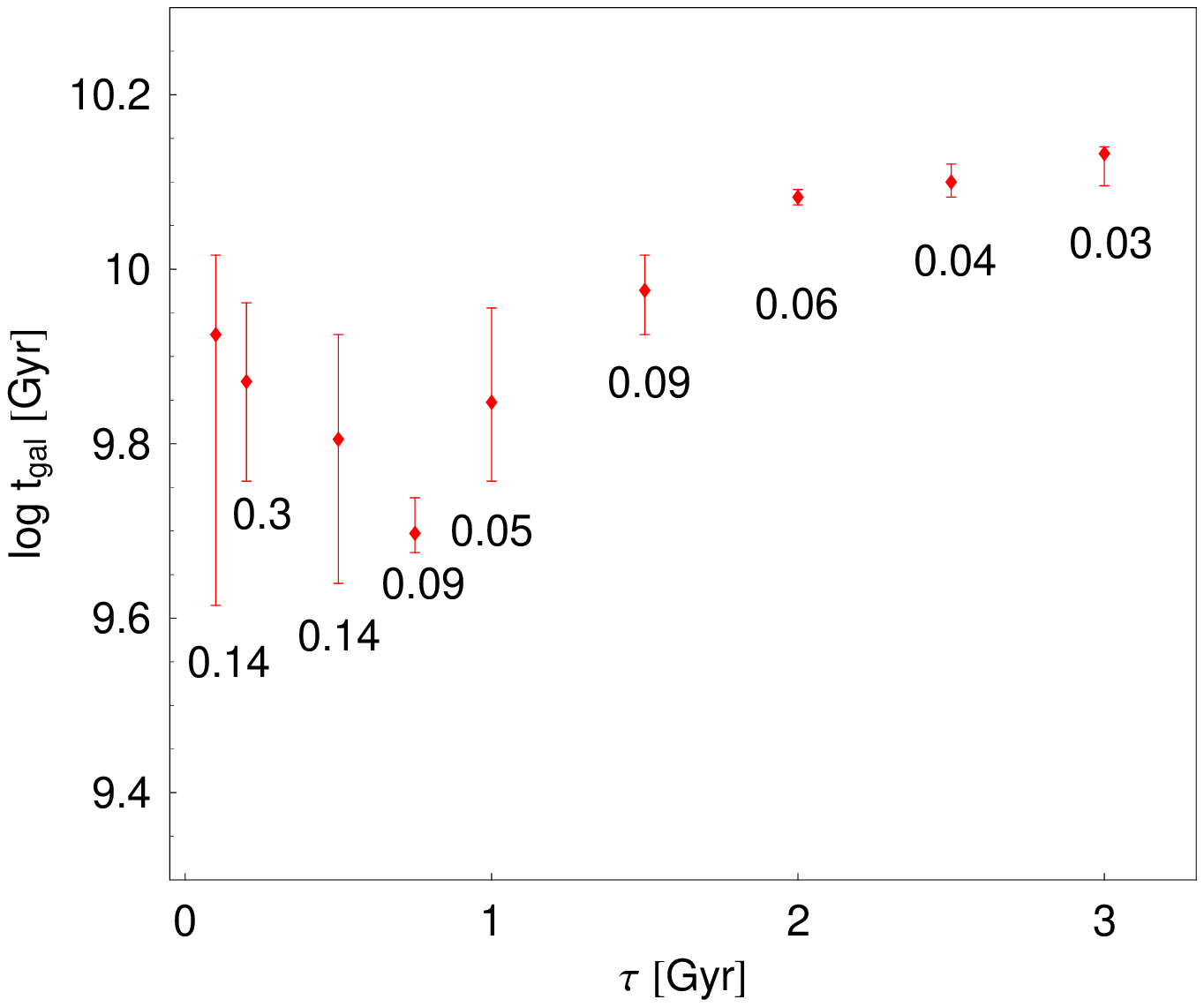, width=0.3\textwidth}\\
\vspace{0.3cm} \psfig{file=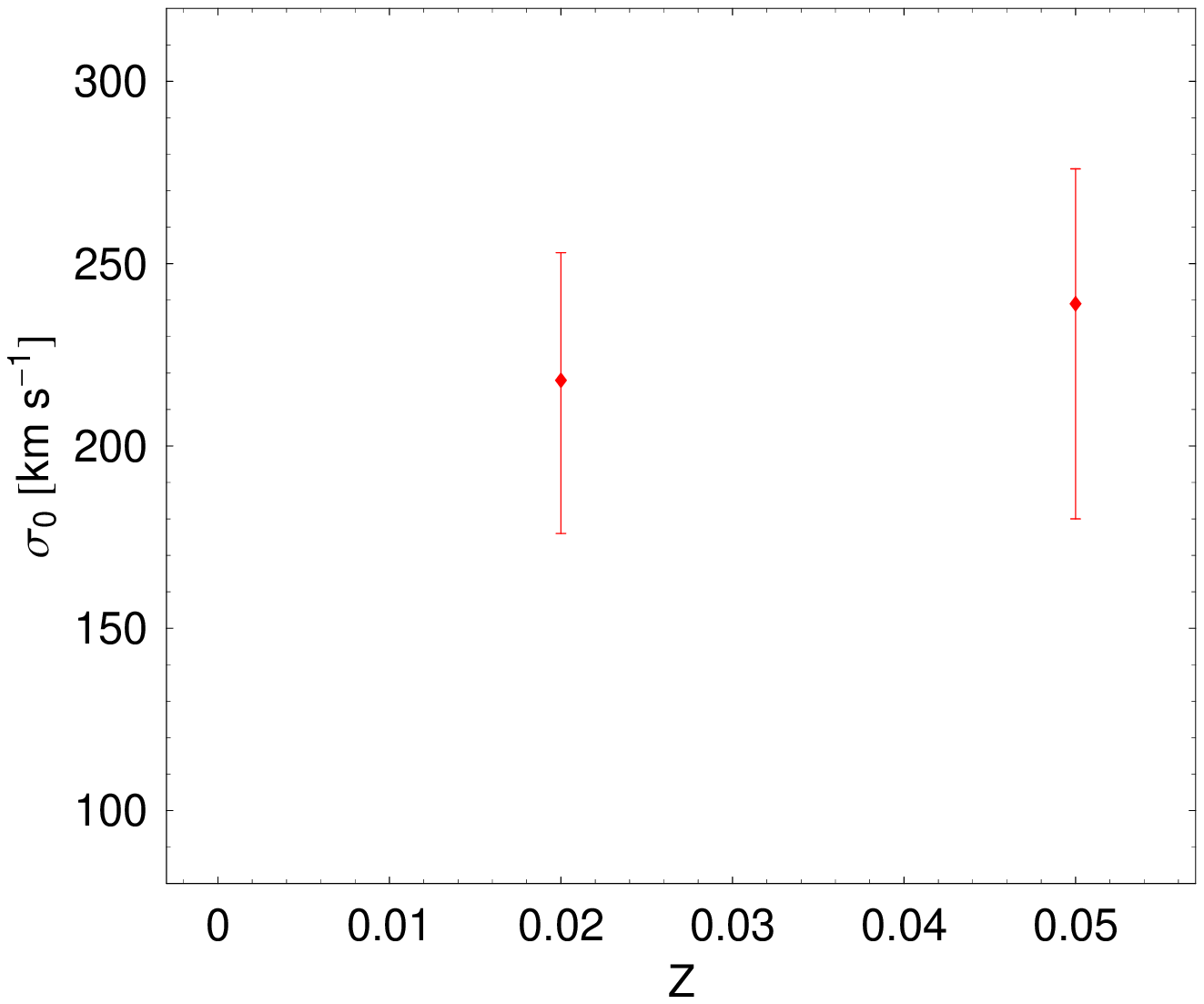, width=0.3\textwidth}
\hspace{1.5cm}\psfig{file=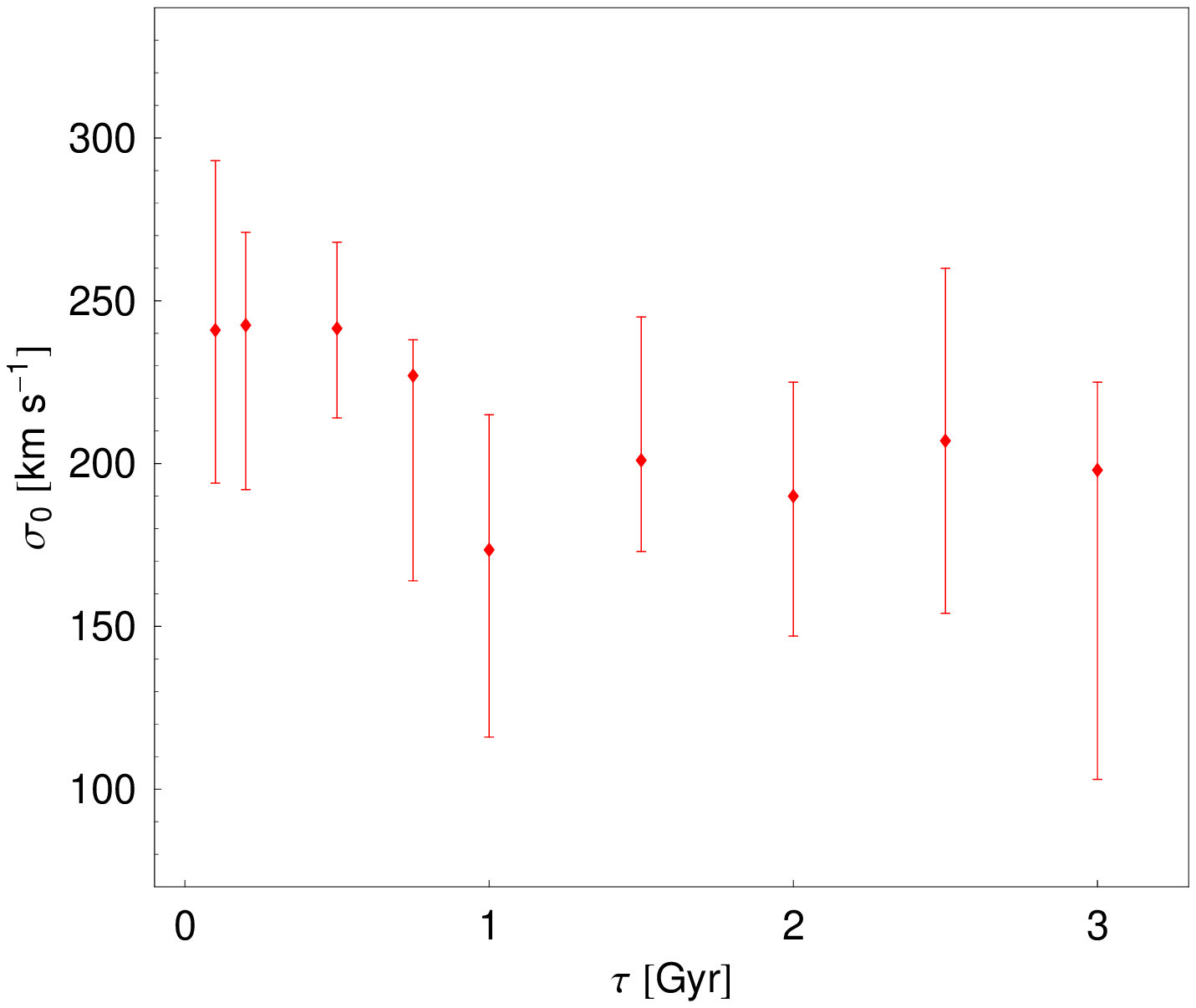, width=0.3\textwidth} \\
\vspace{0.3cm}\psfig{file=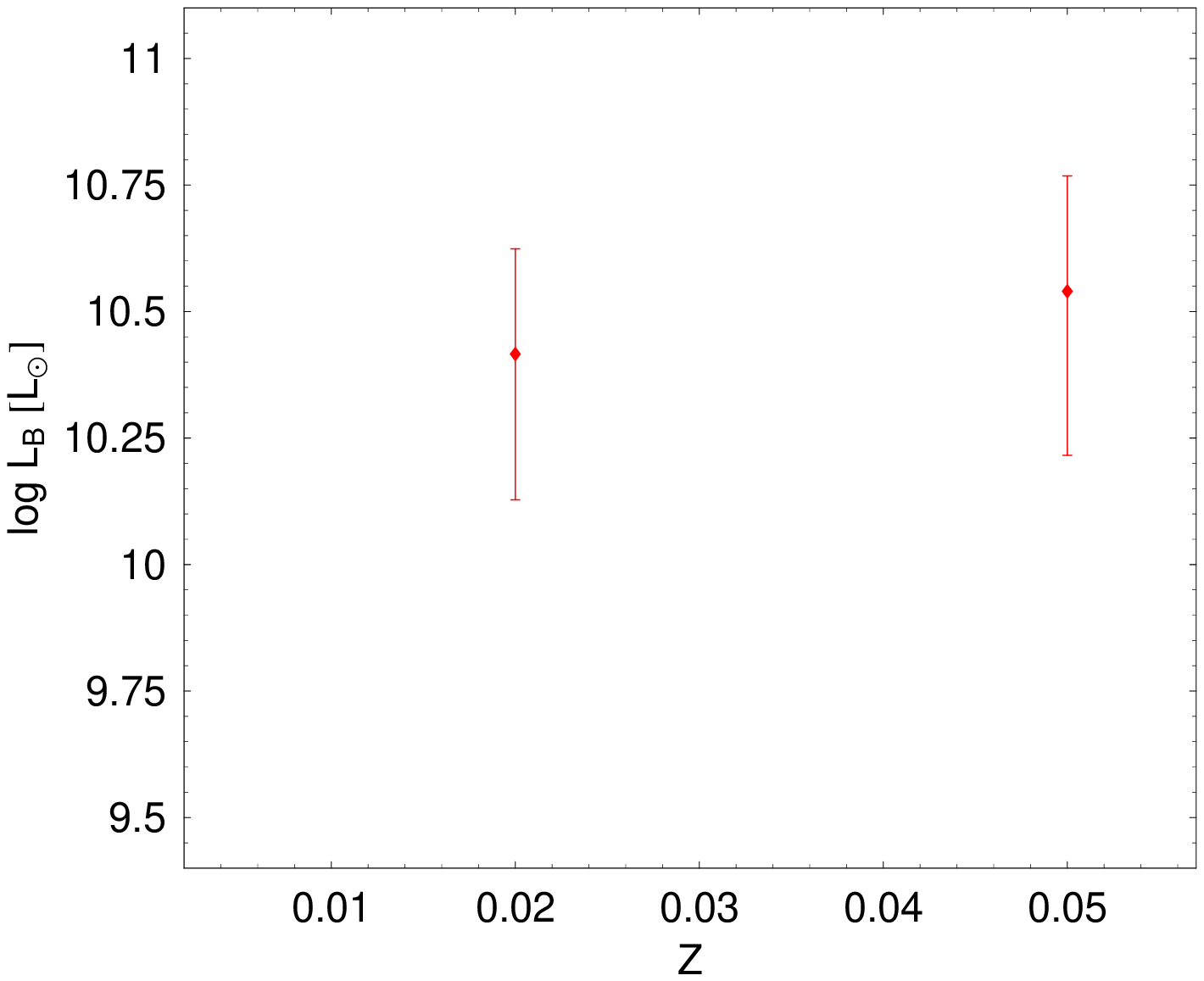, width=0.3\textwidth}
\hspace{1.5cm}\psfig{file=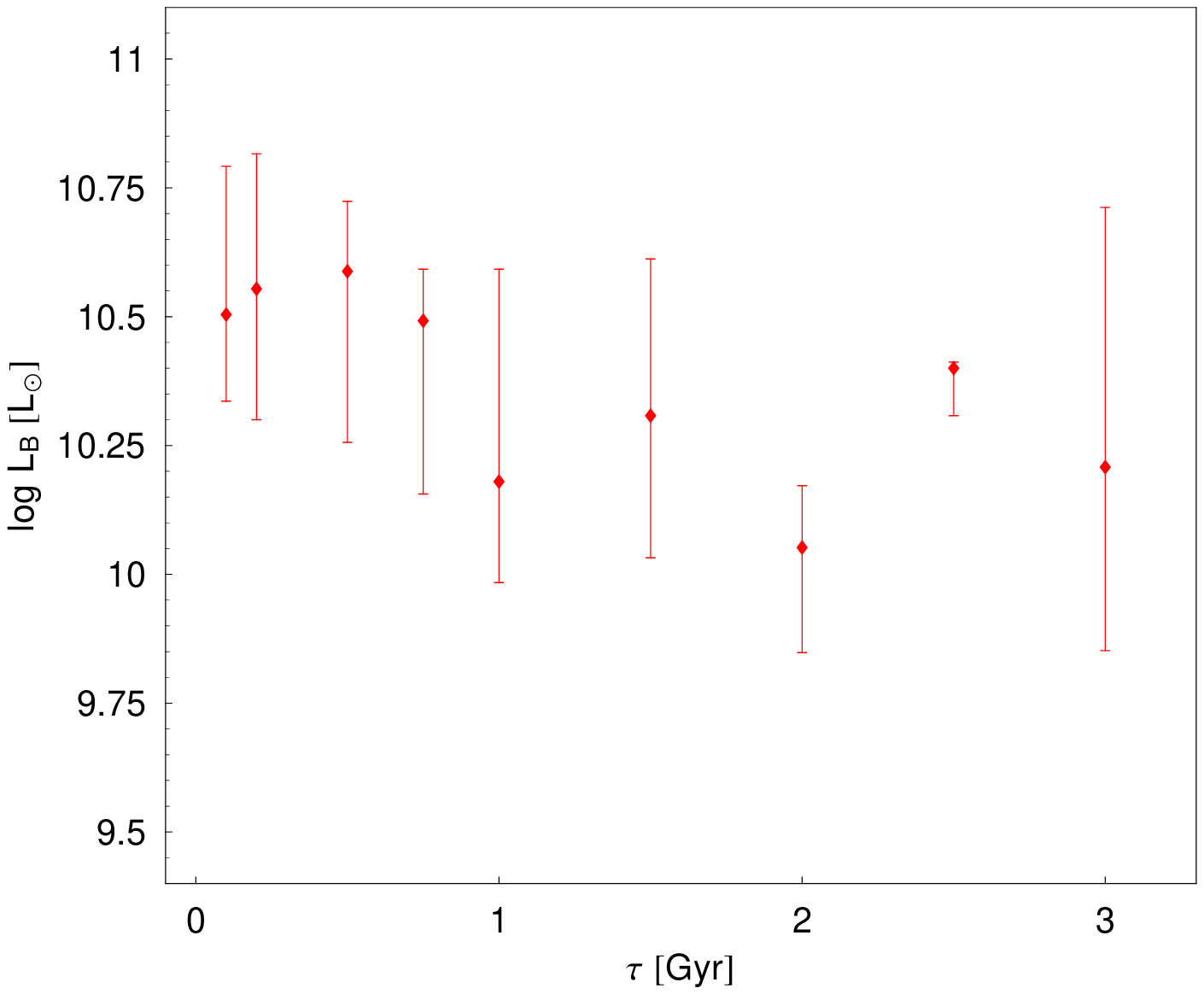, width=0.3\textwidth}
\caption{Stellar populations properties of E sample galaxies:
metallicity $Z$, SF timescale $\tau$, stellar age \age\ and
luminosity \LB. In each panel the data are binned in intervals,
with the median and $\pm 25\%$ scatter shown. The fraction of
galaxies in each bin is reported in the first two panels. For S0s,
the trends with metallicity are identical to the Es, while the
ones with $\tau$ are weaker.}
\label{fig:various_stellar_correlations}
\end{figure*}

After fitting the real data for the complete sample of 335
galaxies, we show some relations between model parameters and
other observed galaxy quantities in
Fig.~\ref{fig:various_stellar_correlations}. From this Figure it
is evident that the metallicity is generally solar or super-solar
($Z \geq 0.02$) and on average only weakly dependent on other
properties such as luminosity -- which is fortunate since the BC03
stellar libraries include only a few reference values for $Z$.
More striking are the strong correlations involving \age\ and
$\tau$, such that the brighter, more massive galaxies formed their
stars on average on shorter timescales than the fainter, less
massive galaxies, while the younger galaxies also had shorter SF
timescales (this does {\it not} mean that the brighter galaxies
are younger, and if the S0s are included, the opposite is clearly
true).

Similar findings on the SFHs of ETGs have been found in
observational (e.g. \citealt{Gavazzi02,Thomas05}) and theoretical
(e.g. \citealt{deLucia06,Romeo+08}) analyses. We will consider
this subject in detail in a subsequent paper (\citealt{Nap09b}, in
prep.), and for now summarize some basic parameters. The bright
ETGs ($M_{B} \leq -20.5$) have a median $\tau \sim 0.5$~Gyr, while
the faint ETGs have $\tau \sim 1$~Gyr. The Es have similar
properties, while the S0s have on average more protracted SFHs
($\tau \sim $1--1.5~Gyr). The median \Yst\ for the ETGs is
$6.9\pm2\, \Ysun$ ($3.8 \pm 1.1\, \Ysun$) for a Salpeter
(Chabrier) IMF, where the quoted errors are the 1~$\sigma$
scatter. The S0s have only slightly smaller median \Yst\ than the
Es: $6.2\, \Ysun$ ($3.4\, \Ysun$) and $7.1\, \Ysun$ ($3.9\,
\Ysun$), respectively. However, the two samples differ more
strongly in the {\it distribution} of \Yst, where the Es have a
fairly symmetric distribution about the mean, while the S0s have a
pronounced tail to low \Yst\ such that the mean value of $5.8\,
\Ysun$ ($3.2\, \Ysun$) differs from the median estimate.

\subsection{Size-mass relations}\label{sec:size_relations}

An indirect way to test our derived \Yst\ values is to check how
the implied scaling relation between size and stellar mass
compares to previously established results. Expressing the
size-{\it luminosity} relation as $\Re\ \propto L^{\alpha_{L}}$,
we find a slope for the Es of $\alpha_{L}=0.70 \pm 0.06$, which is
slightly steeper than some literature findings of $0.54-0.63$
(\citealt{Pahre98}, \citealt{Bernardi2003}, \citealt{ML05}).
However, this slope is very sensitive to the range of luminosities
fitted, since we see a difference between the ``faint'' and
``bright'' subsamples (see Section~\ref{sec:sample} and right
panel of Fig.~\ref{fig:Sample}).

\begin{figure}
\psfig{file= 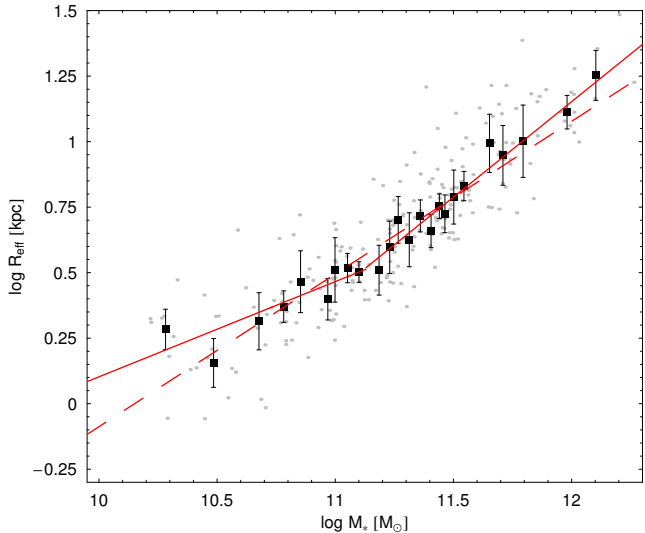, width=0.48\textwidth} \caption{Correlation
between effective radius and stellar mass in E galaxies. The
symbols are as in Fig.~\ref{fig:Sample}. The dashed line is the
best-fit relation over the full range of the data, while the solid
line is the best-fit to the two mass regimes.}
\label{fig:Reff_vs_LB_and_mstar}
\end{figure}

For the size-{\it mass} relation $\Re\ \propto
M_{\star}^{\alpha_{M}}$, we obtain $\alpha_{M} = 0.65 \pm 0.05$
overall (Fig.~\ref{fig:Reff_vs_LB_and_mstar}), consistent with
previous estimates for low-redshift galaxies of typically $\sim
0.6$ (\citealt{Bernardi2003,Shen2003,ML05}; N+05). This
correlation also bends at a characteristic mass scale of $M_* \sim
10^{11.1} \Msun$, with $\alpha_{M}=0.36 \pm 0.13$ and $0.73 \pm
0.12$ for the faint and bright galaxies, respectively (cf.
\citealt{Shen2003}). We find the S0s to have on average smaller
$\Re$ than Es of the same mass, and similar $\alpha_{M}$ at high
masses, but flattening to $\sim 0$ at low masses.

\subsection{Luminosity dependence of \Yst}
\label{sec:stellarML}

\begin{figure}
\psfig{file= 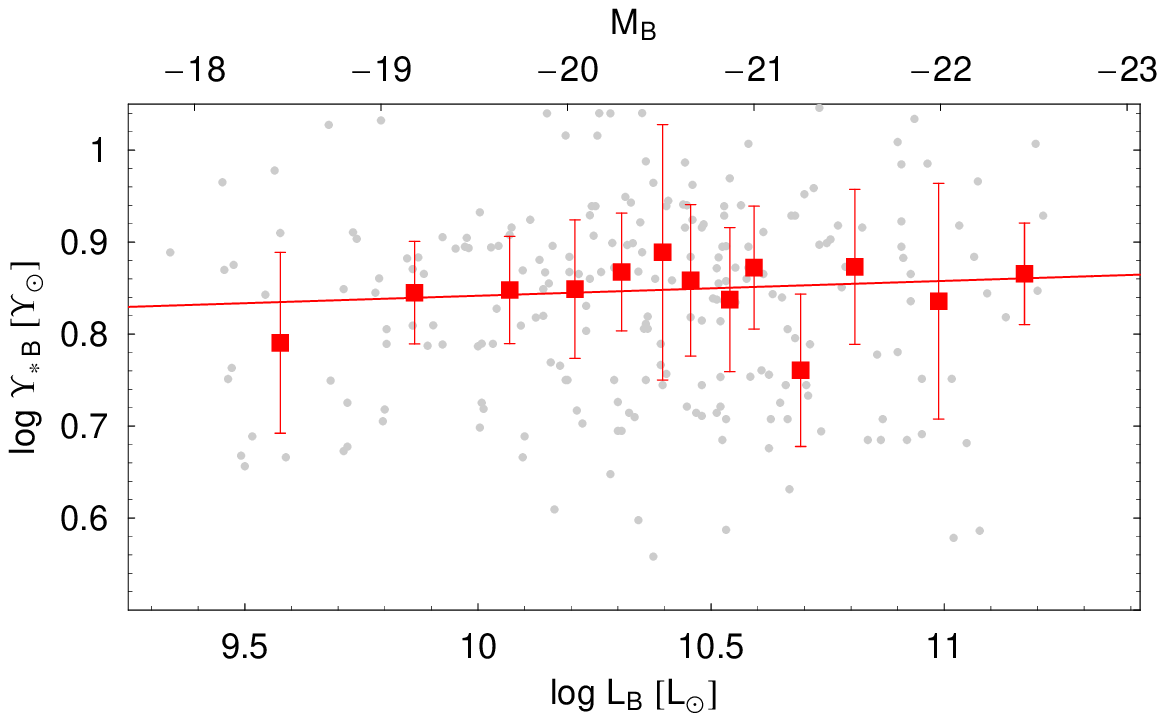, width=0.5\textwidth} \psfig{file=
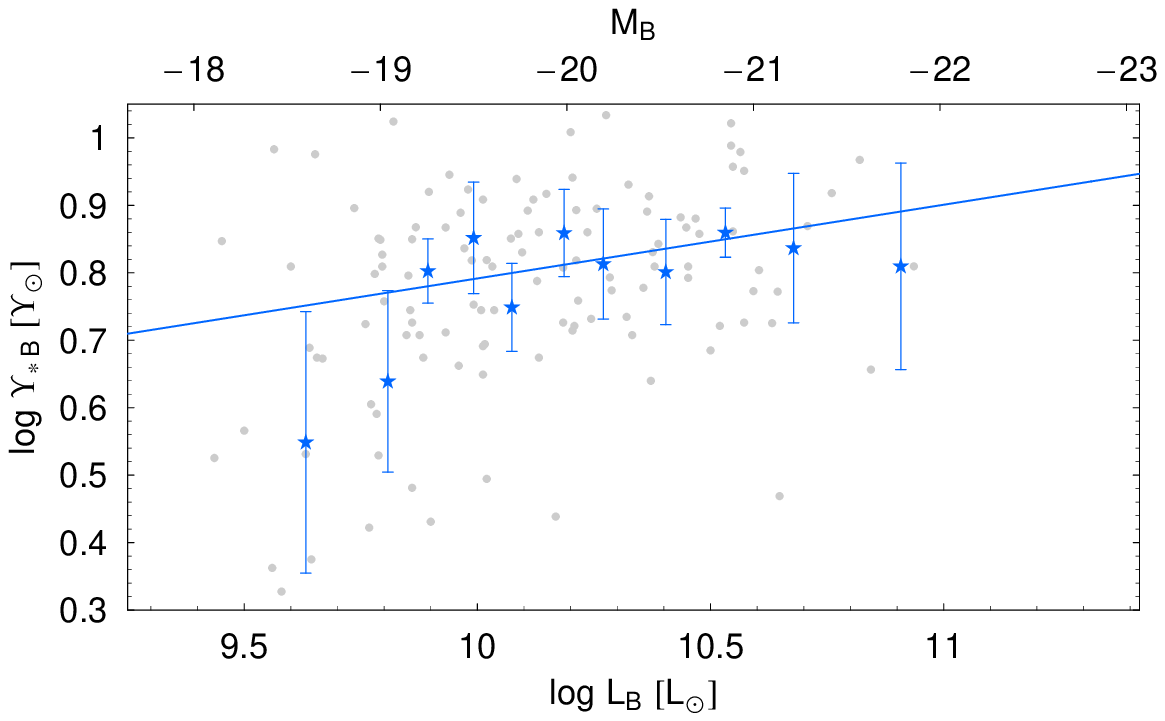, width=0.5\textwidth} \caption{Stellar \ML\ as a
function of $B$-band luminosity, for Es (top panel) and S0s
(bottom panel). Grey points represent individual galaxies, while
points with error bars are binned median values and scatter. Solid
lines are the best fit linear relations in log-log space. A
Salpeter IMF is assumed. }
\label{fig:Ups_star_vs_LB_Zsol_tau_1Gyr}
\end{figure}

The central regions of ETGs are probably dominated by the stellar
mass, so it is of critical importance to ascertain the fraction of
the FP tilt that is connected to the stellar population
properties. We focus on the relation $\Yst\ \propto
\LB^{\gamma_*}$ in log-log space, fitting to weighted medians of
binned data values, with results that are stable to changes in the
binning (Fig. \ref{fig:Ups_star_vs_LB_Zsol_tau_1Gyr}). For the
overall ETG sample, we find a slope of $\gamma_*=0.06\pm 0.01$;
the Es have $\gamma_*=0.02 \pm 0.01$, and S0s have $\gamma_*=0.17
\pm 0.03$, although this steeper slope is driven by the very
faintest galaxies ($M_B > -19$).

Before taking these results at face value, we investigate possible
dependencies on modelling systematics. Changing the IMF from
Salpeter to Chabrier does not affect $\gamma_*$, but only the
overall normalization of \Yst. Adopting a simplified model with
\age\ as the only free parameter, with $\tau \sim 0.75$~Gyr and $Z
\sim \Zsun$ fixed to the median values for the whole sample (see
Section~\ref{sec:stellarmodels}), the \Yst\ steepens for the faint
galaxies and flattens for the bright ones, with an overall result
of $\gamma_* \sim 0.16$ (Appendix~\ref{sec:appA}). An even more
simplified SSP model with $\tau=0$ yields about the same $\gamma_*
\sim 0.18$. Thus, we see that allowing for the variations of SF
timescales (and metallicity) within the sample is critical to
accurately deriving the \Yst\ trends with luminosity. An
additional complication, only partially addressed by our model's
protracted exponential SFH, is multiple bursts of SF in a single
galaxy which, if corrected for, would probably flatten the slope
even further (cf. section~4.7 of C+06).

We next consider alternative stellar populations basis models. As
detailed in Appendix~\ref{sec:appA}, adopting the BJ01 or M05
models with the same assumptions would imply shallower and steeper
\Yst\ slopes, respectively. This turns out to be the dominant
source of uncertainty in the analysis, although the uncertainty is
largest for the faintest galaxies ($M_B \gsim -20.5$), and the
\Yst\ estimates more stable for the brighter
galaxies\footnote{Including all the systematics, the final slope
becomes $\gamma_* = 0.06 \pm 0.01 ^{+0.12}_{-0.04}$.}.

How do our results compare to previous work? As mentioned in
Section~\ref{sec:intro}, most recent studies agree that $\gamma_*$
is a relatively small contributor to the total $\gamma$ of the FP.
E.g. PS96 inferred from the same data set that $\gamma_* \sim
0.1$, and T+04 found using very different data and techniques that
$\gamma_* = 0.07 \pm 0.01$ in the $B$-band\footnote{For the
modelling, PS96 used fitting relations linking single colours and
line-strength index $\rm Mg_{2}$ to the magnitude. T+04 applied
the PEGASE prescription \citep{FR97} to the Sloan Digital Sky
Survey (SDSS) Early Data Release. T+04 also found $\gamma_* = 0.02
\pm 0.01$ in the $K$-band using 2MASS data \citep{Bell2003}, but
one would expect $\gamma_*$ to depend on bandpass because of the
changing contributions from the mix of stellar populations.}. This
general consistency is encouraging, but it should be kept in mind
that as just discussed, there remain significant uncertainties in
the modelling.

\begin{figure}
\psfig{file= 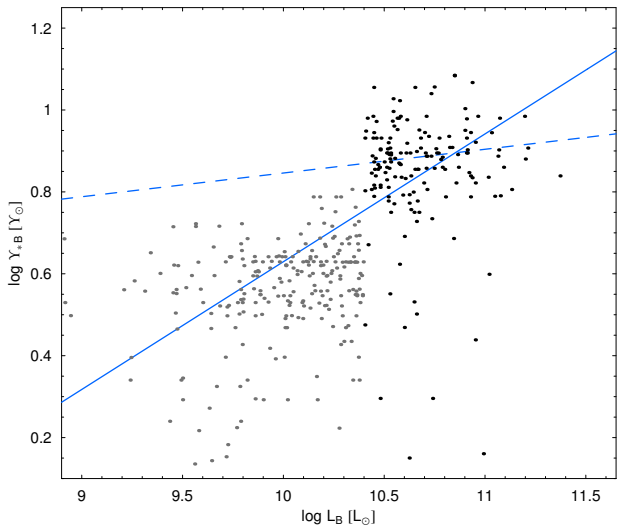, width=0.46\textwidth} \caption{Stellar \ML\
for ETGs as a function of $B$-band luminosity, assuming an
arbitrary change in the IMF at intermediate luminosity $\log
(L/\Lsun)=10.4$. The faint galaxies (grey points) have a Chabrier
IMF and the bright galaxies (black points) a Salpeter IMF. The
blue solid line is the best fit, while the dashed one is the
relation with a constant Salpeter
IMF.}\label{fig:Ups_star_vs_LB_Zsol_tau_1Gyr_IMFchange}
\end{figure}

The foregoing conclusions are based upon a universal IMF, but
there are suspicions that the IMF may vary with time or
environment (e.g.
\citealt{2008ApJ...674...29V,2008MNRAS.385..147D}). \citet{RC93}
pointed out that a variation in the IMF with luminosity could
easily account for the FP tilt. Here we illustrate this point
again with a simple toy model, wherein the faint galaxies have a
Chabrier IMF, and the bright ones a Salpeter IMF (a more realistic
scenario would have the IMF changing smoothly with luminosity).
The implied \Yst\ slope would be $\gamma_* \sim 0.3$ (Fig.
\ref{fig:Ups_star_vs_LB_Zsol_tau_1Gyr_IMFchange}), which as we
will see would be enough to explain the FP tilt with no further
ingredients (e.g. no DM).

\begin{figure}
\psfig{file= 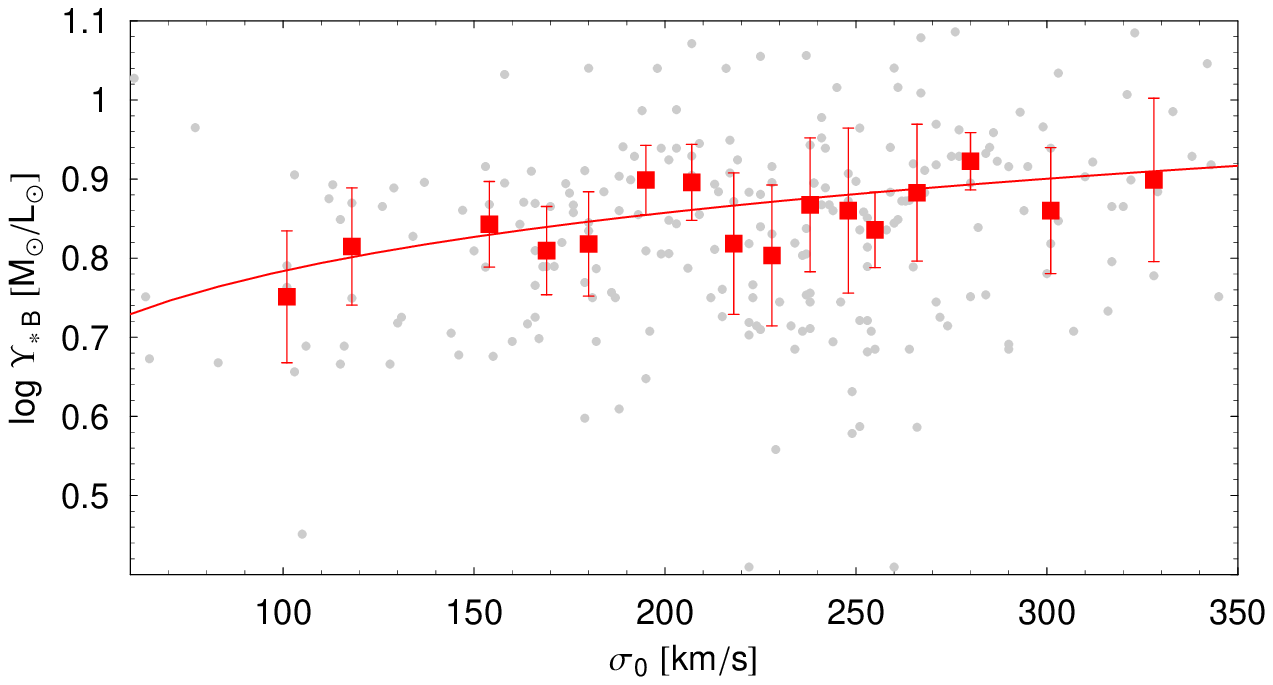, width=0.48\textwidth} \psfig{file=
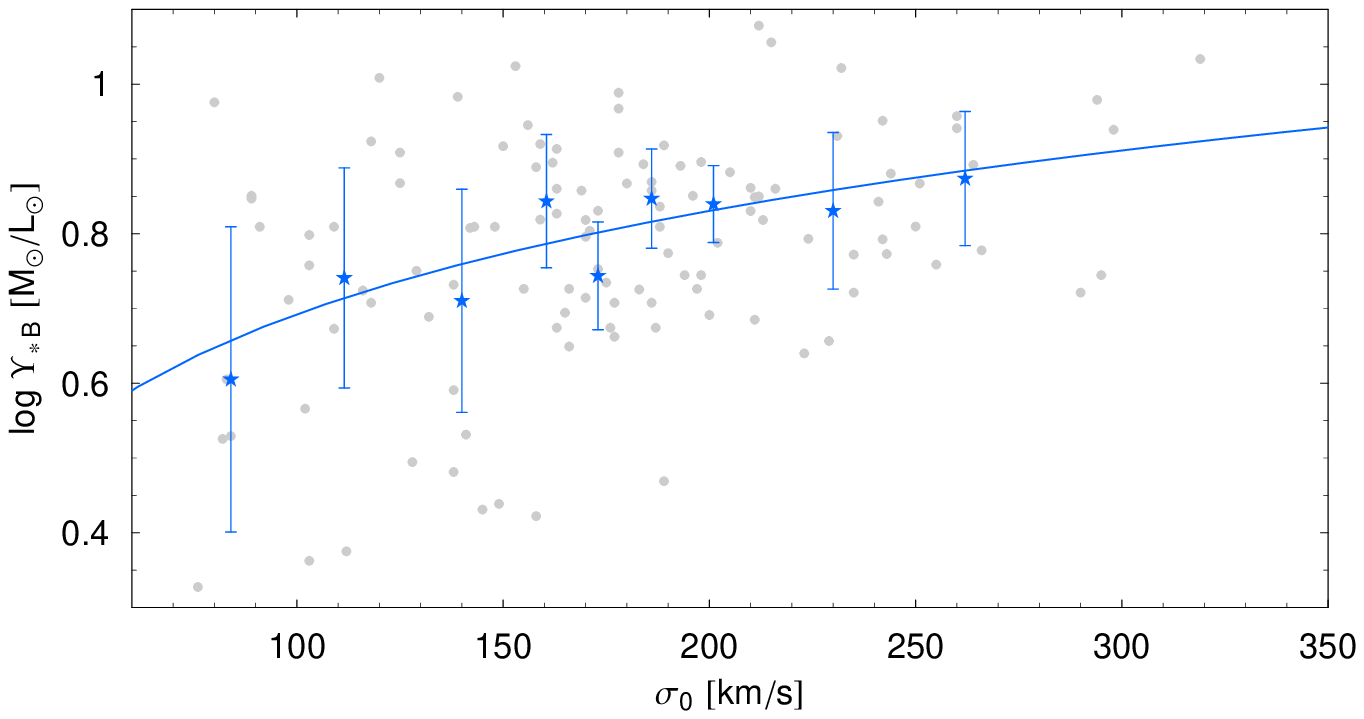, width=0.48\textwidth} \caption{Stellar \ML\ ratio as a
function of velocity dispersion \sigc, using Salpeter IMF for Es
(top panel) and S0s (bottom panel). See Fig.
\ref{fig:Ups_star_vs_LB_Zsol_tau_1Gyr} for colour code and other
details.} \label{fig: MtoLstar_vs_sig0}
\end{figure}

We lastly examine the correlation of \Yst\ with velocity
dispersion: $\Yst \propto \sigc^{\gamma_{\sigma}}$. The fitted
slope is $\gamma_{\sigma} \sim 0.2-0.4$ for both Es and S0s (Fig.
\ref{fig: MtoLstar_vs_sig0}). The steepness of this trend relative
to $\gamma_*$ suggests that the stellar populations of galaxies
are more strongly linked to their dynamical masses than to their
luminosities. This issue will be considered in more detail in the
following Sections, and as part of a separate analysis in
\cite{Nap09b}.

\section{Dynamical mass}\label{sec:dynM}

Besides \Yst, the other fundamental quantity we want to determine
is the total dynamical $M/L$, \Ydyn. The usual way the dynamical
mass is calculated in FP studies is with the virial relation
\begin{equation}
M= \frac{K \sigma_0^{2} r}{G},\label{eq:M_virial}
\end{equation}
where $G$ is the gravitational constant and $K$ is a pressure
correction coefficient (or virial coefficient; e.g.
\citealt{Padmanabhan04,Eke04}; C+06). The $K$ factor subsumes a
host of issues such as the aperture wherein $\sigma_0$ is
measured, and variations or ``non-homologies'' in the viewing
angles, orbital structures, luminosity profile, DM distribution,
etc.

Rather than adopting some approximate value or function for $K$, we will here directly model
the non-homologies as much as possible for every individual galaxy in the sample,
using the Jeans equations to estimate \Ydyn\ within 1~\Re.
We outline the modelling methods in Section~\ref{sec:dynmeth}, present
results based on luminosity profile homology in Section~\ref{sec:hom}
and on more general profiles in Section~\ref{sec:nonhom}.

\subsection{Dynamical methods}
\label{sec:dynmeth}

The basic approach of our dynamical models is to take the observed
luminosity profile for each galaxy, along with a parameterized
mass model, and solve for the projected velocity dispersion
$\sigma_0$ within a central aperture. The mass model parameters
are then optimized to match the observed value for $\sigma_0$.

In detail, the steps are the following:
\begin{enumerate}
\item We parameterize the luminosity profile $j_*(r)$ by either
a (deprojected) \citet{deV48} profile or a more general
\citet{Sersic68} model \citep{Caon93}--which fully takes into
account any non-homologies in the stellar density distributions.
The functional form for $j_*(r)$ is specified in Appendix~B of
PS97.
\item We adopt a simplified form for the total cumulative dynamical mass
profile $M(r)$ which is either a constant-$M/L$ profile $M(r) =
\Upsilon_0 \, L(r)$ (including the cases where DM is missing or
has a cored distribution; see \citealt{Burkert95,Nap08}), or a
singular isothermal sphere (SIS), where $M(r)\propto \sigma_{\rm
SIS}^{2} r$. The latter choice is motivated by evidence from
strong gravitational lensing for near-SIS profiles in the central
regions of ETGs (e.g. \citealt{Kochanek91, SLACS3}). These two
alternatives bracket the plausible range of mass profiles.
\item We solve the Jeans equation:
\begin{equation}
{{\rm d}(j_* \sigma_r^2) \over {\rm d}r} + 2\,{\beta(r) \over r}
\,j_* \sigma_r^2 = - j_*(r)\, \frac{GM(r)}{r^2} \ ,
\label{eq:jeans}
\end{equation}
where $\beta = 1 - \sigma_t^2/\sigma_r^2$ is the anisotropy. This
model assumes spherical symmetry and no rotation (cf.
\citealt{ML05}). For simplicity we also assume isotropy
($\beta=0$), in which case the Jeans Eq.~(\ref{eq:jeans})  can be
transformed to:
\begin{equation}
\sigma_r^2 (r) = \frac{1}{j_*(r)} \int_r^\infty j_*
\frac{GM}{s^2} {\rm d}s \ . \label{eq:iso}
\end{equation}
\item We project Eq. (\ref{eq:iso}) to obtain the line-of-sight velocity dispersion:
\begin{equation}
\sigma_{\rm los}^2 (R) = \frac{2}{I(R)}\,\int_R^\infty \frac{j_*
\sigma_r^2 \,r\,{\rm d}r}{\sqrt{r^2\!-\!R^2}} , \label{eq:siglos}
\end{equation}
where
\begin{equation}
I(R) = 2\,\int_R^\infty \frac{j_*\,r}{\sqrt{r^2\!-\!R^2}}{\rm
d}r \label{eq:IR}
\end{equation}
is the projected density profile.
\item We integrate $\sigma_{\rm los}$ within a fixed aperture $\Re/8$ to obtain the aperture velocity dispersion, $\sigma_{\rm Ap}$
using the Equation:
\begin{equation}
\sigma_{\rm Ap}^2 (R) = \frac{1}{L(R)}\int_0^{R_{\rm eff} /8}
2\pi\,S\,I(S)\,\sigma_{\rm los}^2(S)\,{\rm d}S \ ,
\label{eq:sigap}
\end{equation}
where $L(R) = \int_0^R 2\pi S I(S)\, {\rm d}S$ is the luminosity
within the projected radius $R$\footnote{We check that using an
aperture of \Re/10 would leave the results almost unchanged.
Following T+04, we have also checked that alternatively using
fixed apertures of $1''.6$ and $2''.2$, the median \Ydyn\ values
are overestimated by $3^{+3}_{-2} \%$ and $1^{+3}_{-1} \%$, well
within the typical uncertainty on each estimate and on the
sample's global scatter.}.

\item We fit the model $\sigma_{\rm Ap}$ to the observed $\sigma_0$
and iterate the preceding steps, varying the free parameters in
Eq. (\ref{eq:iso})~ (i.e. $\sigma_{\rm SIS}$ or $\Upsilon_0$). The
resulting best-fit mass profile then provides the total spherical
mass-to-light ratio within an effective radius \Ydyn (\Re) (which
is coincident with $\Upsilon_0$ in the case of the constant-\ML\
model).
\end{enumerate}

This procedure does not take into account certain factors that
could in principle alter the final mass estimates. Firstly, the
mass model does not include a central black hole, but we calculate
the effect to be negligible\footnote{We added an estimated black
hole mass $M_{\rm BH}$ in the Jeans equations, using $M_{\rm BH}$
predicted by the correlation $\sigc - M_{\rm BH}$ in
\cite{FM2000}. The result was to decrease \Ydyn\ by $2 \pm 1
\%$.}. More importantly, real galaxies are neither spherical nor
isotropic in general. We will check the impact of these
simplifications later, but here begin with a first-order
correction to the isotropic results.

Detailed dynamical models of nearby galaxies have shown that their central
stellar parts are close to isotropic {\it after subtracting the rotational component}
(e.g. G+01; C+06; \citealt*{Cappellari07}).
The observed $\sigma_{\rm Ap}$ does incorporate both the projected rotation
and dispersion components of the specific kinetic energy
($\sigma_{\rm Ap}^2 = v_{\rm rms}^2 = v^2+\sigma^2$),
and the Jeans equations could
in principle be reformulated along these lines.
However, for many galaxies the rotation is so dominant that it is preferable to include
it as an additional, separate term, which would require additional assumptions about
the rotation field of each galaxy, and would best entail a non-spherical treatment
anyway -- all of which is beyond the scope of the current paper.

Here we adopt a heuristic correction to the {\it observed}
dispersion in order to approximately account for rotational
effects. Following PS94, we parameterize the corrected
$\sigma_{\rm Ap}'$ by $\sigma_{\rm Ap}' = \sigma_{\rm Ap} \drot$.
To estimate \drot\ we have performed Monte Carlo simulations as in
\cite{Nap01}, beginning with a suite of analytical spherical
stellar+DM models as described in N+05\footnote{The multicomponent
model, in their \S~3, includes a Hernquist (1990) stellar
distributuion and an NFW \citep{NFW96, NFW97} spherical DM halo.}.
For each model with a fixed gravitational potential, we assume
isotropy and an additional rotational component that increases
with radius, then solve the Jeans equations and project to
$\sigma_{\rm Ap}$. Finally we examine the factor \drot\ that
relates the rotating and non-rotating ``measurements''
$\sigma_{\rm Ap}$, finding this simple approximation:
\begin{equation}
\drot \approx 1+0.05 \frac{\vmax}{\sigc} ,
\end{equation}
which is calculated for an aperture of 1~\Re\, and turns out to be
valid for a large range of galaxy masses\footnote{This correction
is much smaller than that found by PS94 because these authors did
not take into account the variations of rotation with radius, nor
of measurements made within an aperture rather than along the
major axis.}. We therefore apply this correction to the observed
$\sigma_{\rm Ap}$ before matching to the models in step (vi)
above. The correction increases the inferred \Ydyn\ values since
rotation at $\sim \Re$ coupled with the $\beta=0$ assumption
depresses the central $\sigma_r$ for a given mass profile: a
rotational component must be subtracted from the right hand side
of Eq.~(\ref{eq:iso}).

\begin{figure}
\psfig{file= 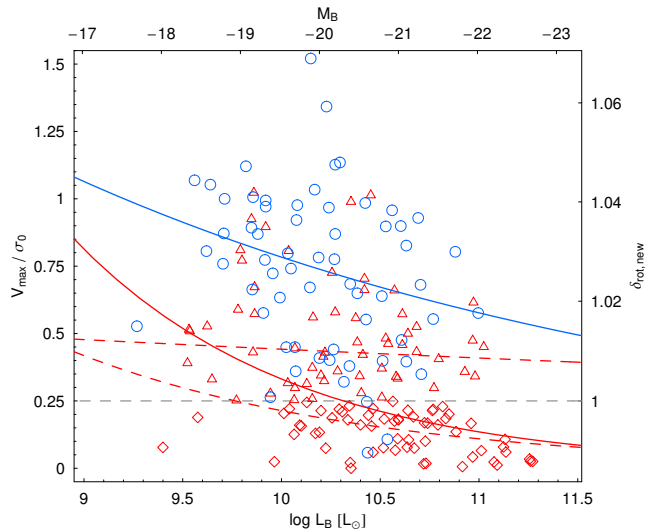, width=0.48\textwidth} \caption{Ratio
between rotation and dispersion of ETG sample, as a function of
luminosity. Red diamonds, red triangles and blue circles are
respectively slow rotator Es, fast rotator Es and S0s. The solid
red and blue curves are functional fits to the binned data, for Es
and S0s respectively, while the red dashed ones are for slow and
fast rotator Es. Finally, the dashed grey line sets the boundary
between slow and fast rotators.} \label{fig:delta_rot_vs_MB}
\end{figure}

The trend with luminosity for the rotation correction in our
galaxy sample is shown in Fig.~\ref{fig:delta_rot_vs_MB}, implying
\Ydyn\ corrections of $\sim 1\%$ for the brightest Es, and $\sim
6\%$ for the faintest S0s (as is well known, rotation is a
stronger factor on average among fainter ETGs). When plotting
results for all galaxies in the sample we use the median
$(\vmax/\sigc)$-\LB\ trend to estimate their \drot. Where
possible, we also classify the Es as fast- or slow-rotators, using
$\vmax/\sigc\ =0.25$ as the demarcation -- a simple scheme that
matches the more robust conclusions of C+06 in more than 90\% of
the overlapping cases.

\subsection{Results from homologous luminosity profiles}\label{sec:hom}

\begin{figure*}
\psfig{file= 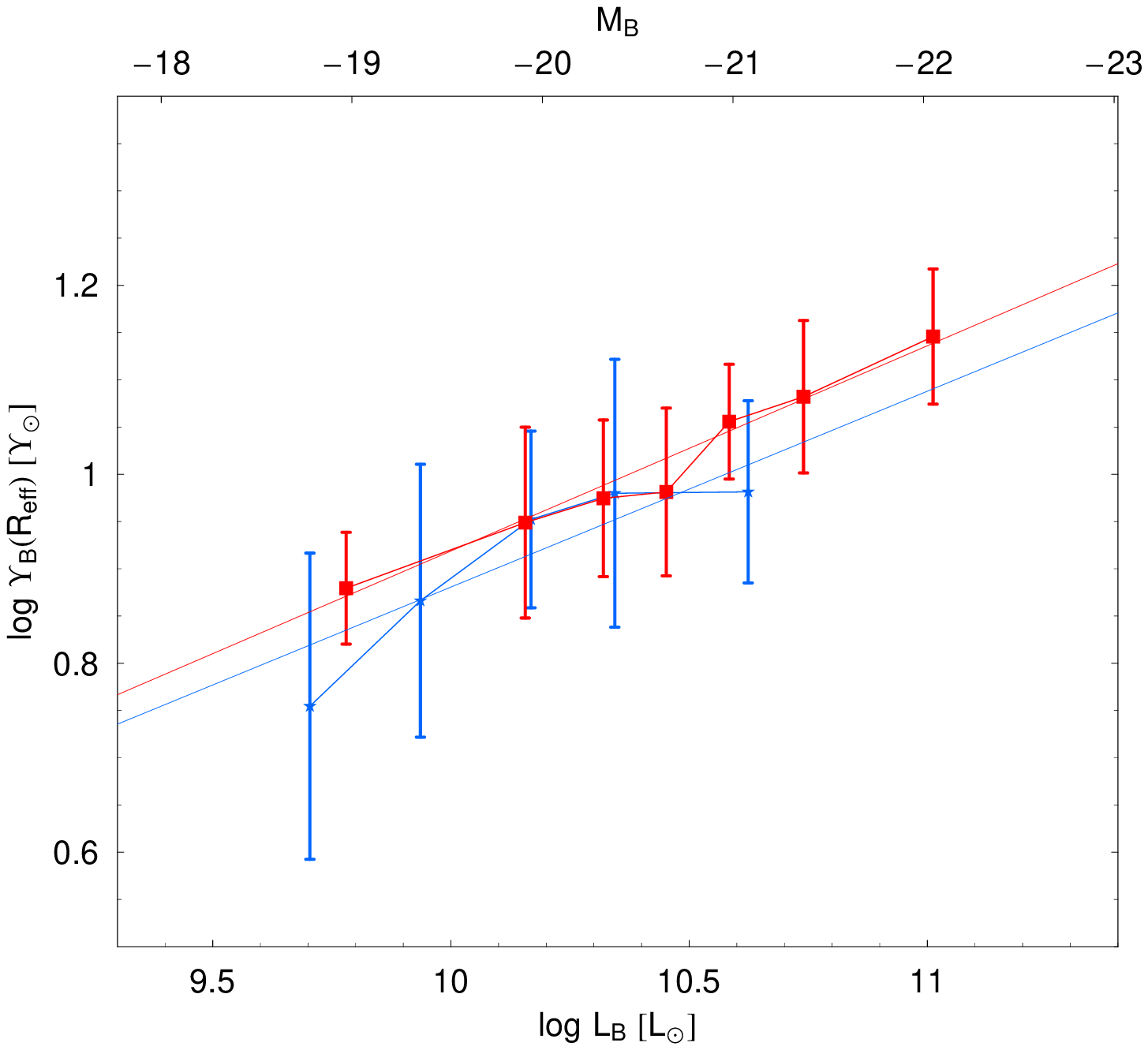, width=0.48\textwidth} \hspace{0.1cm}
\psfig{file= 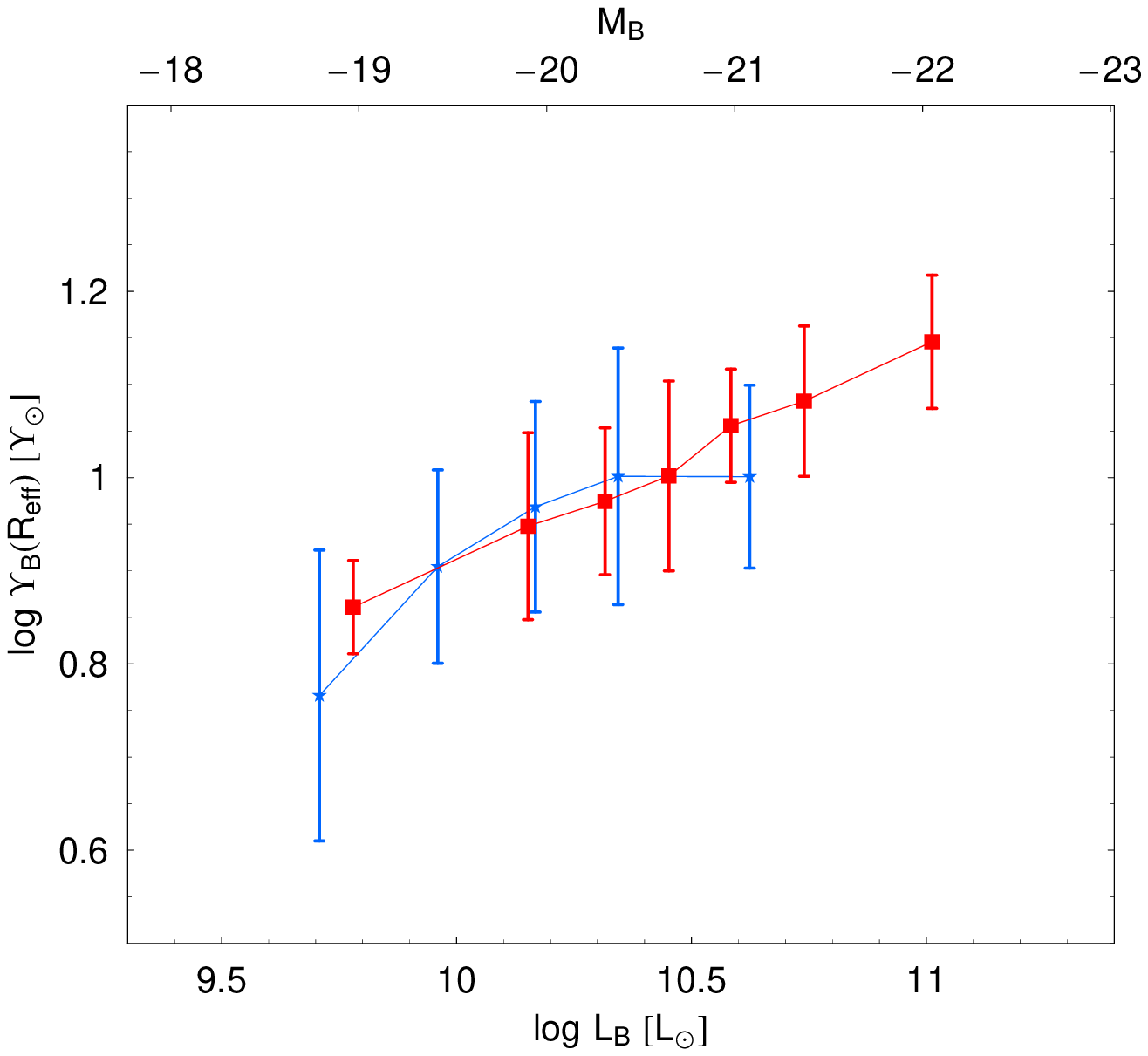, width=0.48\textwidth} \\
\caption{Dynamical \ML\ in $B$-band within \Re\ as a function of
luminosity assuming $j_*$ homology and an SIS total mass profile.
Red squares and blue stars denote E and S0 galaxies, respectively.
Points with error bars are the median values and $\pm25\%$ scatter
for the galaxies in luminosity bins. {\it Left panel:} No rotation
assumed. Linear best fits to the binned data are overplotted as
straight lines. {\it Right panel:} Correction made for rotational
support.} \label{fig:UpsReff_SIS_vs_LB}
\end{figure*}

We can now derive \Ydyn, starting with the simplest case where we
assume no rotation, and a homologous model for the luminosity
distribution $j_*(r)$: an $R^{1/4}$ profile which is completely
determined by the known \LB\ and \Re\ for every galaxy. For the
mass profile we assume initially the SIS model. Fitting the \sigc\
data, we show the mass-luminosity results in
Fig.~\ref{fig:UpsReff_SIS_vs_LB} (left panel). Binning the data,
we fit the median relation $\Ydyn \propto \LB^{\gamma_{\rm dyn}}$
and find $\gamma_{\rm dyn}=0.21 \pm 0.01$. This is identical to
the one for the E subsample alone, $\gamma_{\rm dyn}=0.21 \pm
0.01$ (see Table \ref{tab:slopes}). The S0s show a larger scatter
and have a global slope of $\gamma_{\rm dyn} = 0.18 \pm 0.03$,
which steepens for faint galaxies ($\MB > -20.5$, $\gamma_{\rm
dyn} \sim 0.3$) and appears to flatten or even {\it decrease} at
higher luminosities ($\gamma_{\rm dyn} \leq 0$). These slope
results are scarcely changed by including the rotational
correction (see Table~\ref{tab:slopes}), although the {\it
normalization} of \Ydyn\ is increased for the S0s (see
Fig.~\ref{fig:UpsReff_SIS_vs_LB}, right), an issue to which we
will return in \S\ref{sec:DM}.

\begin{table}
\centering \caption{Slope of \ML--\LB\ relation for Es and S0s and
different dynamical models (the Sersic profiles assume the
$n$-\LB\ relation discussed in the text). The first five rows are
the slope for the total (dynamical) mass, and the last row due to
stars only, as derived with our stellar-populations model
(Section~\ref{sec:MtoL_star}). $^{*}$The faintest S0s have a
steeper slope than the brighter ones (see
Fig.~\ref{fig:Ups_star_vs_LB_Zsol_tau_1Gyr}). Uncertainties on
slopes are the 1$\sigma$ scatter computed by a bootstrap
method.}\label{tab:slopes} \scriptsize
\begin{tabular}{lccc} \hline \hline
  Model & $\gamma_{\rm E}$ & $\gamma_{\rm S0}$ & $\gamma_{\rm tot}$  \\
  \hline
$R^{1/4}$+SIS & $0.21\pm0.01$ & $0.18\pm0.03$ & $0.21\pm0.01$\\
$R^{1/4}$+SIS+rot & $0.20\pm0.01$ & $0.18\pm0.03$ & $0.20\pm0.01$\\
$R^{1/4}$+const-\ML+rot & $0.20\pm0.01$ & $0.18\pm0.03$ & $0.20\pm0.01$\\
S\'ersic+SIS+rot & $0.21\pm0.01$ & $0.20\pm0.03$ & $0.21\pm0.01$\\
S\'ersic+const-\ML+rot & $0.10\pm0.01$ & $0.10\pm0.03$ & $0.13\pm0.02$ \\
\hline Stars & $0.02\pm0.01$ & $0.17\pm0.03$~$^{*}$ &
$0.06\pm0.01$  \\ \hline \hline
\end{tabular}
\end{table}

Previous dynamical studies of ETGs using $j_*$ homology have
found a variety of tilt slopes, ranging from $\gamma_{\rm dyn} \sim 0.1$
to $\sim 0.3$
(e.g. \citealt{1993ApJ...411...34J,Bernardi2003,Padmanabhan04}; T+04).
The average of the literature $B$-band values in Table 1 of PS96
yields $\gamma_{\rm dyn}=0.25\pm0.05$, consistent with our result.

Now considering the other extreme assumption for the mass model,
constant-\ML, the extrapolation to \Re\ after fitting to \sigc\
changes the \Ydyn\ normalization, corresponding to $K=2.05$ at
\Re\ in Eq.~(\ref{eq:M_virial}) for SIS, and $K=1.93$ for
constant-\ML . The {\it slope} of the \Ydyn-\LB\ relation is on
the other hand unchanged (see Table~\ref{tab:slopes}). The $K$
difference does raise the interesting possibility of mass profile
non-homology, e.g. a systematic change with luminosity. As with
the IMF toy model in Section~\ref{sec:stellarML}, we can consider
an arbitrary case where the faintest galaxies have constant-\ML\
profiles, and the brightest ones have SIS.  This would increase
\Ydyn\ by $\sim 0.02$, i.e. mass non-homology does not appear to
be a significant contributor to the FP tilt, assuming $j_*$
homology.

\subsection{Results from generalized luminosity profiles}\label{sec:nonhom}

We next relax the $j_*$ homology assumption,
allowing for more realistic luminosity profiles based on
the S\'ersic law, with surface brightness profiles expressed as:
\begin{equation}
\mu(R) \propto C-(R/R_{\rm eff})^{1/n},
\end{equation}
where $C$ is a constant and $n$ is an index of profile curvature
which correlates with luminosity, such that the brighter galaxies
have higher $n$ (less curved profiles; e.g.
\citealt{Caon93,Graham98,GG03,ML05,2008arXiv0810.1681K}). As
illustrated by Eq.~(\ref{eq:jeans}), for a given dispersion
profile, changing the shape of $j_*(r)$ will affect the inferred
mass. Thus it is important to explore the impact of $j_*$
non-homology on \Ydyn, which may be expressed as a trend with
luminosity $K=K(n|\LB)$. This is all a fancy way to say that {\it
accurate dynamical results require accurate luminosity profile
models.}

The $n$-\LB\ correlation has been investigated for our galaxy sample
by PS97.
From the overall ETG sample in their fig.~5, we define a simple relation where
$n \sim L_B^{0.2}$ for $M_B > -20$, and $n = 4$ for all the brighter galaxies\footnote{PS97
noted that at least one other study found higher values of $n$ for the brightest galaxies,
but commented that those results were more sensitive to the outer profiles than to
the central regions of relevance here.
Similar concerns might apply to the recent smaller galaxy sample of \citet{2008arXiv0810.1681K},
but it is beyond the scope of our paper to re-investigate $n$ dependencies in detail.
If $n$ were systematically higher for the brighter galaxies, then these systems' \Ydyn\ results
would be {\it lower} (cf. next footnote).
Note also that the \Re\ values that we use were obtained by $R^{1/4}$ fitting
in PS96 rather than the self-consistent S\'ersic values, which could in principle
affect the results for the fainter galaxies.
}.
This relation also applies for the E subsample, and we assume that it does for the S0s as well.

\begin{figure*}
\psfig{file= 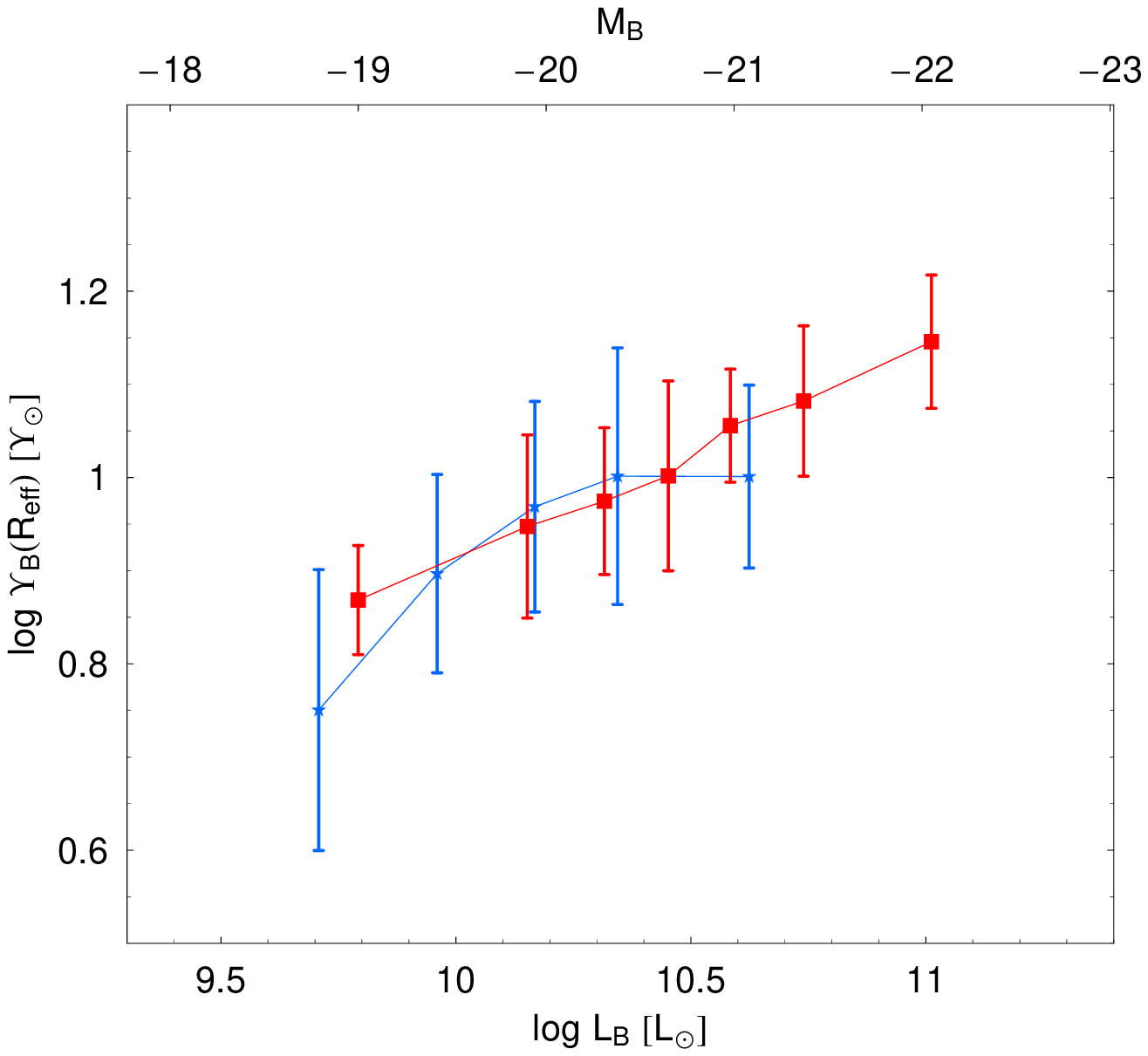, width=0.48\textwidth} \hspace{0.1cm}
\psfig{file= 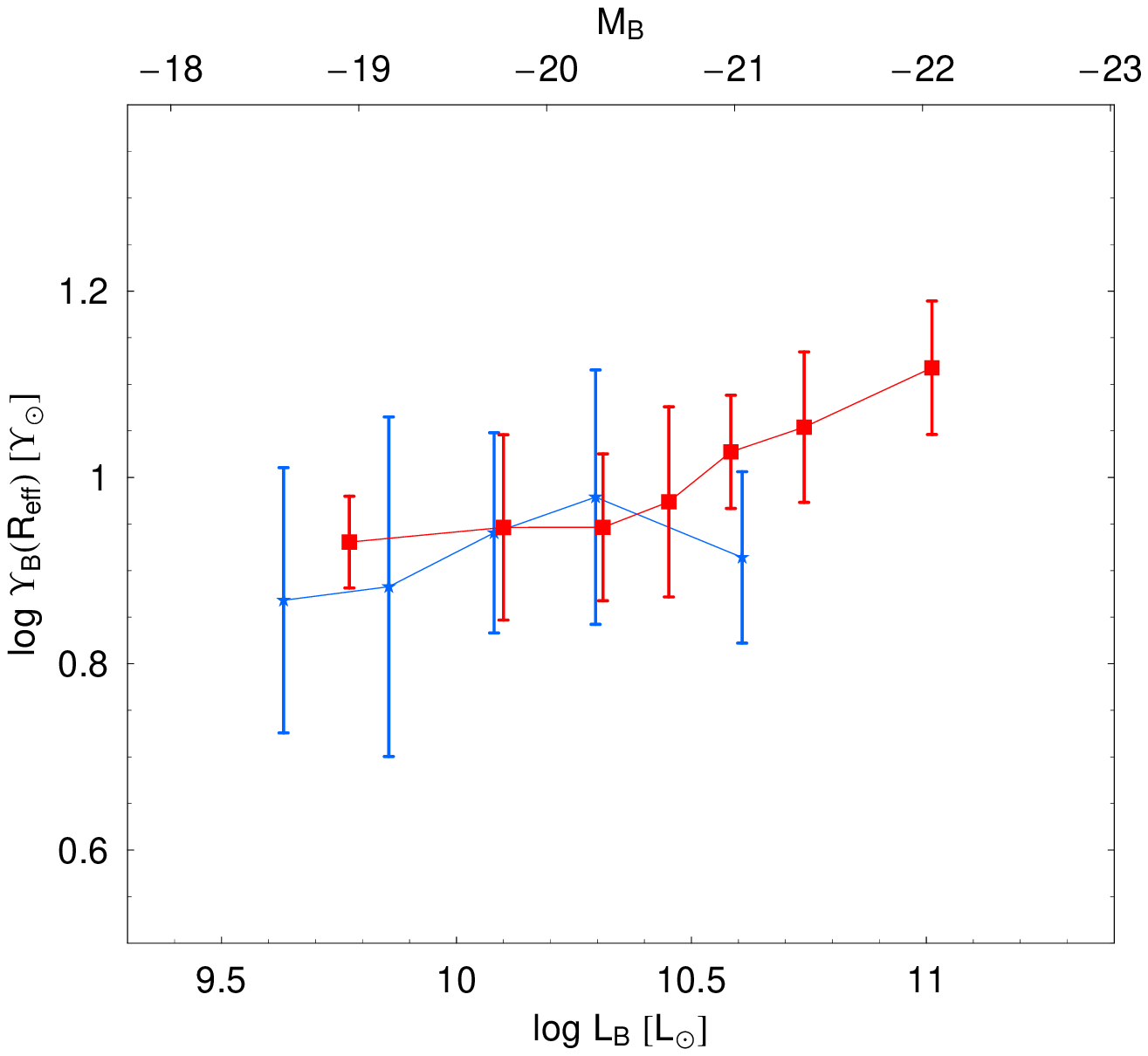, width=0.48\textwidth} \\
\caption{Dynamical $\ML$ as in Fig.~\ref{fig:UpsReff_SIS_vs_LB},
corrected for rotational support and using the S\'ersic luminosity
profile. {\it Left panel:} SIS mass model. {\it Right panel:}
constant--\ML.} \label{fig:MtoL_nonhom}
\end{figure*}

We now use the $n$-\LB\ relation to construct the $j_*(r)$
S\'ersic profile for each galaxy as needed for the dynamical
modeling (Section~\ref{sec:dynmeth}). Since we have examined the
effects of rotation in Section~\ref{sec:hom}, we will here skip
over the simplified case of no rotation. The resulting \Ydyn\
values for both SIS and constant-\ML\ cases are summarized in
Table~\ref{tab:slopes} and Fig.~\ref{fig:MtoL_nonhom}. For the SIS
case, relaxing the $j_*$ homology slightly changes the slope
$\gamma_{\rm dyn}$. However, in the constant-\ML\ case, both the
luminosity {\it and} mass profiles are affected, and significant
differences arise. The masses are increased for the fainter
galaxies\footnote{This is because lower $n$ for the fainter
galaxies implies shallower central profiles of both luminosity and
mass density, and therefore lower central velocity dispersions at
a fixed mass, finally requiring higher model masses to match the
observations. This effect might be somewhat reversed by the
central cusps of light generally found in faint Es to be
superimposed on their overall S\'ersic profiles, in many cases on
spatial scales comparable to the \sigc\ measurement (e.g.
\citealt{2008arXiv0810.1681K}), but it is beyond the scope of this
paper to consider this possibility in detail.}, causing the \Ydyn\
slope to become shallower overall ($\gamma_{\rm
dyn}=0.13\pm0.02$), and even constant at lower luminosities
($\gamma_{\rm dyn} = 0.05\pm0.04$ and $0.23\pm0.02$ for the faint
and bright Es, respectively). The Es and S0s are again not
noticeably different in their region of luminosity overlap. In
Appendix~\ref{sec:appB} we investigate systematic uncertainties in
these results, whose impact we will consider in the next section.

\begin{figure*}
\psfig{file= 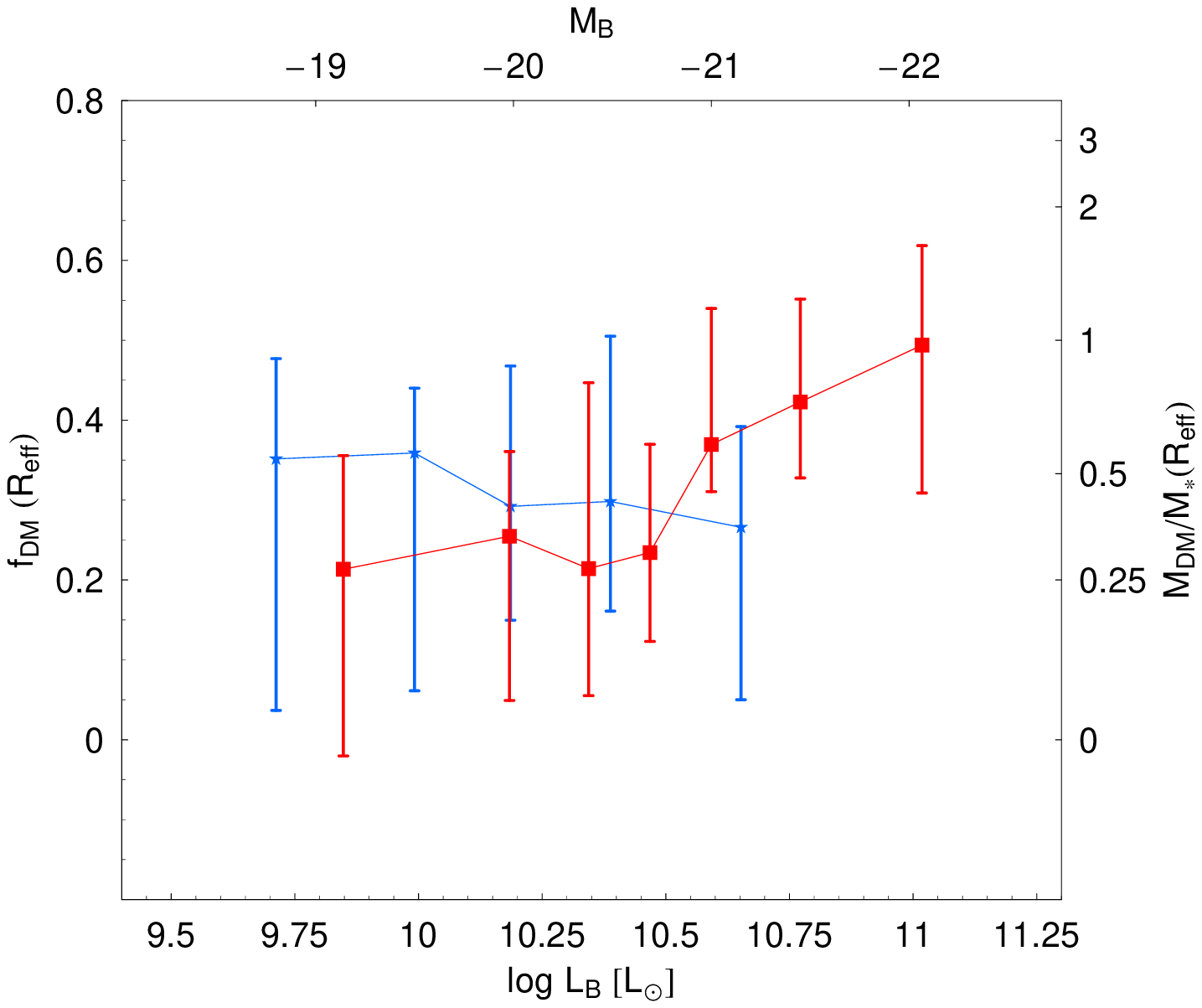, width=0.48\textwidth} \hspace{0.1cm}
\psfig{file= 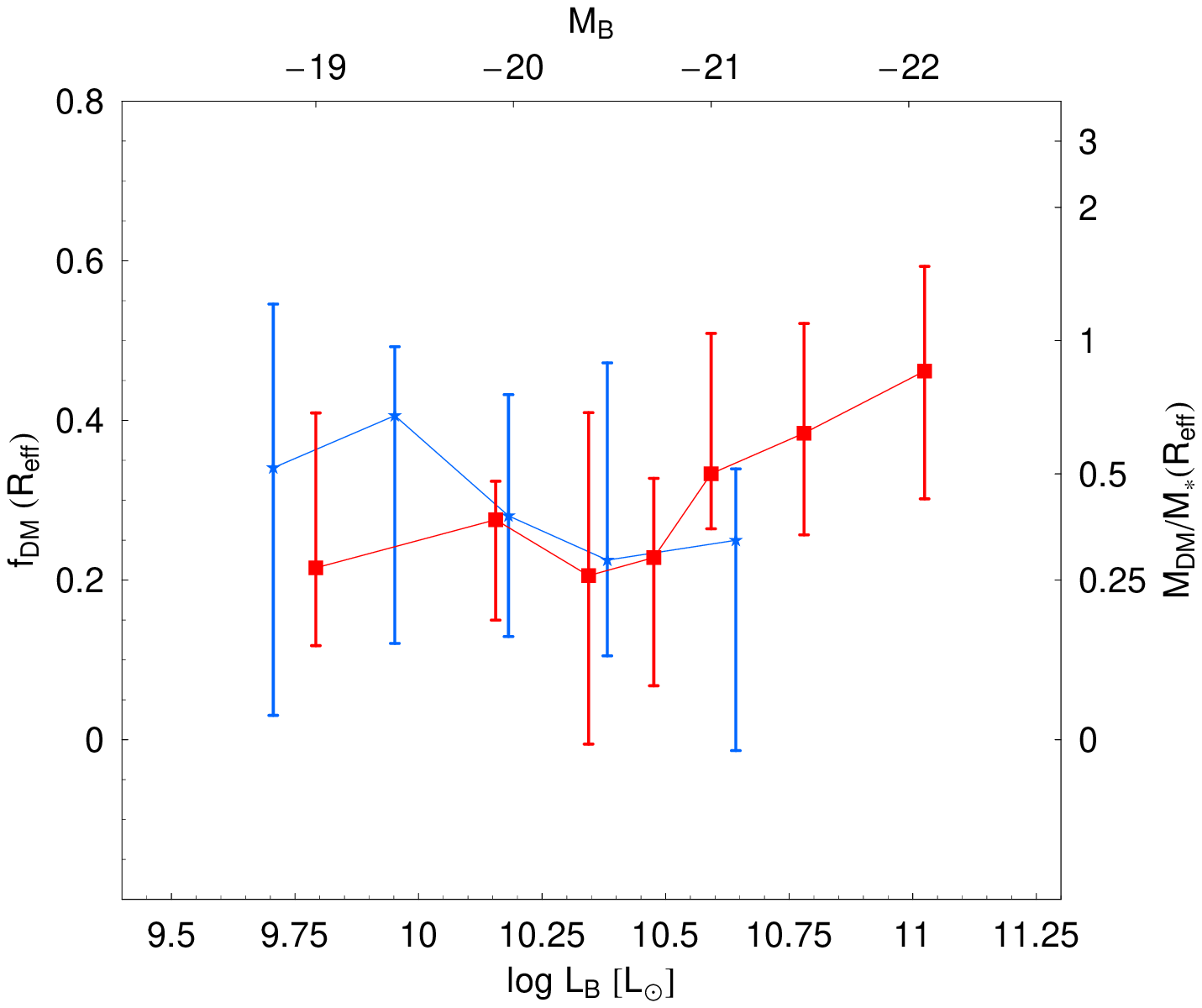, width=0.48\textwidth} \hspace{0.1cm}
\caption{ Trends of DM fraction with luminosity, using SIS and
constant-\ML\ mass models (left and right panels, respectively).
Symbols are as in
Figs.~\ref{fig:UpsReff_SIS_vs_LB}--\ref{fig:MtoL_nonhom}.}
\label{fig:fDM_nonhom}
\end{figure*}

\begin{figure*}
\psfig{file= 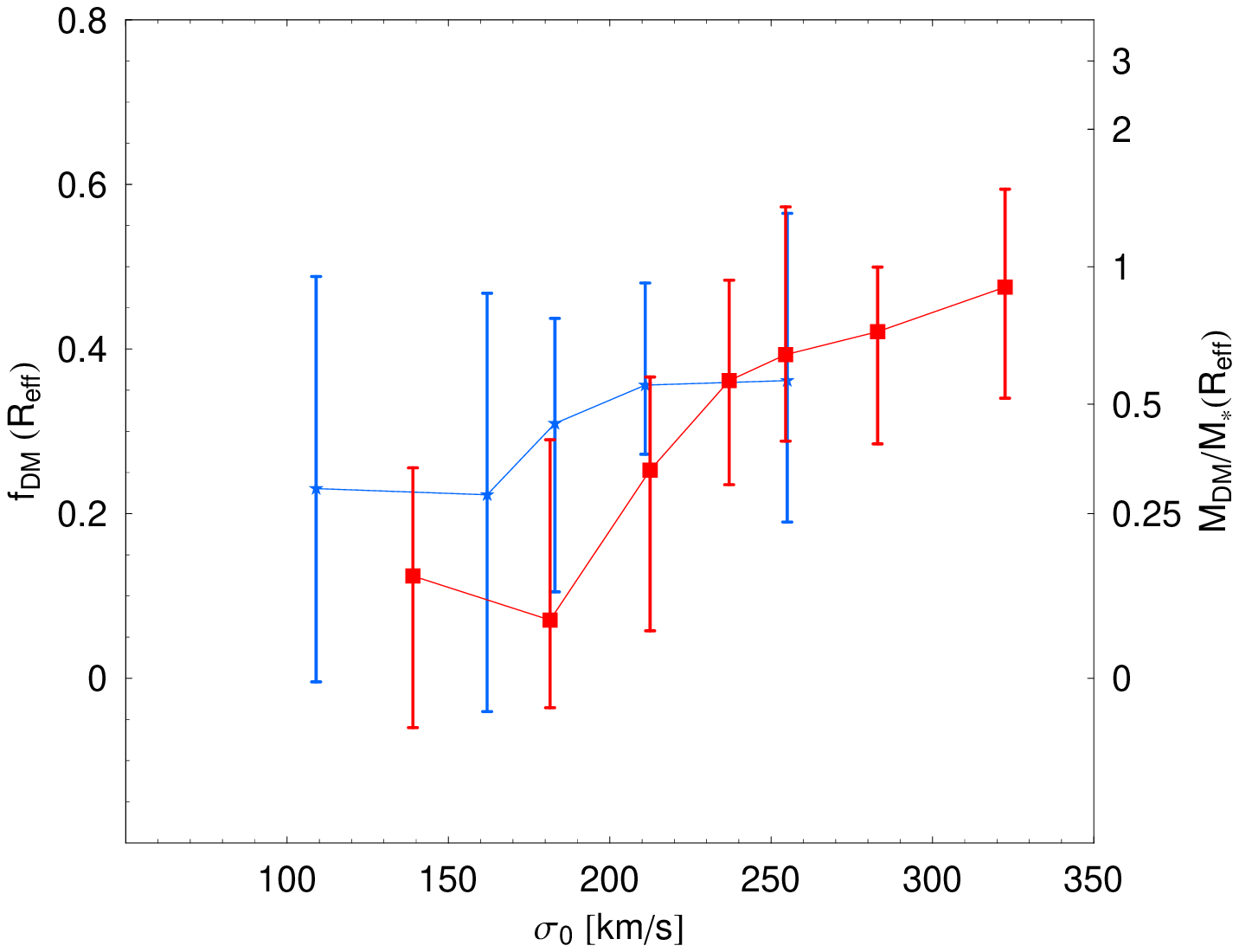, width=0.48\textwidth} \hspace{0.1cm}
\psfig{file= 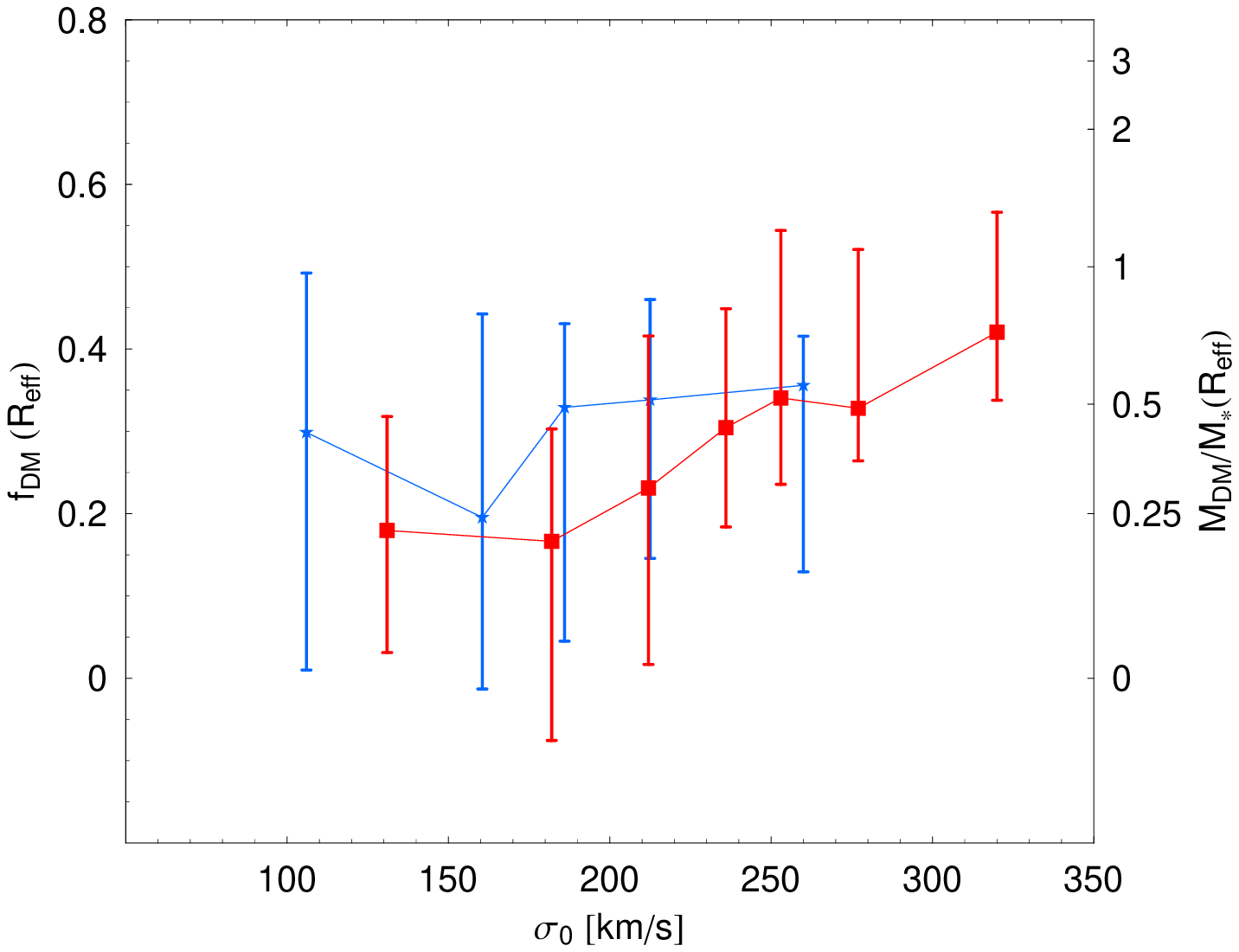, width=0.48\textwidth} \hspace{0.1cm}
\caption{Trends of DM fraction with velocity dispersion, using SIS
and constant-\ML\ mass models (left and right panels,
respectively). Symbols are as in
Figs.~\ref{fig:UpsReff_SIS_vs_LB}--\ref{fig:MtoL_nonhom}--\ref{fig:fDM_nonhom}.}
\label{fig:fDM_vs_sig0_nonhom}
\end{figure*}

Now reviewing the results of this and the previous Sections,
with the $j_*$ homology assumption, the steep \Ydyn\ slope relative
to \Yst\ ($\gamma_{\rm dyn} = 0.20$ vs $\gamma_*=0.06$)
would imply that $\sim75\%$ of the FP tilt is related to DM content or
some other factor.
Including the (realistic) $j_*$ non-homologies changes the picture
somewhat: if all galaxies have SIS mass profiles, the previous
conclusion is unchanged.  If they have steeper mass profiles,
then the dynamical contribution to the tilt decreases, and for the
fainter galaxies may even disappear.

Thus our results suggest overall that DM contributes to the tilt
for the brightest galaxies, while the contribution for the fainter
galaxies is unclear but probably less. This conclusion differs
from that of T+04, who found using similar S\'ersic models and
{\it assuming constant-M/L}, no need for a correlation between DM
fraction and luminosity.

Their galaxy sample is fainter and much smaller, so their results
are actually consistent with ours in general. The exception is for
the brightest galaxies, where the higher $n$-values of T+04 lead
to less tilt than we find. In any case, it should be noted that
reproducing the FP tilt without DM variation is not a unique
solution, and as we have shown, DM could still be a primary driver
of the tilt.

\section{Dark matter fractions}\label{sec:DM}

Having analyzed the trends for stellar and total mass in our galaxy
sample, we now examine the implications for DM content.  We define
the DM fraction within the three-dimensional radius $r=1~\Re$ by:
\begin{equation}
\fDM=\frac{\mtot-\mst}{\mtot}=1-\frac{\Yst}{\Ydyn} ,
\end{equation}\label{eq:fDM}
where for physically meaningful results we should have $\Yst\ \leq \Ydyn$
and thus $\fDM\ \geq 0$. Strictly speaking, our derived \Yst\ should be
{\it deprojected} before computing \fDM, but we do not have the information
necessary to do so.  Given the negative colour gradients in ETGs, we expect
the deprojected \Yst\ to be somewhat higher than in projection, and thus the
true \fDM\ to be somewhat lower. For this reason and especially because of
the large IMF uncertainty, {\it the absolute values for \fDM\ are not
definitive, but instead the relative variations are more robust and are
the focus of our study.}

We now consider the \fDM\ trends found for our galaxy sample,
taking as a default the \Yst\ estimates from the generalized
($\age,\tau,Z$) BC03-based stellar populations model, and the
\Ydyn\ estimates from the dynamical models using generalized
luminosity profiles (Section~\ref{sec:nonhom}). As shown in
Fig.~\ref{fig:fDM_nonhom}, \fDM\ increases with luminosity in the
E galaxy subsample, but is constant or even {\it decreasing} for
the S0s. The combined ETG sample has \fDM\ increasing overall, but
with the hint of a slope change at $M_B \sim -20.5$, from roughly
constant at faint magnitudes to steeply increasing for brighter
objects; the trends for the S0s and the correspondingly fainter Es
are roughly consistent. These conclusions are valid for both
bracketing mass profile cases (SIS and constant-\ML), although the
slope change is less apparent for the SIS model. To quantify this
breakdown, we have measured the slopes $\gamma_{DM}$ of \fDM--\LB\
relation for the two models, and found that $\gamma_{\rm DM} \sim
0.5$ for $\log \LB \gsim 10.4 \, \Lsun$ and $\gamma_{\rm DM} \sim
0$ for $\log \LB \lsim 10.4 \, \Lsun$, clearly inconsistent within
the errors.

\begin{figure*}
\centering \psfig{file= 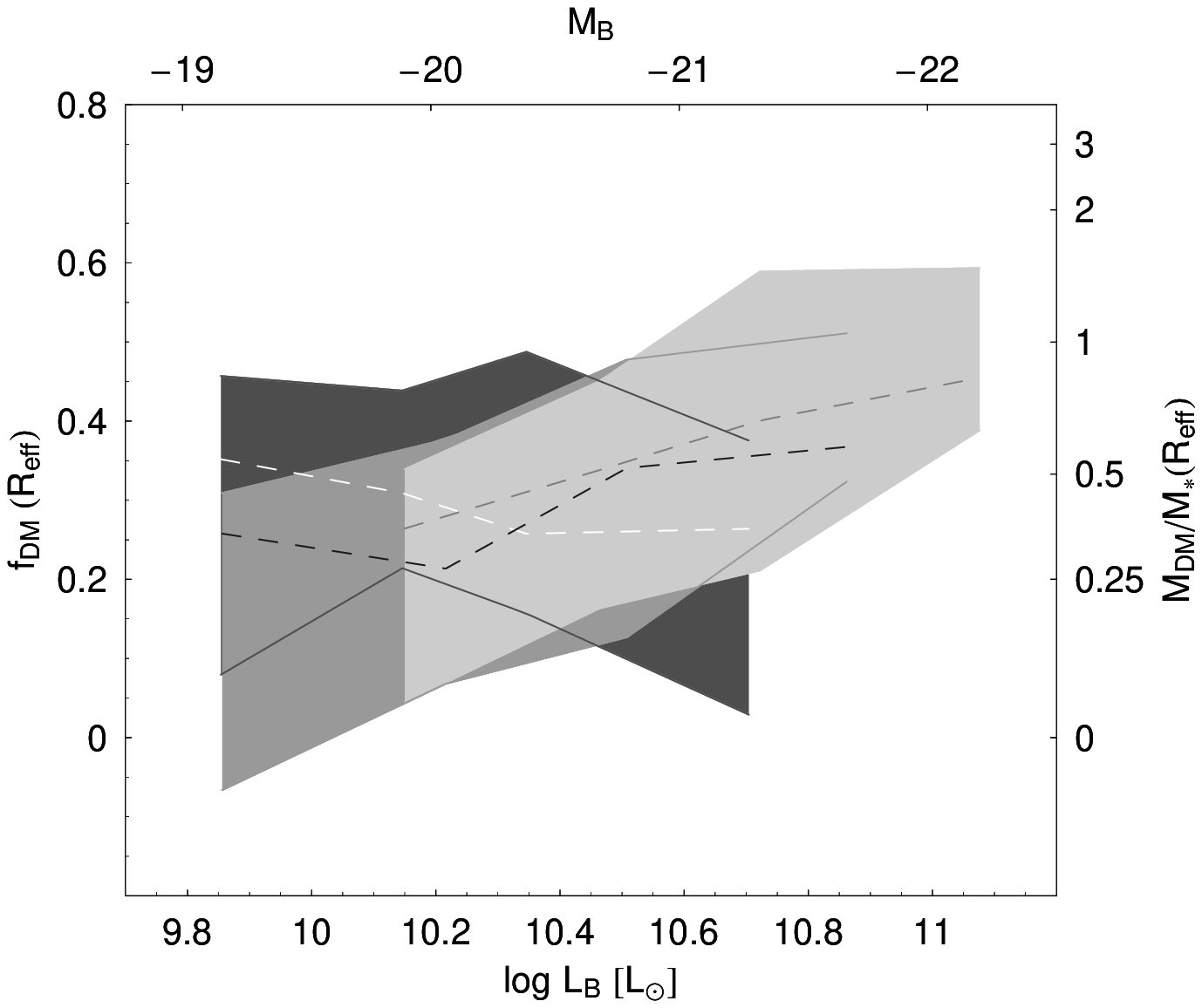, width=0.47\textwidth}
\psfig{file= 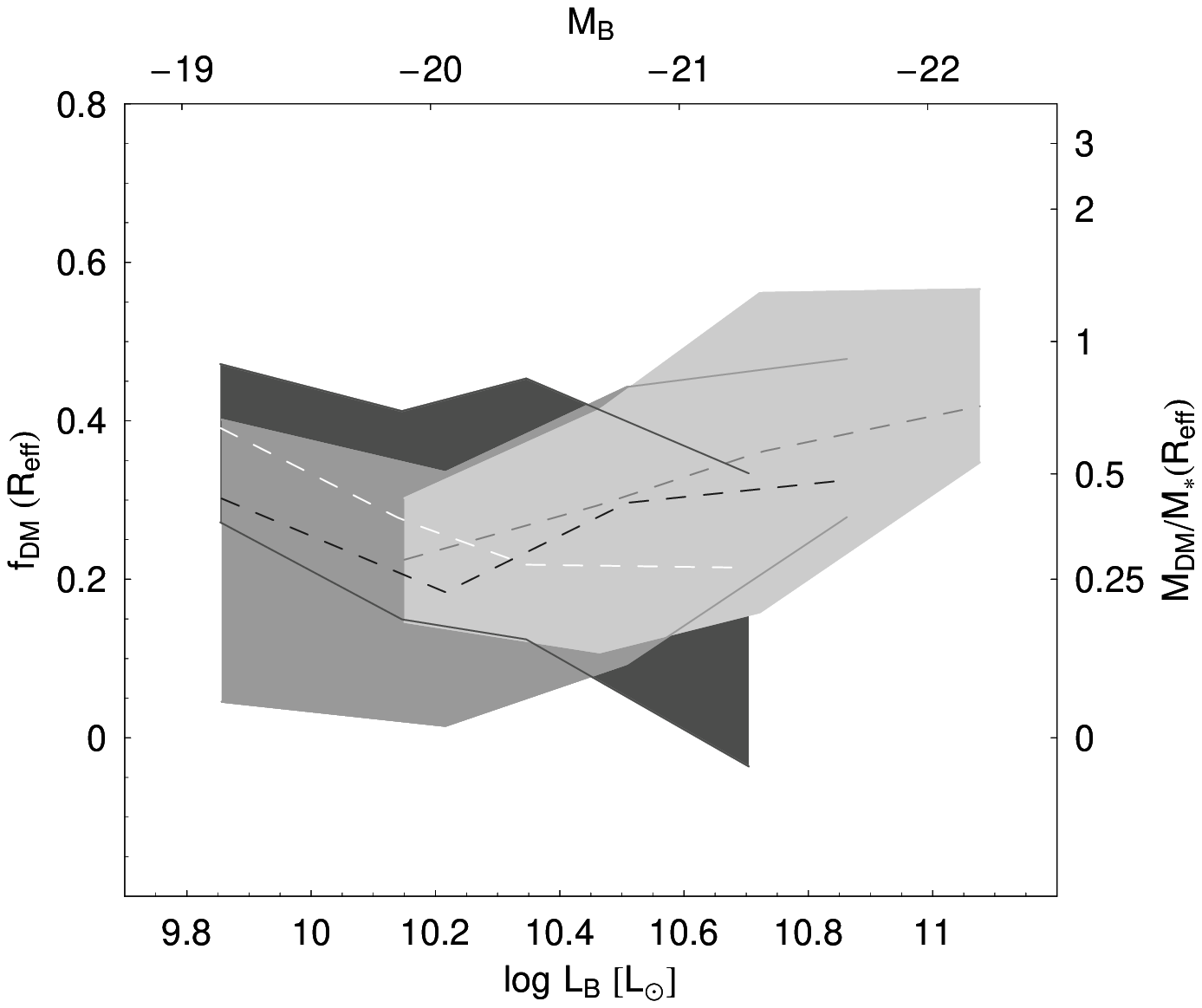, width=0.47\textwidth} \caption{DM
fraction trends for different galaxy types, using SIS and
constant-\ML\ mass models (left and right panels, respectively).
Dashed lines show median binned values and shaded areas show
$\pm25\%$ scatter. Light grey shows slow rotator Es, medium grey
shows fast rotator Es and dark grey shows S0s. } \label{fig:fDM_1}
\end{figure*}

The trends with luminosity are mirrored by similar correlations
with the velocity dispersion as we show in Fig.
\ref{fig:fDM_vs_sig0_nonhom}. Here the slope is steeper in general
because of the combined effect of the \Yst\--\sigc\ correlation
shown in Fig. \ref{fig: MtoLstar_vs_sig0} and the stronger
dependence between \Ydyn\ and \sigc.

The DM fraction is typically $\fDM\ \sim 0.3$ assuming a Salpeter
IMF, with a broad range for individual galaxies from $\sim 0$ to
$\sim 0.9$ (rms scatter of $\sim 0.15$). About 15\% of the
galaxies have, within the errors, $\fDM\ < 0$ (typically those
with low surface brightness $\mu_{\rm eff}$), an unphysical result
which may indicate that the Salpeter IMF is inaccurate (cf. C+06);
adopting a Chabrier IMF would imply more DM, with $\fDM\ \sim 0.6$
typically, and only a tiny handful of galaxies with $\fDM\ < 0$.
Changing the IMF also flattens slightly the luminosity dependence
of \fDM, since this quantity is not directly proportional to \Yst.

We next look for any DM differences between the fast-rotator and
slow-rotator Es, following the classification in
Section~\ref{sec:dynmeth}. However, as shown in
Fig.~\ref{fig:fDM_1}, there is no discernible difference; slow and
rotators typically have $\fDM\ \sim 0.35$ and $\sim 0.25$,
respectively, but this is consistent with a simple luminosity
effect, since fast rotators are fainter on average than slow
rotators. This result appears contrary to the finding of C+06
(based on more detailed dynamical models and somewhat different
stellar populations constraints for a much smaller galaxy sample)
that there is an \fDM\ discontinuity between slow and fast
rotators.

We also compare S0s in Fig.~\ref{fig:fDM_1}, where it appears that
their declining trend of \fDM\ with luminosity is inconsistent
with the Es in the same luminosity range. However, we caution that
our spherical dynamical models are most questionable for the S0s,
so the overall situation appears consistent with a continuous
trend of \fDM\ with luminosity for all ETGs, independent of
morphology and rotation. For the rest of the paper, we will
therefore generally lump all these ETG sub-classes together as one
population.

Before continuing further, we check once more the effects of
systematic uncertainties, as detailed in Appendix~\ref{sec:appC}.
Despite the uncertainties, our default model is consistent with
results on DM content at larger radii, and we therefore consider
the overall mild increase of \fDM\ to be robust, with the
inflection at intermediate luminosities perhaps less so.

How do our results compare to previous studies of DM trends in ETG centres?
The analysis most similar to ours is from T+04.
As discussed in Section~\ref{sec:nonhom}, they found no indication of a
correlation between \fDM\ and luminosity, but their sample was primarily
of faint galaxies, where we also found the correlation is weak.
If we adopted higher S\'ersic indices for the brightest galaxies,
the correlation would also weaken for them, but this scenario would seemingly
be inconsistent with large-radius tracers of DM
(see App.~\ref{sec:appC}).

\cite{borr+03} modelled a large sample of ETGs dynamically and
claimed that the flatness of the FP would not permit
centrally-concentrated DM halos as predicted by cosmological
models.  Their results imply $\gamma_* = 0.27 \pm 0.04$, thereby
explaining all the tilt through the stellar populations -- in flat
contradiction to our $\gamma_* = 0.06 \pm 0.01$. This is mainly
the consequence of their choices to not allow for a systematic
variation of the virial DM fraction with luminosity in their
model, and to use homologous $j_*(r)$ profiles which we have seen
produce misleading results.

Finally, \cite{Padmanabhan04} analyzed a large sample of SDSS ETGs,
using a combination of stellar populations and dynamical models.
Although the different redshift ranges make comparisons
not straightforward, their results do appear roughly equivalent
to ours, with $\gamma_* \sim 0$ and $\gamma_{\rm dyn} \sim 0.17$,
and even a hint of a flattening of the \fDM\ slope at lower
luminosities.  Note however that they used an inaccurate homologous $j_*(r)$
profile.

\section{Implications: dark matter and galaxy formation}\label{sec:av_DM}

The trends we have seen for \fDM\ as a function of luminosity
could provide fresh clues to galaxy formation. The most basic
interpretation of central DM variations is that they reflect
variations in {\it total} DM within the virial radius. Assuming
that the Universal baryon fraction is roughly conserved from
galaxy to galaxy, the implication is then that higher \fDM\ means
lower efficiencies of star formation \eSF . In this respect, the
trends we find are qualitatively expected. Both observations and
theory point to a universal U-shaped trend of \eSF\ (or
equivalently virial \ML) with luminosity, and a peak efficiency at
$M_* \sim 10^{11} M_\odot$ (e.g. \citealt{Benson2000,MH02}; N+05;
\citealt{vdeB07}).

Physically, the lowest-mass galaxies are least able to retain
their primordial gas content long enough to form many stars, since
their gravitational potential wells are not deep enough to prevent
ejection from supernovae feedback. More massive galaxies are
increasingly able to inhibit feedback and form more stars, but at
a certain mass scale, additional processes kick in such as AGN
feedback, inhibiting gas cooling and decreasing \eSF\ again (e.g.
\citealt{2006MNRAS.370.1651C,2006ApJ...643...14S,Kaviraj07,
Tortora2009}). Thus, the lowest-mass and the highest-mass galaxies
are the most DM-dominated. Our current galaxy sample does not
extend faint enough to discern any U-shape, but the change we see
in the \fDM\ trend below scales of $M_B \sim -20.5$ or $M_* \sim
10^{11} M_\odot$ does coincide with the generically expected
minimum of DM content. The consistencies of the trends for the ETG
sub-types (S0s, fast-rotator Es, slow-rotator Es) suggest that the
dominant driver of star formation is mass, not angular momentum.

It is of course a stretch to draw firm conclusions about virial
quantities based on data from scales $\ll \Re$. The central DM
content that we are actually probing may be decoupled from the
overall DM content in several ways: the central DM density
reflects the ambient density at the time of initial halo collapse;
the baryons could have interacted with the DM and changed its
distribution; and the \fDM\ quantity that we measure is somewhat
dependent on the particular values of \Re\ for the stars rather
than simply probing the DM properties. To allow for such effects,
and to provide quantitative marks for comparison to cosmological
theory, we now consider the properties of the DM alone, in terms
of its average density within some small radius, \rhoDM.

\begin{figure}
\hspace{-0.75cm} \epsfig{file=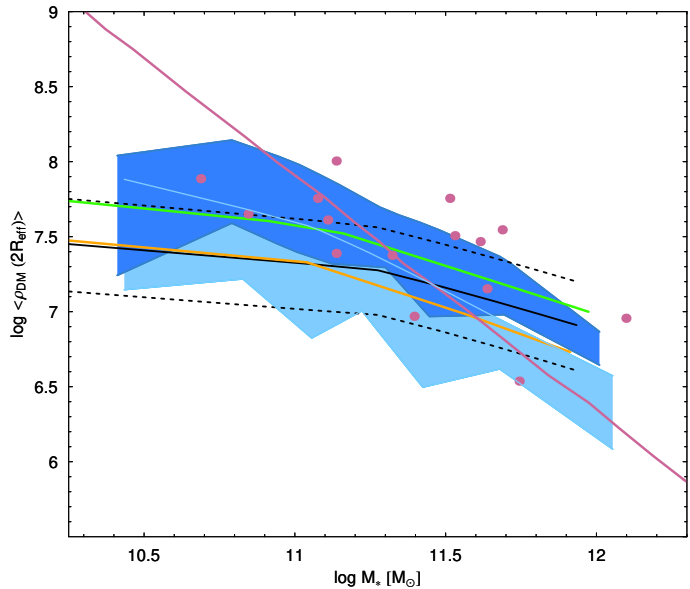, width=0.52\textwidth}
\caption{Central DM density in ETGs, versus stellar mass.  The
density is averaged within a radius of 2~\Re, and is in units of
\Msun/kpc$^3$; a Salpeter IMF is assumed. The 25$^{\rm
th}$-75$^{\rm th}$ percentiles are shown as darker and lighter
blue for the S\'ersic+SIS and const$-\ML$ models respectively.
Dots are the T+08 galaxy sample with the best fit shown as purple
solid line. Orange and green solid lines are the $\Lambda$CDM
predictions using $\eSF=60\%$ and 6\%, respectively. The black
solid line is the model assuming a varying \eSF\ (see text for
details). The dotted curves represent the range of model
predictions due to the scatter in the \Re-\mst
relation.}\label{fig:rho_dm}
\end{figure}

In order to make critical comparisons with a literature study
discussed below, we estimate \rhoDM\ within 2~\Re, extrapolating
our usual models outwards in radius. We present this result vs
stellar mass for the whole ETG sample in Fig.~\ref{fig:rho_dm},
using a Salpeter IMF and alternatively the SIS or constant-\ML\
mass profile. Although these bracketing mass profiles gave similar
results for \fDM\ within 1~\Re, at 2~\Re\ they start to diverge
more, giving noticeably different results for \rhoDM. The less
massive galaxies have increasingly dense DM halos, apparently
reaching a plateau of $\rhoDM\ \sim 0.05$~$M_\odot$~pc$^{-3}$ at
masses below $\log M_*/M_\odot \sim 11$.

As a reality check, we compare the results for flattened ETGs in
the Coma cluster from \citet[hereafter T+08]{Thomas08}, who used
detailed three-integral axisymmetric dynamical models of stellar
kinematics to decompose the galaxies into their stellar and DM
mass components. Their \rhoDM\ values from their NFW halo model
match up remarkably well with our SIS-based results. T+08 fitted
their data with a logarithmic density-mass trend which would imply
very high central \rhoDM\ for the faintest galaxies.  However, as
we can see in the Figure, such conclusions would involve
extrapolating outside the mass range covered by the data, and in
fact the T+08 results do show some sign of the density plateau at
small masses which we find.

Now we calculate predictions from $\Lambda$CDM cosmological
models, adopting an NFW density profile, and the \citet{bull01}
mass-concentration relation, as discussed in N+05. The final
parameter in this model is the mass ratio between stars and DM
within the virial radius, taking plausible values of alternatively
$M_*/M_{\rm vir} = 0.1$ or 0.01 (corresponding to $\eSF \sim 6\%$
or $\sim 60\%$, respectively, for a baryon fraction of 0.16; see
N+05 and \citealt{WMAP}). The results for \rhoDM\ are shown in
Fig.~\ref{fig:rho_dm}, where the model predictions are seen to be
fairly consistent with the observations, including the bend in the
\rhoDM\ trend at similar galaxy masses.  This bend is not caused
by anything intrinsic to the DM itself, but by the radius adopted
for measuring the density. As shown in Fig.~\ref{fig:Sample}
(right panel), the mass-\Re\ relation for ETGs has a bend, which
probably explains not only the density trends seen in this section
but also the \fDM\ results of the previous section. For bright
galaxies, \Re\ increases rapidly with mass, probing quickly into
regions contain more DM, at lower averaged densities. Fainter
galaxies have less quickly varying \Re\, which thus tracks the
slowly-varying DM scale radius more closely, so that the
observable DM properties are roughly constant.

In more detail, the $\eSF\ = 60\%$ theoretical case coincides
roughly with our observational findings assuming a constant-\ML\
profile, and the $\eSF\ = 6\%$ case coincides nicely with our
SIS-based findings. However, neither of these cases is plausible
observationally or theoretically for the full range of galaxy
masses. Our final case invokes a transition from high \eSF\ for
the faint galaxies to low \eSF\ for the bright galaxies, which is
generically expected from various lines of evidence. More
specifically, motivated by the findings of N+05 based on
radially-extended dynamical studies of ETGs (see also e.g.,
\citealt{Nap08}), we assume that the bright galaxies have \eSF\
decreasing steadily from 60\% to 16\% in the mass range of $\log
\mst\ = $~11--12~$M_\odot$; the faint galaxies have a constant
$\eSF\ = 90\%$. As shown in Fig.~\ref{fig:rho_dm}, this model
would be consistent with our SIS findings at the bright end, and
with constant-\ML\ at the faint end. However, the NFW-based models
themselves would have roughly SIS profiles for the entire range of
luminosity, which means this set of model assumptions is not
self-consistent. It is beyond the scope of this paper to explore
the possible combinations of DM parameters that would be fully
consistent with the data, but we speculate that the low-luminosity
objects have low-concentration DM haloes. Note that changing the
IMF to Chabrier would not significantly change these conclusions,
since the data curves would shift up and to the left in
Fig.~\ref{fig:rho_dm}.

\section{Conclusions}\label{sec:conclusions}

The relative amounts of dark and luminous mass in ETGs is
crucial information for understanding the internal structure of these
systems and their formation mechanisms. In this paper we have
analyzed both the stellar and dynamical \ML\ in the central
regions of one of the largest homogeneous samples of local
early-type galaxies, provided by PS96.

We estimate the stellar content by accurate stellar population
synthesis models of several observed colours using the BC03
prescription. We measure dynamical masses using the observed
central velocity dispersion \sigc\ and several simplyfing
assumptions in the Jeans equations.

We find that the stellar \ML, $\Yst$,
 has a shallow trend with luminosity with a
slope $\sim 0.06$ for the whole ETG sample (with S0s showing a
steeper trend than the Es: see Table \ref{tab:slopes}).
Dynamical \ML, on the other hand, have a slope for the \MLga\ relation of
is $0.21\pm0.01$ when considering ETGs as a (photometrical and
kinematical) homologous galaxy family, i.e. fully consistent with
results derived in local galaxies' $B$-band FP.

For the non-homology case (i.e. assuming the S\'ersic profile for
the light distribution and differential rotation within \Re), we
find that using the SIS model as the total mass distribution does
not much affect the \ML\ slope and thus not the FP tilt either. On
the contrary, non-homology can account for as much as $\sim 40\%$
if considering the constant--\ML\ model, and even more (up to
80\%) for the faint systems.

A further $30\%$ (i.e. 0.06/0.21) is provided by the \Yst\ slope.
The residual contribution to the \MLga\ slope ($\sim 70\%$ for the
SIS model and $30\%$ for constant--\ML) is mainly due to a
variation with luminosity of their DM fraction.

It must be stressed that this {\it average} budget of $\gamma$
contributions masks a more complicated distribution with
luminosity. For instance, for the bright/massive galaxies (i.e.
$\log \LB \gsim 10.4 \, \Lsun$ and \ $\log \mst\gsim 11.3 \,
\Msun$) which have a quasi-$R^{1/4}$ profile and little or no
rotation, the effect of the non-homologies is minimal and the
slope of the \MLga\ remains steeper than the faint systems where
non-homologies can account for almost all the slope $\gamma$.
This, obviously, relates to the trend of the DM fractions
discussed in Section~\ref{sec:DM}.

Here we have seen that \fDM\ is strongly varying with luminosity
and mass. In particular, we observe a dichotomy in DM content of
bright and faint Es: galaxy brighter than $M_{B}\sim -20.5$ and
more massive than $\log \mst \sim 11-11.3 \, \Msun$ have an
increasingly larger \fDM\ while galaxies lying below these
luminous and mass scales invert the trend, such that \fDM\ is
constant or marginally decreasing with luminosity and mass. When
separating the E sample into ``slow'' and ``fast'' rotators it is
evident that this two-fold trend is mainly found in the fast
rotator systems (see Fig. \ref{fig:fDM_1}). These two kinematical
varieties do not show large differences in their \fDM\ properties.
In particular, we do not find significant evidence for
systematically lower \fDMRe\ for the fast rotator variety (C+06),
although with a large scatter one might make such a conclusion
using a small statistical sample. The inclusion of the ellipticity
and orbital anisotropy would increase the steepness of the
faint/less massive sample, but would leave unaffected the
bright/massive galaxy range, still maintaining the dichotomy (Fig.
\ref{fig:fDM_2}, bottom right). As an alternative to a variable DM
content, we have briefly analyzed the effect of a change of IMF as
a function of luminosity (see Fig.
\ref{fig:Ups_star_vs_LB_Zsol_tau_1Gyr_IMFchange}), which could
also explain the FP tilt.

The \fDM\ dichotomy adds to other well known ETG correlations as
found in the $\mu_e-\Re$ relation, FJ, size-luminosity (or
size-mass) relations and in the correlations of S\'ersic index
with both galaxy size and luminosity, as discussed in
Sections~\ref{sec:sample}, \ref{sec:size_relations} and
\ref{sec:dynM} (\citealt{1992MNRAS.259..323C}, \citealt{PS97},
\citealt{Shen2003}, \citealt{MG05}, etc.). Our results mirror the
DM content in the outskirts of galaxies, where variations of
virial $M/L$ as a function of mass and luminosity have been found
both in simulations and observational analysis
(\citealt{Benson2000}, \citealt{MH02}, \citealt{vdeB07}). A
similar dichotomy in DM content is not observed for S0s, which are
generally fainter and less massive than Es and are strongly
affected by rotational support (influencing the normalization of
\Ydyn). They have a slightly higher DM fraction and show a
monotonically decreasing trend with mass and luminosity,
consistent with what is known for spiral galaxies
(\citealt{Persic93}).

A continuity in DM content of galaxy as a function of amount of
rotational support is possibly shown in Fig. \ref{fig:fDM_1},
where we plot DM fractions as a function of luminosity for slow
and fast rotators and lenticulars.

Looking at the average central DM density, \rhoDM, we
have found that this quantity has a fairly small scatter within
the ETG sample. Albeit model dependent -- the S\'ersic+SIS model
providing \rhoDM\ which are $0.2-0.4$ dex larger than the ones
obtained with the const$-\ML$ -- the overall trend of the galaxy
distribution decreases monotonically with the stellar mass and
luminosity in good agreement with independent results obtained
by Thomas et al. (2008) for ellipticals in the Coma cluster. Our
larger statistical sample, though, has allowed us to discern the
presence of a ``knee'' in the distribution (around the usual
mass/luminosity scale at \Mcrit\ and \Lcrit) where the relation of
the more massive/luminous galaxies bends to a steeper slope than
the one followed by the less massive/luminous systems. We have
shown that this ``knee'' can be explained with the change of the
slope in the $\Re-\mst$ relation at \Mcrit.

As a robust estimator of the central DM density, \rhoDM\ can
be compared against the expected values for standard NFW
profiles. The match found is broadly good, with the results
obtained assuming the S\'ersic+SIS model favoring high
dark-to-luminous mass ratios, i.e. lower star formation
efficiencies, while the constant-\ML\ models fit lower $M_{\rm
vir}/\mst$ values, i.e. higher efficiencies.
In order to match up with the picture where galaxies have star
formation efficiencies varying with the stellar mass
(\citealt{Benson2000}, \citealt{2006MNRAS.368....2D}), we have
shown that the DM density characteristics should change with the
mass with low mass systems being surrounded by more ``cored''
haloes (well approximated by the constant-\ML\ models) and high
mass systems by ``cusped'' haloes (here reproduced by the
S\'ersic+SIS profile).

This {\em DM non-homology} could be a possible explanation of the
``anomalously'' low halo concentration parameters recently found
modeling intermediate luminosity galaxies, compared to the giant
ellipticals showing ``regular'' concentration as expected from the
$\Lambda$CDM simulations (\citealt{Nap09b}). In this respect a
model like the Einasto profile (\citealt{Einasto65}, but see also
\citealt{Navarro04}, 2008, \citealt{Cardone05},
\citealt{Graham06}) or a phenomenological model including a wide
range of innermost density slopes (\citealt{Tortora2007}) provides
suitable working hypotheses to test on larger data sample with
extended kinematics (e.g. Atlas3D\footnote{ {\tt
http://www-astro.physics.ox.ac.uk/atlas3d/}}; or the PN.S
Elliptical Galaxy Survey:
\citealt{D+07,Nap08,Coccato09}).

\section*{Acknowledgments}
We thank the anonymous referee for his/her kind report. We also
thank Michele Cappellari for fruitful discussions and Claudia
Maraston for providing us with her synthetic spectral models. NRN
has been funded by CORDIS within FP6 with a Marie Curie European
Reintegration Grant, contr. n. MERG-FP6-CT-2005-014774, co-funded
by INAF.
AJR was supported by the National Science Foundation Grants
AST-0507729 and AST-0808099.

\appendix
\section{Systematic effects in the stellar populations models}
\label{sec:appA}

\begin{figure*}
\epsfig{file=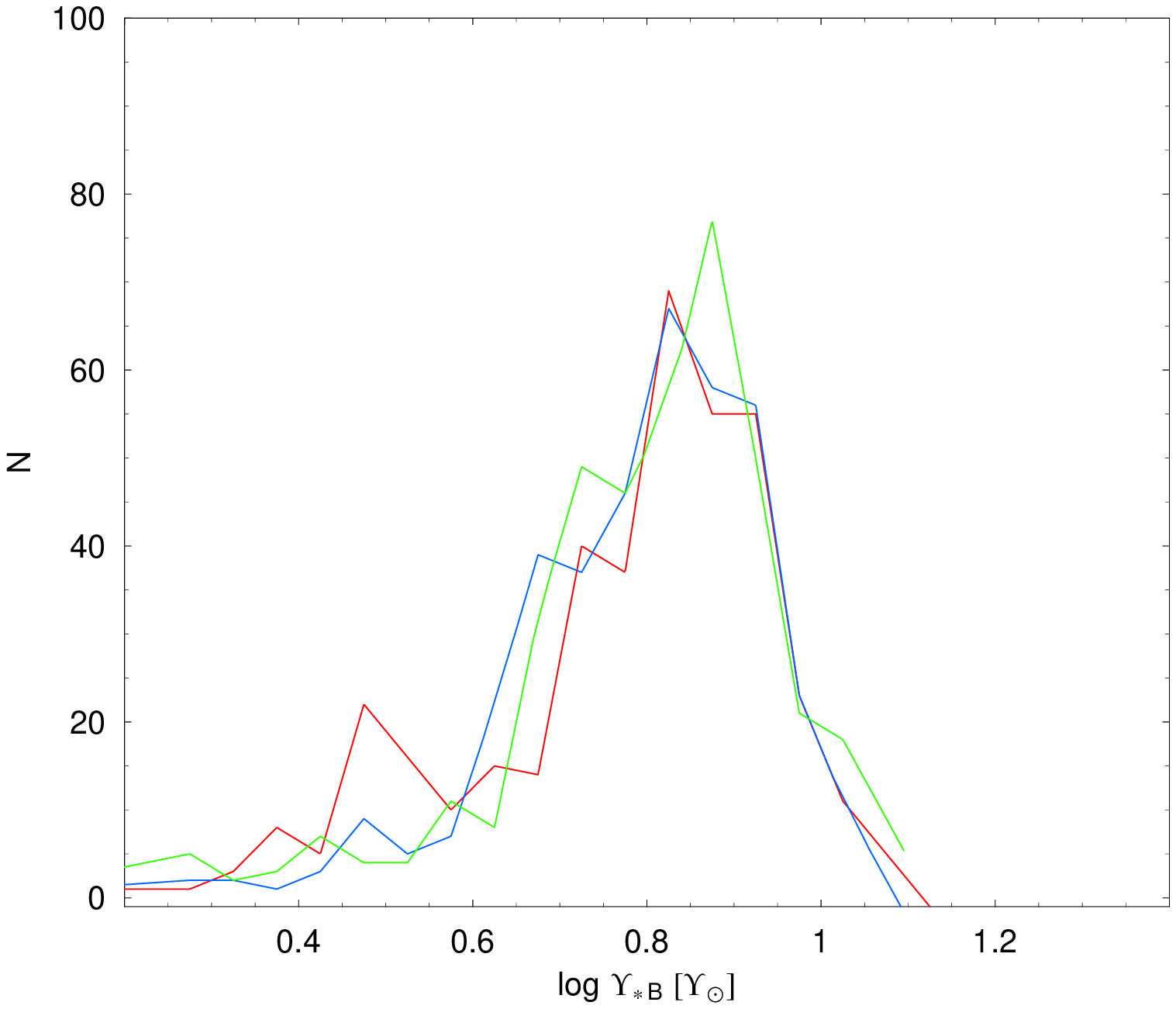, width=0.4\textwidth}
\epsfig{file=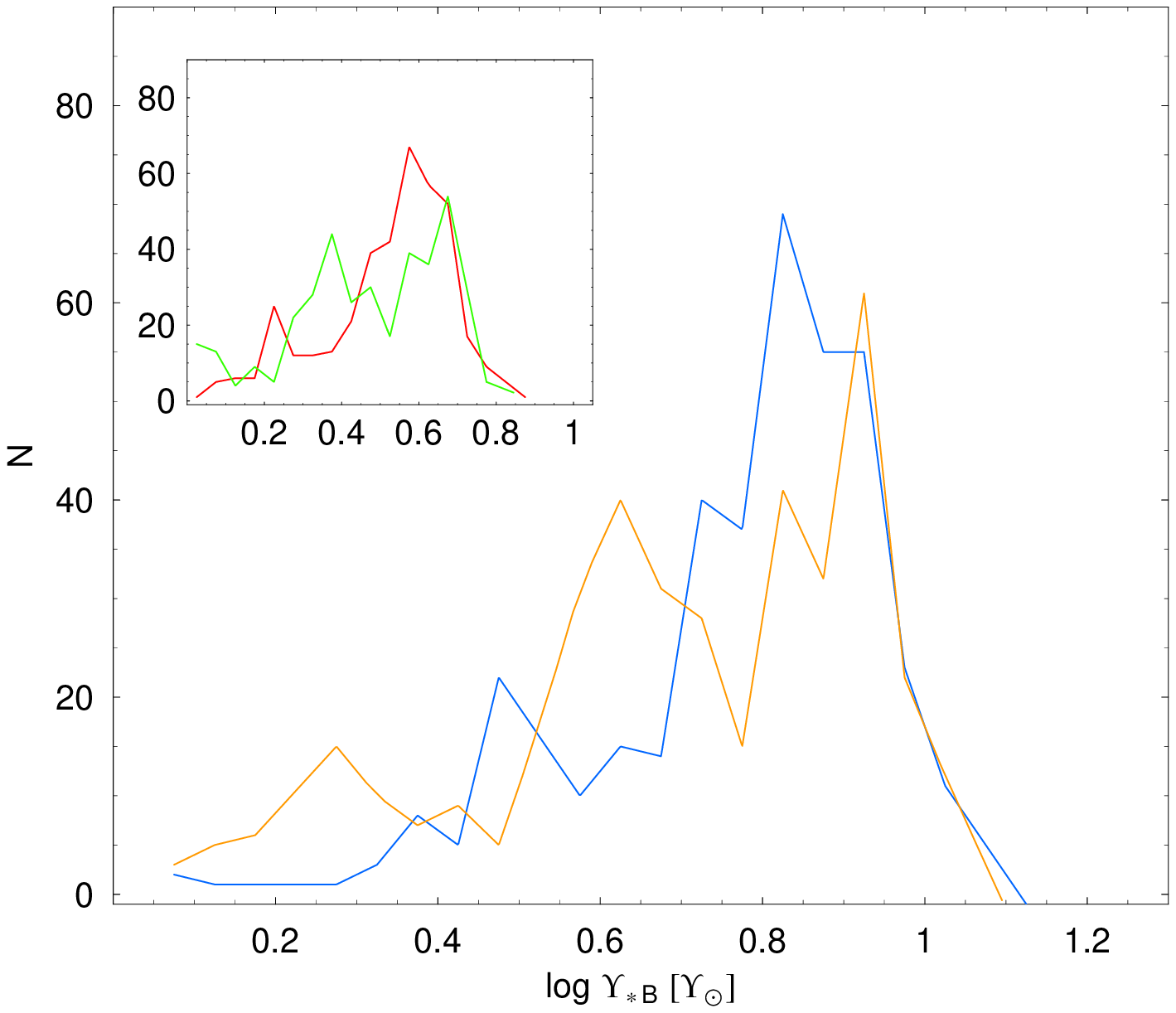, width=0.4\textwidth}\\
\vspace{0.5cm} \epsfig{file=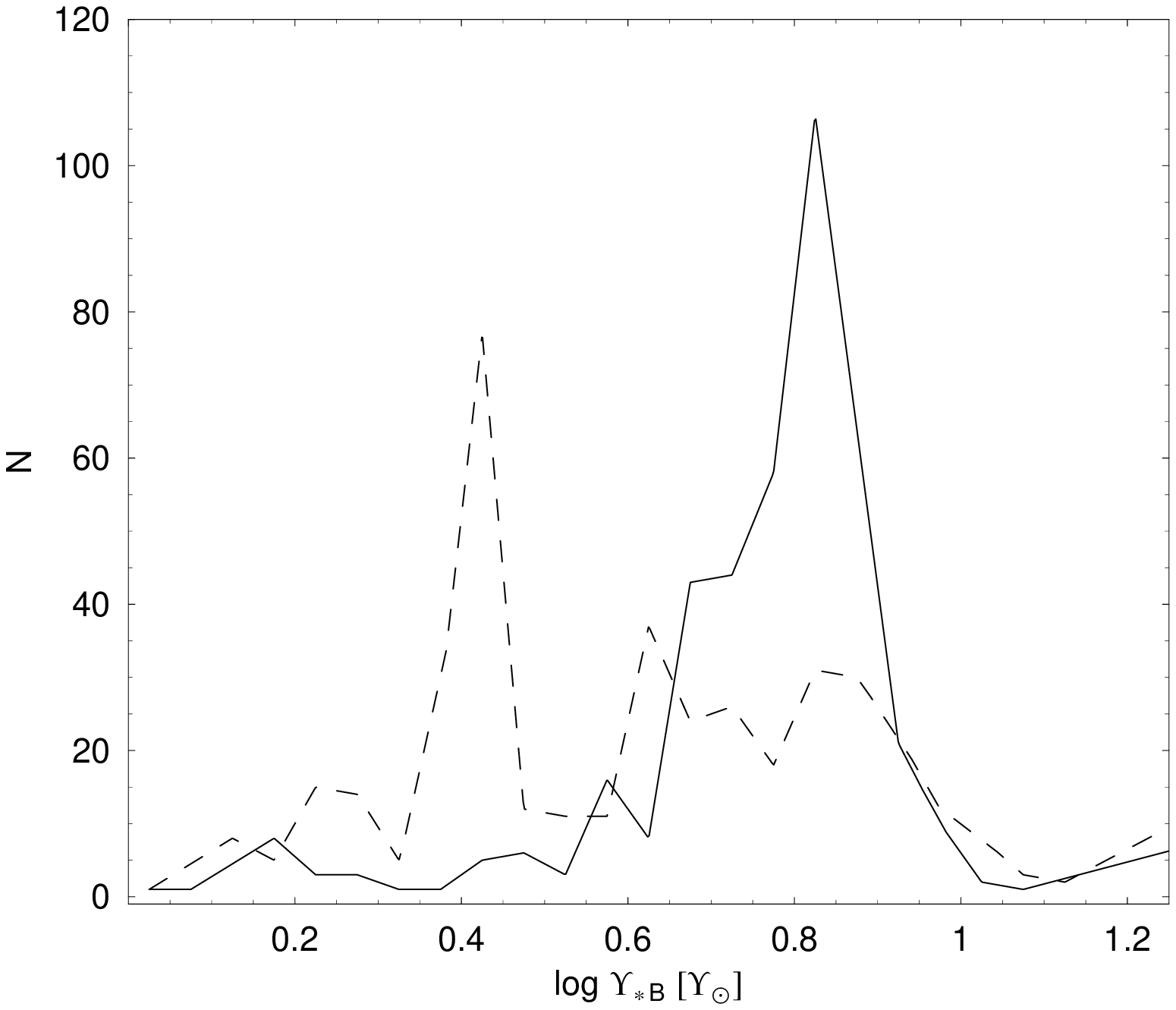,
width=0.39\textwidth}\hspace{0.5cm} \epsfig{file=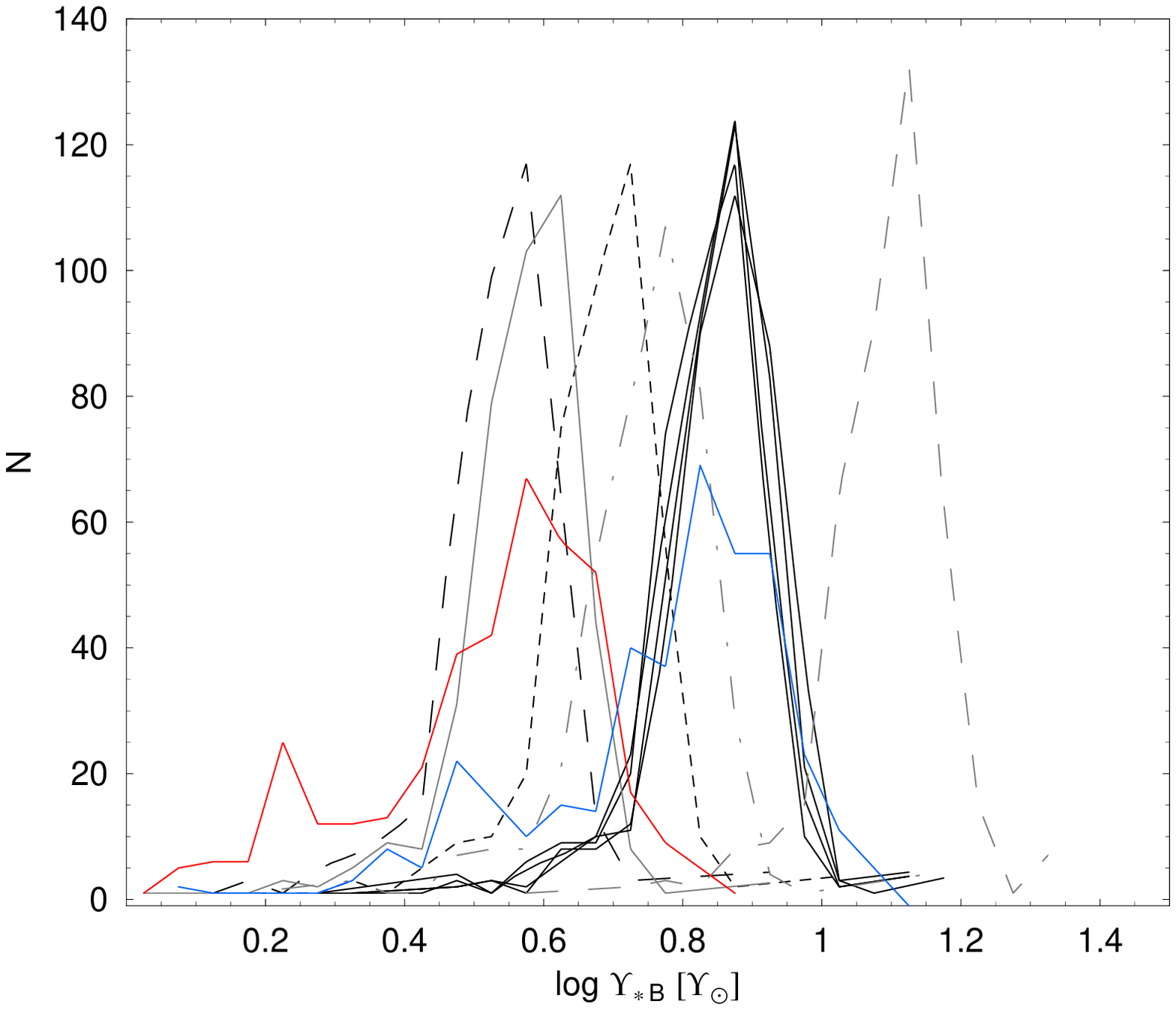,
width=0.4\textwidth}\hspace{0.5cm} \caption{Distributions of
recovered $\YstB$ from PS96 galaxy sample, for different stellar
populations models and assumptions. {\it Top left}: BC03+Salpeter
model with different parameter assumptions: $\tau = 0$ and
$Z=\Zsun$ (red); $\tau = 0.75 \, \rm Gyr$ and $Z=\Zsun$ (blue);
and $\tau$ and $Z$ free to vary (green). {\it Top right}: BC03 and
M05 models compared, assuming $\tau=0$ and $Z=\Zsun$. The main
panel shows a Salpeter IMF, with BC03 in blue and M05 in orange;
the inset panel shows a lower-mass IMF, with BC03+Chabrier in red,
and M05+Kroupa in green. {\it Bottom left}: BC03 and M05 models
compared (solid and dashed lines, respectively), with $Z$ free to
vary. {\it Bottom right}: BC03 and BJ01 models compared, assuming
$\tau=0$ and $Z=\Zsun$. The solid black lines show BJ01, and blue
shows BC03, in both cases with a Salpeter IMF. Lower-mass IMFs are
also shown: red is BC03 with Chabrier IMF, short-dashed black is
BJ01 with ``scaled'' Salpeter IMF, long-dashed black is BJ01 with
``modified'' Salpeter IMF, solid dark grey is BJ01 with Scalo IMF,
dashed light grey is BJ01 with ``top-light'' IMF slope, and
dot-dashed light grey is BdJ01 with ``top-heavy'' IMF slope (the
latter two cases using the PEGASE prescription). }
\label{fig:Distr_Upsilon}
\end{figure*}

\begin{figure*}
\psfig{file= 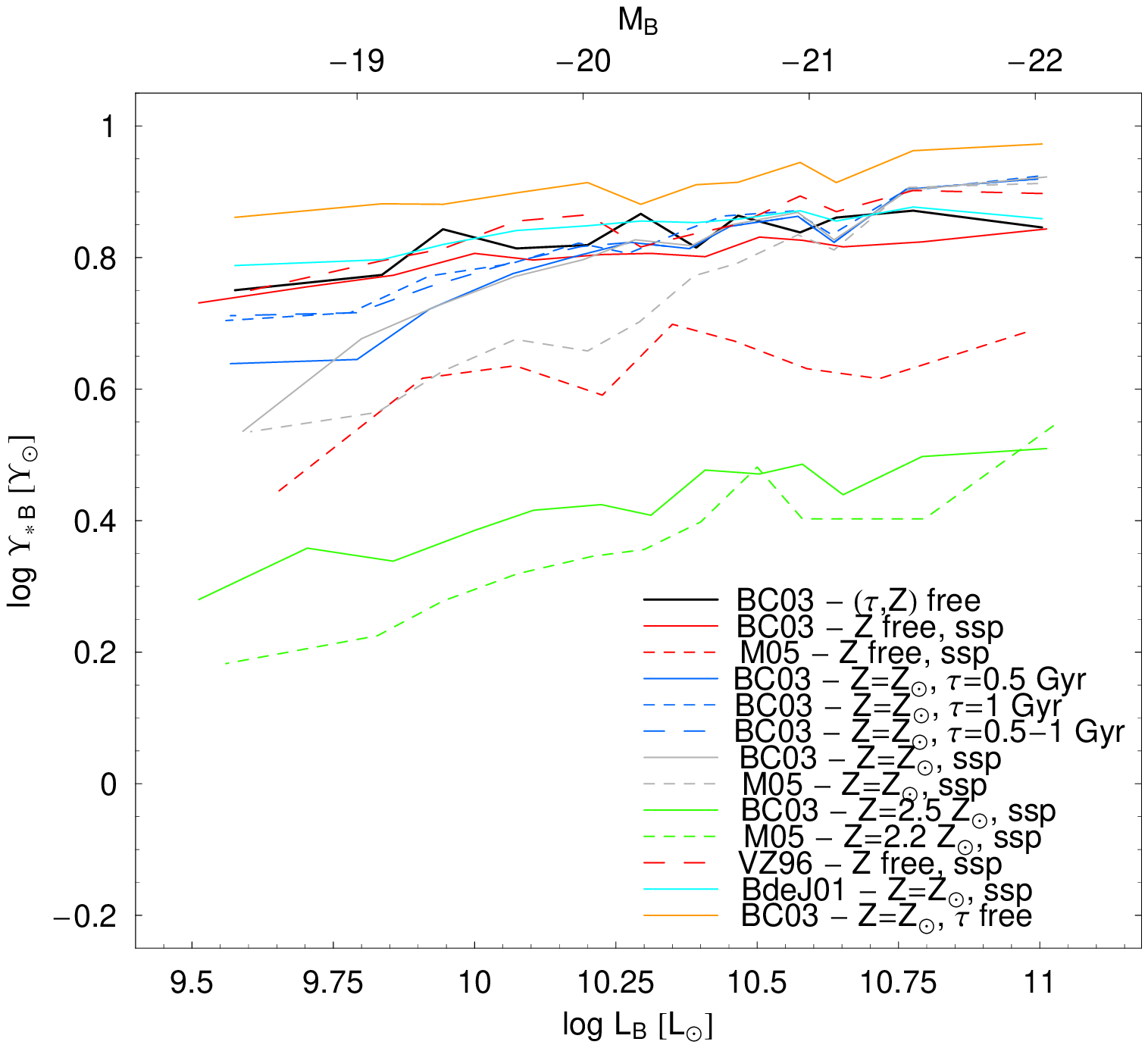, width=0.60\textwidth} \psfig{file=
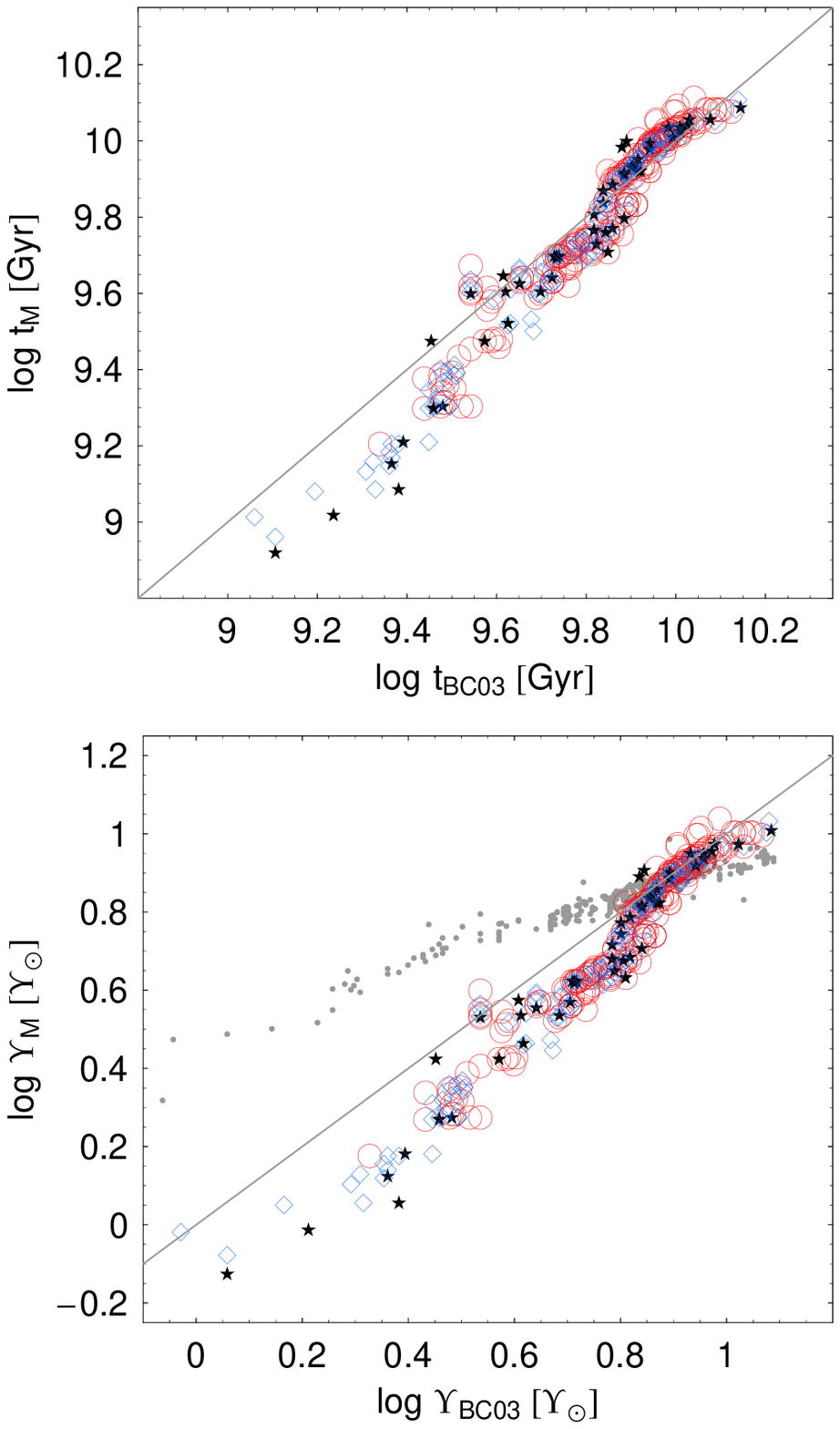, width=0.29\textwidth} \caption{{\it Left panel}:
Median binned stellar \ML\ as a function of luminosity, for our
ETG sample and under various modelling assumptions. Different
stellar populations prescriptions are shown (VZ96; BJ01; BC03;
M05), as well as different parameter assumptions -- see figure
legend for details. {\it Right panels}: Comparisons of model
parameters derived from the same data using different basis
populations models. The top right panel shows the age, and the
bottom right is stellar \ML, with the horizontal axes providing
the results from BC03 models, and the vertical axes M05 results --
in both cases assuming a Salpeter IMF. The red circles are E
galaxies, blue diamonds are S0s, and black stars are galaxies of
other classifications not used in our main study. Solid lines show
the one-to-one consistency relation. Grey squares in the bottom
right panel show BJ01 modelling results.
}\label{fig:Distr_Upsilon_2}
\end{figure*}

Here we examine the role of systematic uncertainties in the
stellar populations results, using different assumptions and basis
models (see, also, similar analysis in \cite{Rettura06},
\cite{KG07} and \cite{Conroy08}). First we consider our default
model based on BC03, using three different parameterizations for
the SFH. In our reference model, $\tau$ and $Z$ (as well as \age)
are free parameters fitted to each galaxy; a more simplified model
has fixed $\tau=1$~Gyr and $Z=Z_\odot$ corresponding to typical
values for the whole sample; an even simpler SSP model has
$\tau=0$ and $Z=Z_\odot$. As shown in Fig.~\ref{fig:Distr_Upsilon}
(left panel), the distributions of \Yst\ in all these models are
fairly similar, except in the simplest case which shows a stronger
tail to low values of \Yst. The impact of these differences is
shown in Fig.~\ref{fig:Distr_Upsilon_2} (left panel), where it can
be seen that overly restrictive modelling assumptions compensate
with large variations in \Yst\ and thus steeper values for
$\gamma_*$.

Next we compare basis model variations, starting with BC03 and
M05. The different input stellar models and treatments of the
thermally-pulsing asymptotic giant branch phase (TP-AGB) lead to
different \Yst\ predictions for the same colours
\citep{Maraston06}\footnote{Recent preliminary updates of the BC03
models have included an improved TP-AGB treatment
\citep{Bruzual07, Eminian2008}. The colours and \Yst\ predictions
are altered, particularly in the near-infrared, but not
substantially for $Z \geq 0.4$~$Z_\odot$. These new models are
more similar to M05 than BC03 but for old ages resemble BC03
\citep{McGrath07}.}. Comparisons of some parameters derived from
the two models with solar metallicity, given the same colour data,
are made in Fig.~\ref{fig:Distr_Upsilon_2} (right panels). The
inferred ages agree very well for older populations, with M05
returning ages up to $\sim$~10\% higher than BC03, while for
younger populations, up to $\sim$~30\% lower. The difference stems
from M05 predicting $V-R$ and $V-I$ to be redder for young
populations and bluer for old, while $B-V$ is redder for all ages.
The implications for \Yst\ are that agreement is good for $\Yst
\gsim \, 6$~\Ysun, while for lower values the M05 predictions are
smaller by up to a factor of two. The extension of the M05 results
to smaller values of \Yst\ can also be seen in
Fig.~\ref{fig:Distr_Upsilon}. The final impact of these systematic
differences on the trend of \Yst\ with luminosity is shown in
Fig.~\ref{fig:Distr_Upsilon_2} (left panel): M05 yields a more
steeply increasing trend.

We finally examine the SSP models from BJ01. BJ01 predict a tight
correlation  between \Yst\ and galaxy colours\footnote{This was
obtained for spiral galaxies but it has been proven to work for
ETGs as well (BJ01, \citealt{Bell2003})}: using relations for the
three colours $B-V$, $B-R$ and $V-I$ and minimizing a $\chi^{2}$
function we determine the best fitted \Yst\ for the BdJ01
pescriptions which are shown in
Fig. \ref{fig:Distr_Upsilon} (left panel).\\
Assuming a Salpeter IMF, the use of different stellar
prescriptions (BC03 vs BdJ01) has a negligible effect on the bulk
of the \Yst\ distribution (e.g. solid black lines and blue ones in
the Figure). Some of the assumed IMFs in BdJ01 (``scaled'' and
``modified'' Salpeter and the \citealt{Scalo86}, for further
details see BJ01) predict lower \Yst, with the Scaled Salpeter and
Scalo IMFs giving similar results of the BC03+Chabrier one (red
curves in the same Figure). Finally, using BJ01 results for PEGASE
(\citealt{FR97}) prescription we obtain that: 1) a top-heavy IMF
with a slope $-0.85$ gives \Yst\ values which are in the between
of the Kroupa (2001) IMF (or Chabrier or Scalo) and Salpeter IMF
predictions, while 2) a top-light IMF with a slope $-1.85$ give
much larger \Yst\ values. Note that distributions using directly
BC03 (red and blue lines in left panel of Fig.
\ref{fig:Distr_Upsilon})
have a larger spread around the peak distribution than the BdJ01 results.\\

BJ01 results have been plotted in middle panel of Fig.
\ref{fig:Distr_Upsilon_2} like grey points. The slope of the
relation shown in this figure is unchanged if we use the various
prescriptions analyzed in the paper above (see distributions using
a Salpeter IMF in left panel of Fig. \ref{fig:Distr_Upsilon}); on
the contrary, a little change in the zero point is observed.

As a final test, we compare results using different models {\it
and data} on the same galaxies. As a stellar synthesis model, C+06
fit single burst models (using stellar prescription in
\citealt{V96}), to some line-strength indices. Their estimates are
on average $20 \%$ larger than ours (with a scatter of $17\%$).
This discrepancy could not be fixed by fitting \cite{V96} or BC03
SSP models to our galaxy colours. Some systematics can be
ascribed, partially, to the extrapolation of line-strength indices
(and velocity dispersion) from the very central regions to the
effective radius, if some change in the average stellar population
is present and unaccounted.

\section{Independent checks on dynamical masses}
\label{sec:appB}

Given the simplifications of our Jeans models used to derive the
dynamical masses (spherical quasi-isotropic models), we test here
using independent results whether our methods have introduced any
systematic bias for \Ydyn. Our first cross-check is with C+06, who
constructed detailed two-dimensional models of nearby ETGs. Our
sample has 18 galaxies in common with theirs\footnote{Another 7
from their sample did not have measured $B-I$ colours available
for making a proper comparison between their $I$-band and our
$B$-band \Ydyn\ results.}. The main differences between the two
datasets are: 1) our distance moduli are on average larger ($0.05$
mag) than C+06 but consistent within the scatter; 2) our effective
radii are on average $5\%$ larger with a median scatter of $16\%$;
3) the central velocity dispersions from C+06 are lower than the
PS96 values by $6 \pm 15 \, \rm km s^{-1}$ (see
\citealt{Emsellem04}).

\begin{figure*}
\centering
\psfig{file=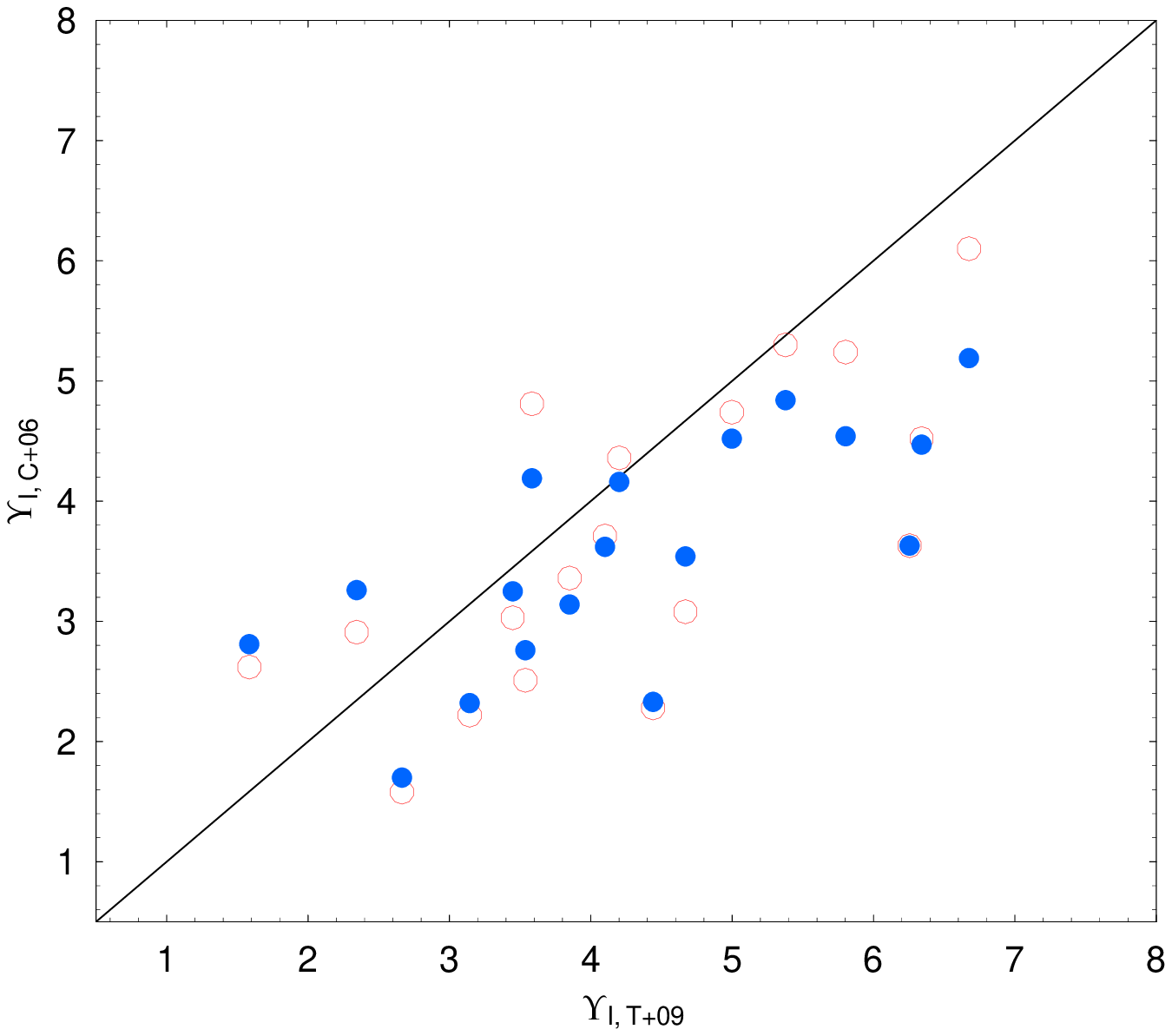,width=0.47\textwidth}\hspace{0.4cm}
\psfig{file=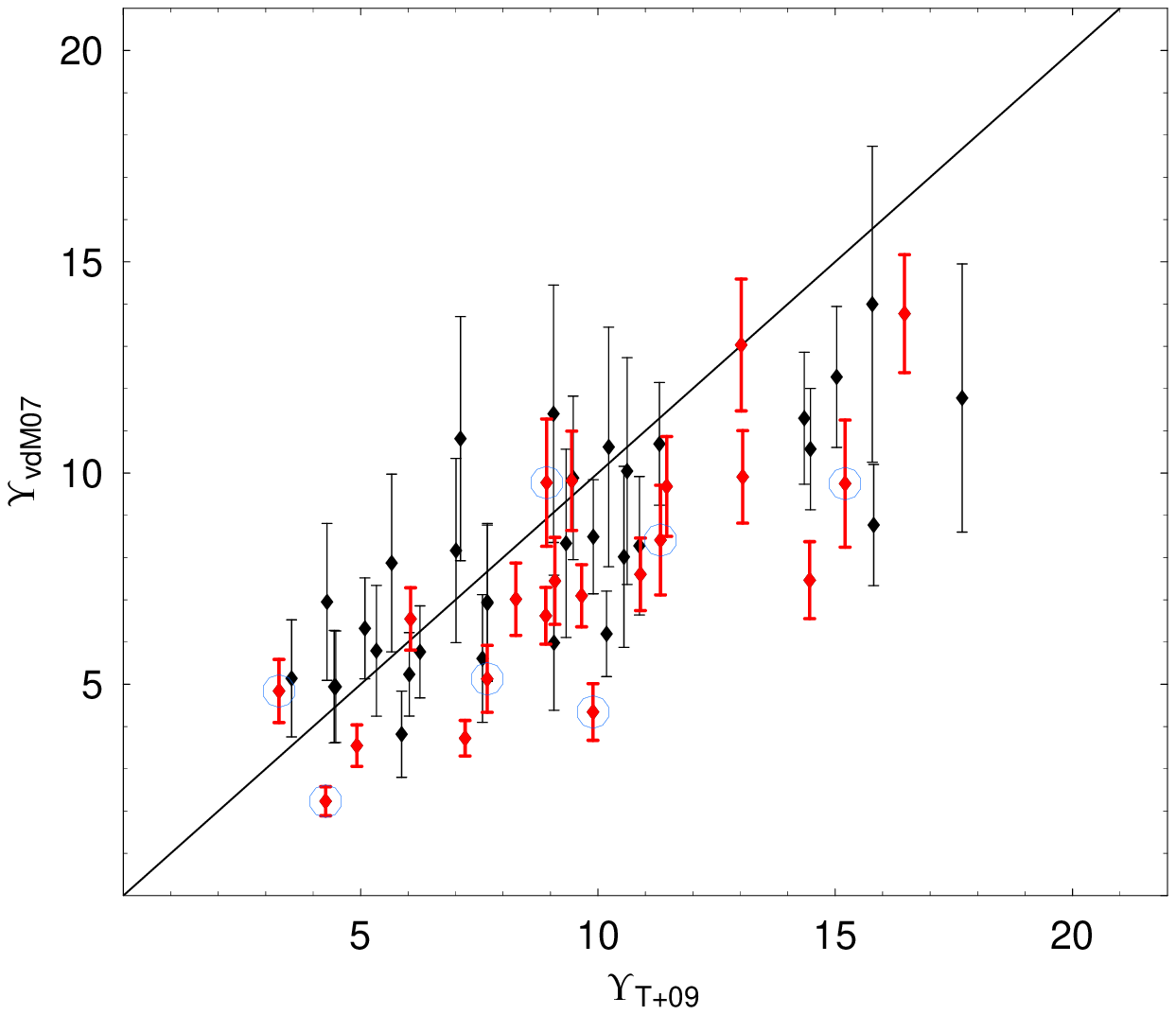,width=0.47\textwidth} \caption{Comparison
of dynamical \ML\ estimates from our models (horizontal axes) and
those from the literature (vertical axes). The diagonal lines show
the one-to-one relation. {\it Left panel:} Comparison to C+06
results in the $I$-band. Open red and filled blue symbols are for
three-integral Schwarzschild and two-integral Jeans dynamical
models respectively. {\it Right panel:} Comparison to MD07 results
in the $B$-band. Red symbols represent galaxies with results
including C+06; blue circles show those based solely on also data
from C+06, while those with blue circles use only results from
C+06.}\label{fig:Comp_Cap06}
\end{figure*}

C+06 constructed flattened, axisymmetric, constant-\ML\ dynamical
models, using both two-integral Jeans models and three-integral
\citet{Schwarzschild79} orbit models. Their luminosity models are
multi-Gaussian expansions of the observed surface brightness
profiles, and thus quite non-homologous. Converting our
constant-\ML\ S\'ersic-based \Ydyn\ results to the $I$-band and to
the C+06 distances, we compare to their results in
Fig.~\ref{fig:Comp_Cap06}. The masses are broadly consistent, with
a systematic trend for ours to be higher by $\sim$~20\%.

There are several possible reasons for this residual discrepancy,
including rotation, orbital anisotropy variations, and galaxy
flattening -- all of which were handled in rigorous detail by C+06
but not by our models. Based on the results of
\citet{Cappellari07}, the anisotropy effect should not correlate
strongly with luminosity, but rotation and flattening probably do.
We also compare our {\it modelled} values of $\sigma_e$ (the
velocity dispersion integrated over a 1~\Re\ aperture, folding in
the rotational contribution) with their {\it observed} values, to
see if our extrapolation from the central aperture could be
generating the discrepancy. However, our $\sigma_e$ values turn
out to be {\it lower} by $13^{+9}_{-8}\, \rm km \, s^{-1}$, which
goes the wrong way to explain our {\it higher} masses.

Next we consider the detailed spherical dynamical models of G+01,
with 16 galaxies in common.  After shifting to the same distance
scale, our \Ydyn\ values at \Re\ are $27\%\pm8\%$ {\it lower} on
average than theirs. Since their sample was focussed on round
galaxies, we suspect again that flattening is playing a key role
in the accuracy of our results, but that we have been able to
largely compensate for its effects in our simplified modelling.

Finally we turn to the dynamical results of \citet[hereafter
MD07]{MD07}, who compiled \Ydyn\ for 60 local galaxies from the
literature (\citealt{vdM91,Magorrian98,K00,Gebhardt03}; C+06). The
original works made use of various types and quality of data and
dynamical models, but should in general be superior to ours. The
\Ydyn\ values are combined after homogenizing the distances and
cosmology, and converting to the $B$-band. As shown in Fig.
\ref{fig:Comp_Cap06} (right panel), there is good agreement in
general, but again a tendency for our results to overestimate the
masses by $\sim20\%$, which appears to be an effect of the
fainter, flatter galaxies. Note that our extrapolated $\sigma_e$
shows no systematic offset from MD07. In order to potentially
correct for a systematic error in our dynamical modelling, we
adopt a heuristic correction of $66\%$ and $90\%$ for faint and
bright galaxies separately.

\section{Systematic uncertainties for dark matter fraction}
\label{sec:appC}

\begin{figure*}
\centering \psfig{file=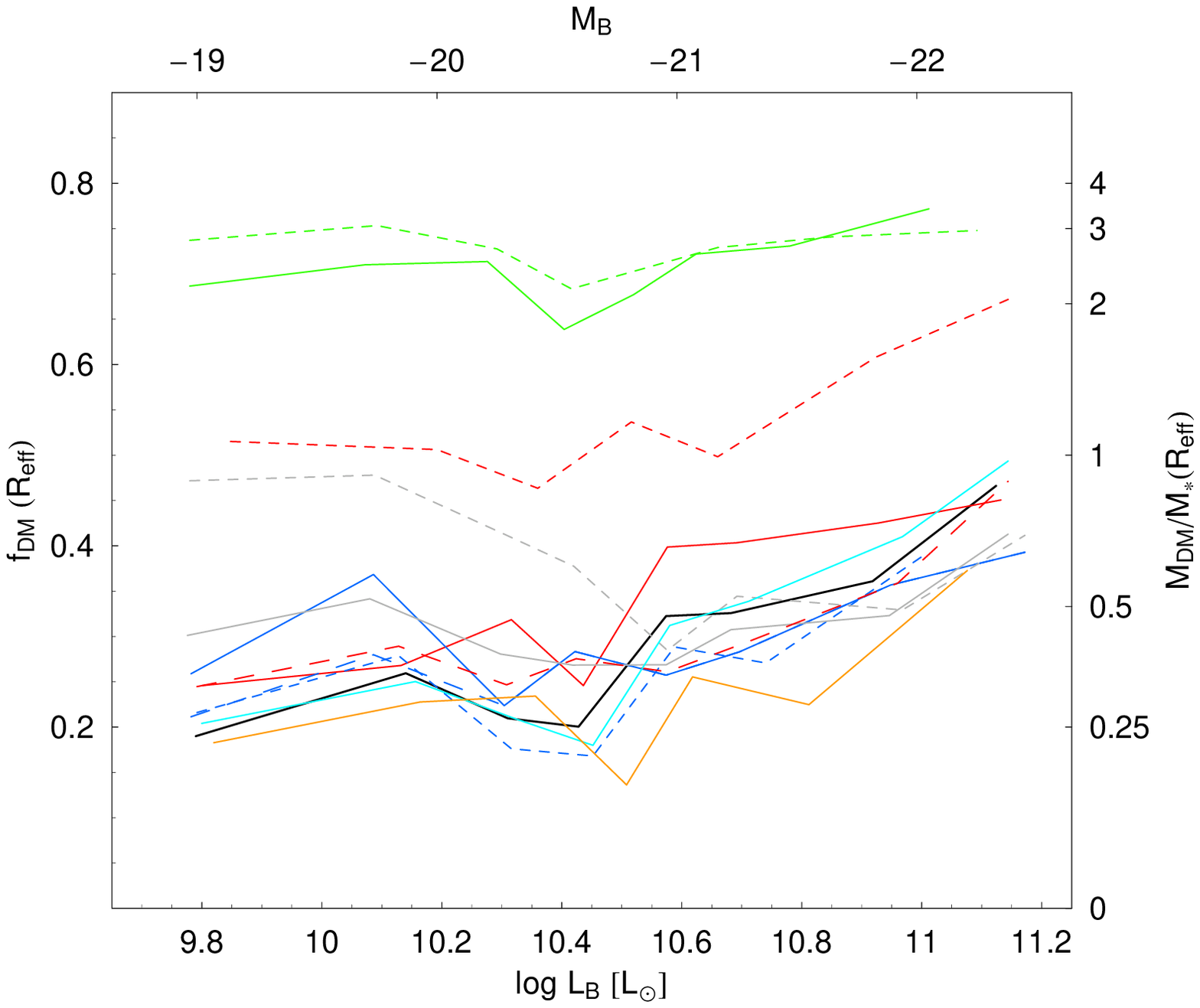,
width=0.47\textwidth}\hspace{0.4cm}
\psfig{file=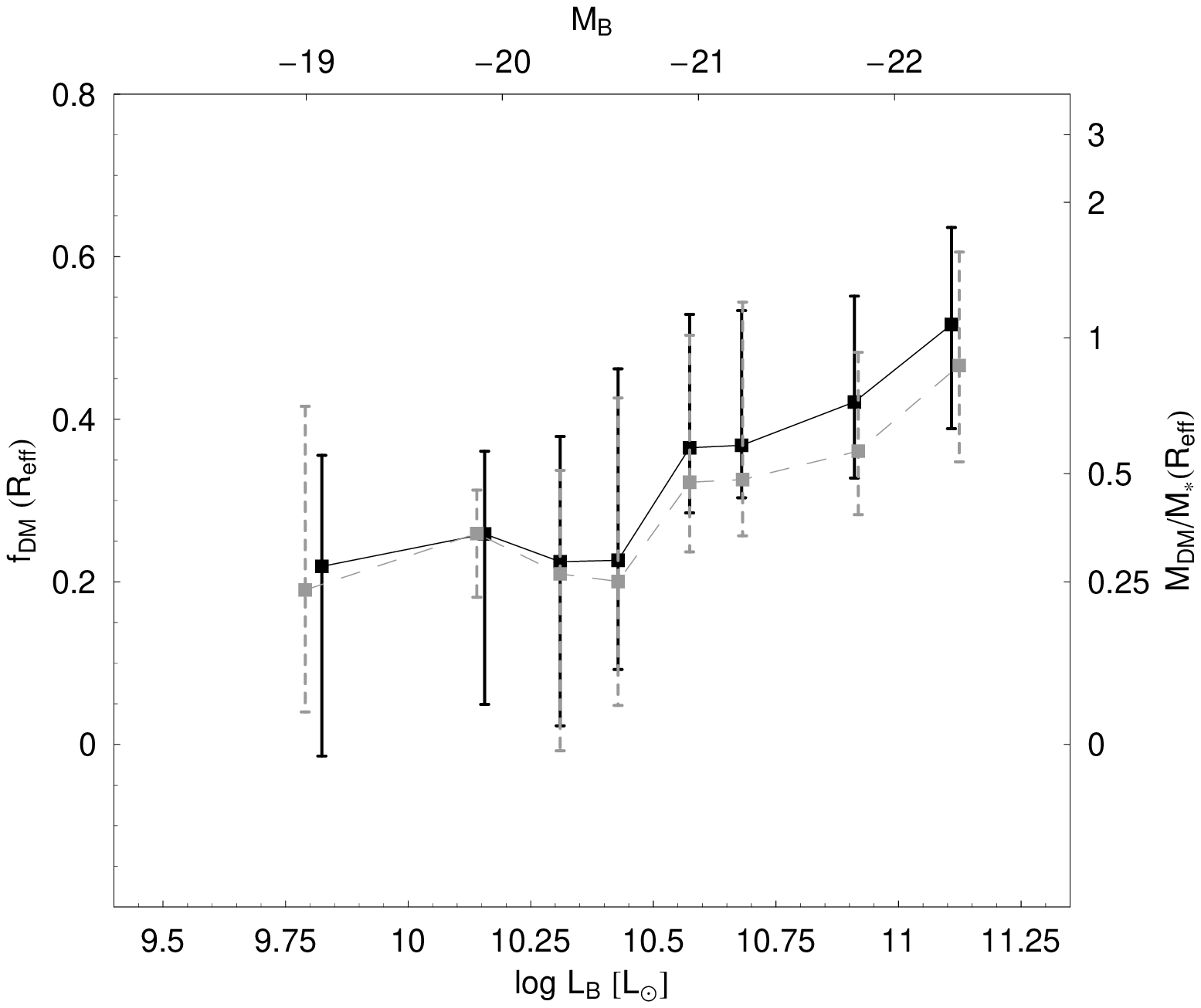, width=0.47\textwidth} \\
\psfig{file=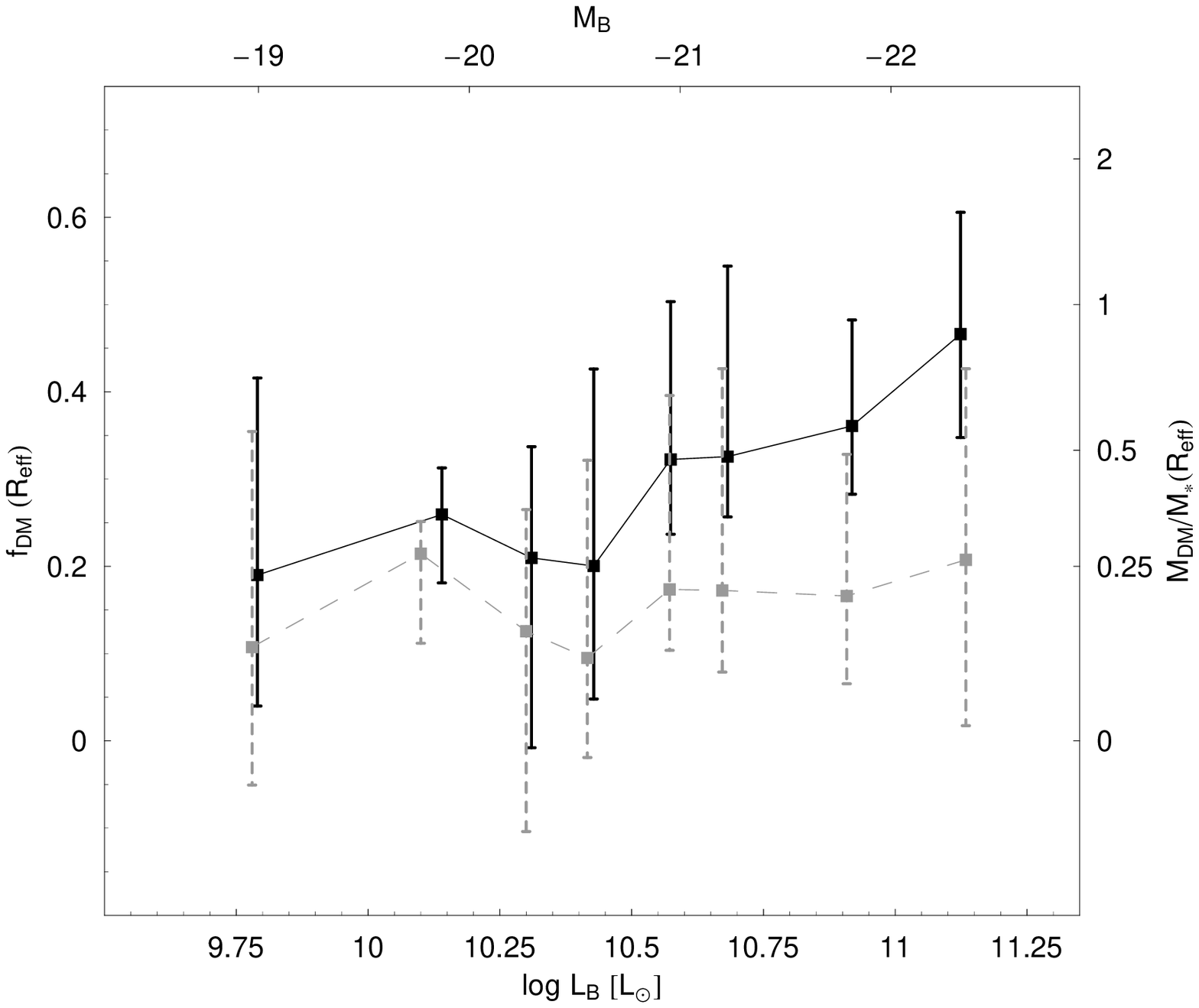, width=0.47\textwidth}\hspace{0.4cm}
\psfig{file=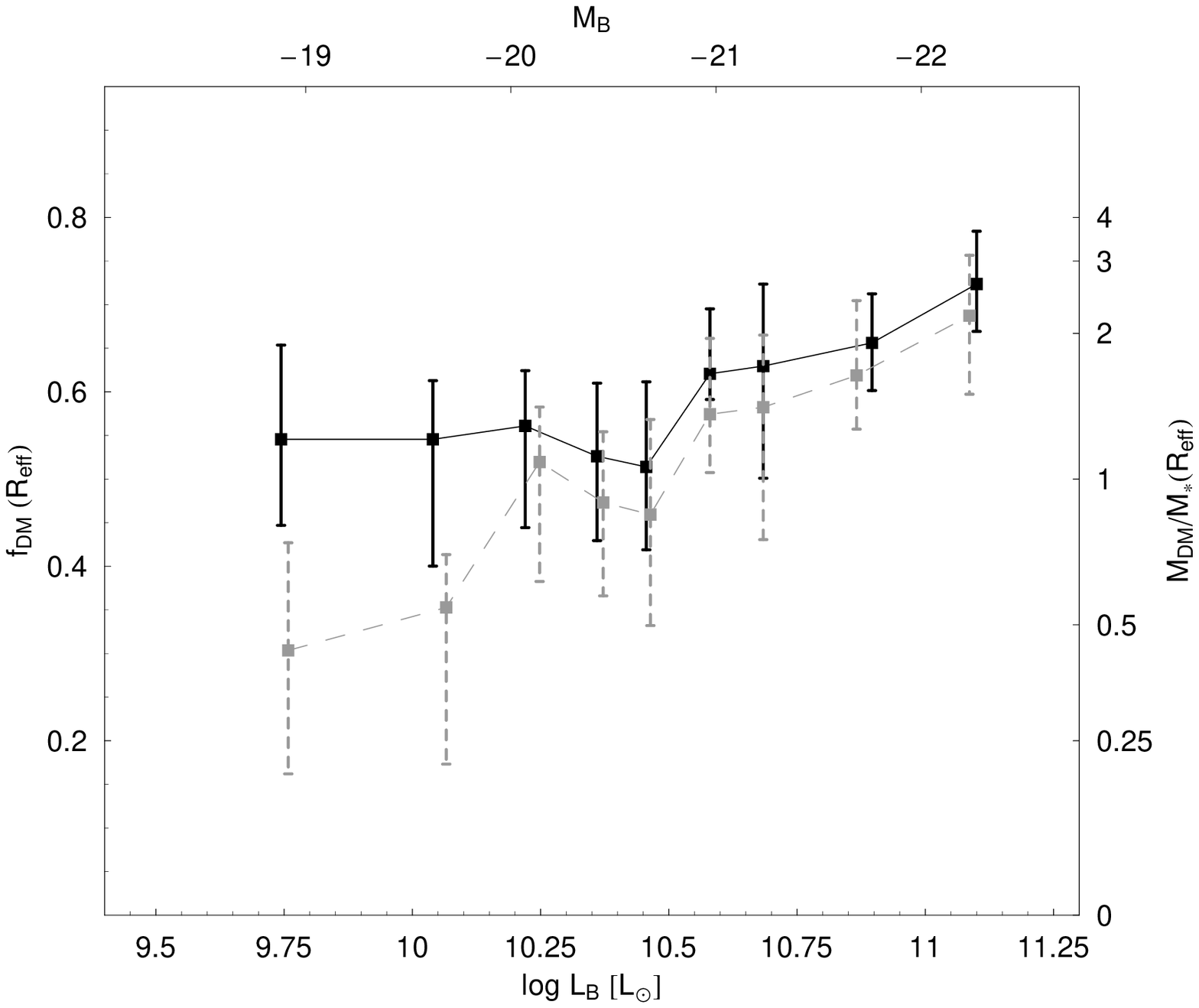, width=0.47\textwidth} \caption{Effects of
systematic modelling uncertainties on DM fractions, for the
overall EG sample. Unless otherwise stated, the mass model is
constant-\ML, and the IMF is Salpeter. {\it Top left panel:}
Changing the stellar populations basis model (see left panel in
Fig.~\ref{fig:Distr_Upsilon_2} for line definitions). {\it Top
right panel:} Changing the dynamical mass model from SIS (black)
to constant-\ML\ (grey). {\it Bottom left panel:} Changing the
S\'ersic index $n$ in the dynamical modelling (black: PS97 values;
grey: \citet{Caon93} values). {\it Bottom right panel:}
Calibrating the dynamical models using MD07 (black: original;
grey: recalibrated). Here a Chabrier IMF is used to avoid negative
\fDM\ values. } \label{fig:fDM_2}
\end{figure*}

We consider finally how various systematic uncertainties could
impact the \fDM\ determinations. We first consider the stellar
populations models. As detailed in Appendix~\ref{sec:appA}, the
model prescription that is used can have a noticeable effect on
the \Yst\ trends. We show in Fig.~\ref{fig:fDM_2} (upper left
panel) the differences engendered in \fDM\ by adopting different
models.  Among the most plausible models, the results are roughly
consistent, but the trend of \fDM\ with luminosity tends to
flatten or steepen with the use of M05 or BJ01, respectively,
rather than BC03.

We next consider uncertainties in the dynamical models, starting
with the assumed mass profile.  As shown in Fig.~\ref{fig:fDM_2}
(upper right panel), the bracketing models of SIS and
constant-\ML\ produce similar results for \fDM. Testing the
possibility that our $n=4$ S\'ersic index for the bright Es is
inaccurate, we alternatively take the higher $n-M_B$ relation from
\citet{Caon93} as reported in PS97, and find that in a
constant-\ML\ case, \Ydyn\ for the brighter galaxies decreases and
the overall \fDM\ trend is constant with luminosity
(Fig.~\ref{fig:fDM_2}, bottom left panel). However, an SIS profile
is probably a better match for these galaxies, and in this case,
changing $n$ would not affect the results. Finally, we try to
calibrate out the inaccuracies in our simplified Jeans modelling,
based on the MD07 results, and find that the \Ydyn\ values for the
fainter galaxies might actually be lower, and the \fDM\ slope with
luminosity therefore steeper (Fig.~\ref{fig:fDM_2}, bottom right
panel).

To quantify the effect of ellipticity ($\epsilon$) on our
estimates, we have also selected E galaxies with $\epsilon < 0.3$
(as derived by RC3). For these systems, the results are still
consistent with an increasing (flat) trend of \fDM\ with
luminosity for bright (faint) galaxies.

\begin{figure}
\centering
\psfig{file=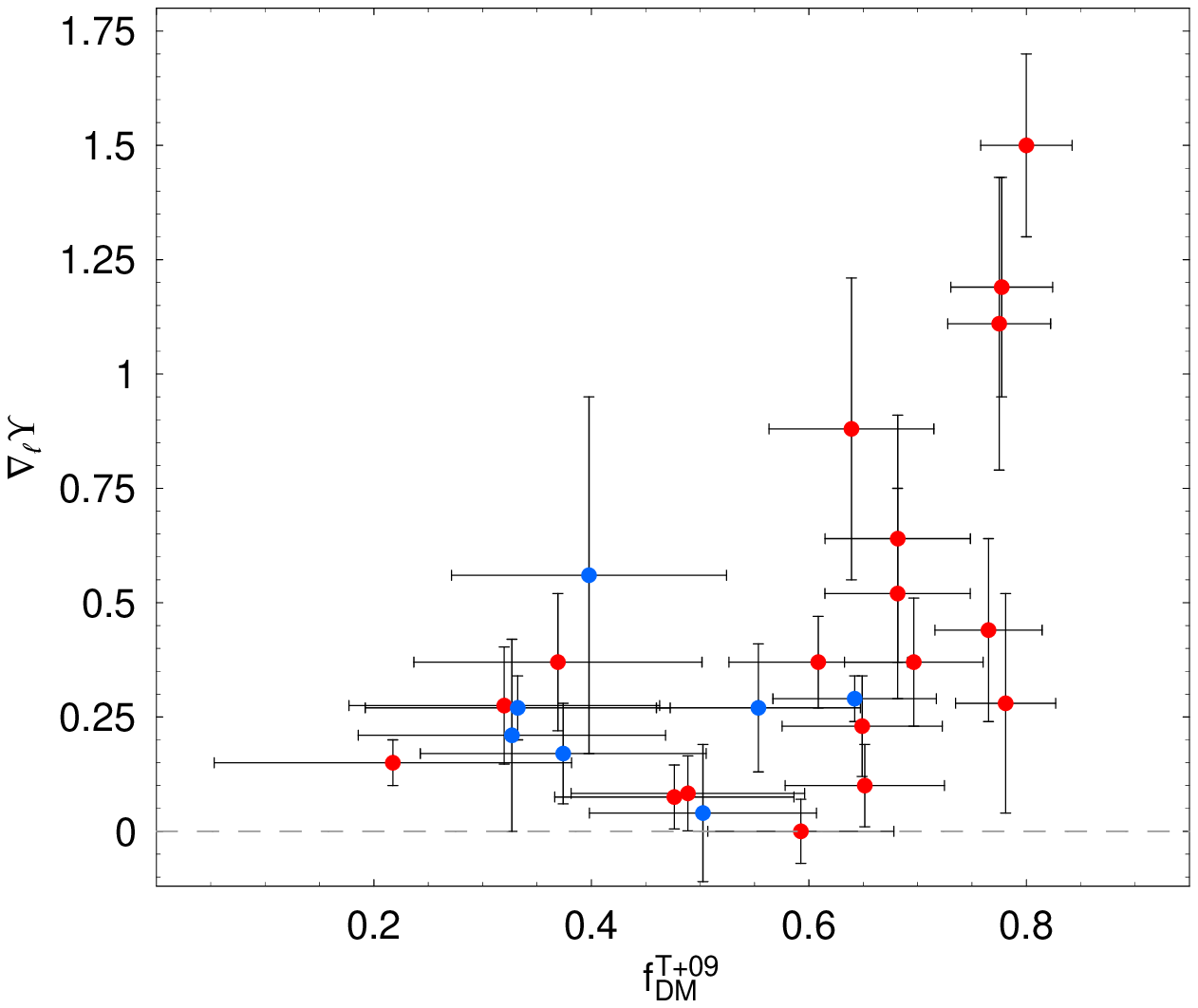, width=0.45\textwidth}\\
\caption{\ML-gradient parameter based on extended dynamics,
compared to central DM fraction (default SIS model with Chabrier
IMF). Red and blue dots are E and S0 galaxies. Most of the data
are taken from N+05, with several updates and additions from more
recent literature
(\citealt{Teodorescu05,Schuberth06,D+07,2008MNRAS.383.1343W,Nap08,R+09,Kumar09}).
} \label{fig:grad}
\end{figure}

In summary, there are several potential competing systematic
effects, and it is not clear which one might win out in biassing
the \fDM\ slope. Given this uncertainty, we carry out a different,
critical test of confidence in our results.  Finding results in
the literature for the mass content of galaxies in our sample {\it
at large radii}, we construct the \ML-gradient parameter \dML\
introduced by N+05. This simple but powerful metric is calculated
from dynamical measurements of \ML\ at inner and outer radii by
the following formula:
\begin{equation}
\dML\ = \frac{\Re}{R_{\rm out}-R_{\rm in}}\left(\frac{\Upsilon_{\rm out}}{\Upsilon_{\rm in}} - 1\right) .
\end{equation}
Given the longer lever arm, \dML\ when available tells us with
greater security whether or not an object is rich or poor in
DM\footnote{No attempt is made here to decompose the \ML\
measurements into stars and DM, i.e. to determine \fDM. Instead,
the broad premise is that \Ydyn\ increases more rapidly with
radius in galaxies with higher \fDM.}. We compare \fDM\ and \dML\
in Fig.~\ref{fig:grad}, and confirm that high-\fDM\ objects from
the current paper generally have high halo DM content in the
literature while low-\fDM\ have small \dML\, consistent with a
lower global DM content.

\end{document}